\documentclass[twocolumn]{aa} 

\usepackage{graphicx}
\usepackage[varg]{txfonts}
\begin{document} 

\title{Formation and dynamics of water clouds on temperate sub-Neptunes: The example of K2-18b}

   \author{B. Charnay
          \inst{1}
          \and
          D. Blain\inst{1}
          \and
          B. B\'ezard\inst{1}
          \and
          J. Leconte\inst{2}       
          \and
          M. Turbet\inst{3}   
          \and
          A. Falco\inst{2}                                   
          }

   \institute{LESIA, Observatoire de Paris, Universit\'e PSL, CNRS, Sorbonne Universit\'e, Universit\'e de Paris, 5 Place Jules Janssen, 92195 Meudon, France. \\
              \email{benjamin.charnay@obspm.fr}
         \and
             Laboratoire d'astrophysique de Bordeaux, Univ. Bordeaux, CNRS, B18N, all\'ee Geoffroy Saint-Hilaire, 33615 Pessac, France
         \and
             Observatoire Astronomique de l'Universit{\'e} de Gen{\`e}ve, 51 chemin de P{\'e}gase, 1290 Sauverny, Switzerland
             }



\abstract
   {Hubble Space Telescope (HST) spectroscopic transit observations of the temperate sub-Neptune K2-18b were interpreted as the presence of water vapour with potential water clouds.
1D modelling studies also predict the formation of water clouds in K2-18b's atmosphere in some conditions. However, such models cannot predict the cloud cover, which is driven by atmospheric dynamics and thermal contrasts, and thus neither can they predict the real impact of clouds on spectra.}
   {The main goal of this study is to understand the formation, distribution, and observational consequences of water clouds on K2-18b and other temperate sub-Neptunes.}
   {We simulated the atmospheric dynamics, water cloud formation, and spectra of K2-18b for a H$_2$-dominated atmosphere using a 3D general circulation model (GCM). We analysed the impact of atmospheric composition (with metallicity from 1$\times$solar to 1000$\times$solar), concentration of cloud condensation nuclei, and planetary rotation rate.}
   {Assuming that K2-18b has a synchronous rotation, we show that the atmospheric circulation in the upper atmosphere essentially corresponds to a symmetric day-to-night circulation with very efficient heat redistribution. This regime preferentially leads to cloud formation at the sub-stellar point or at the terminator. Clouds form at metallicity $\geqslant$100$\times$solar with relatively large particles (radius=30-450 $\mu$m). At 100-300$\times$solar metallicity, the cloud fraction at the terminators is small with a limited impact on transit spectra. At 1000$\times$solar metallicity, very thick clouds form at the terminator, greatly flattening the transit spectrum. 
The cloud distribution appears very sensitive to the concentration of cloud condensation nuclei and to the planetary rotation rate, although the impact on transit spectra is modest in the near-infrared.
Fitting HST transit data with our simulated spectra suggests a metallicity of $\sim$100-300$\times$solar, which is consistent with the mass-metallicity trend of giant planets in the Solar System.
In addition, we found that the cloud fraction at the terminator can be highly variable in some conditions, leading to a potential variability in transit spectra that is correlated with spectral windows. This effect could be common on cloudy exoplanets and could be detectable with multiple transit observations.
Finally, the complex cloud dynamics revealed in this study highlight the inherent 3D nature of clouds shaped by couplings between microphysics, radiation, and atmospheric circulation.
}
   {}


\maketitle

\section{Introduction}

Detection surveys revealed a high abundance of exoplanets with intermediate masses between the Earth and Neptune: these are called super-Earths and sub-Neptunes. Statistical analysis of Kepler data suggests a transition between these two populations at $\sim$1.8 R$_{Earth}$ \citep{fulton17}, which is compatible with models of photo-evaporation of H$_2$-dominated atmospheres surrounding rocky cores \citep{owen17, lehmer17}, or  a water world formation model as proposed by \cite{zeng19}. Following the observed trend in giant planets of the Solar System and a prediction of planetary formation model, one would expect the fraction of heavy elements (metallicity) in primary atmospheres to decrease with planetary mass \citep{kreidberg14b,fortney13,kral20}. Sub-Neptunes are thus expected to be enriched in heavy elements reaching typically 100-1000$\times$ solar metallicity. Measuring the atmospheric composition in particular, the water abundance of sub-Neptunes would place major constraints on planetary formation and evolution. Unfortunately, past transit spectroscopic observations of warm sub-Neptunes have been unsuccessful in measuring molecular abundances because of the presence of high and thick clouds or hazes \citep{kreidberg14a, knutson14a, benneke19a}.

The planet K2-18b is a moderately irradiated (1.06$\pm$0.06~$\times$ that of the insolation on Earth) sub-Neptune \citep{foreman-mackey15, montet15, benneke17}. It is on an orbit slightly inside the inner limit of the classical habitable zone  \citep{kopparapu14}. It is, however, probably in the habitable zone for slow synchronous rotating terrestrial planets around M-stars \footnote{To date, the maximum insolation for which a sub-Neptune-like planet can retain surface oceans (i.e. the inner edge of the habitable zone for planets endowed with a thick H$_2$-dominated envelope) has never been directly evaluated for a large range of planetary masses, gravities and types of host star. We acknowledge some recent progress in this regard \citep{koll19}. Finally, the cloud feedback proposed by \cite{yang13} may be less efficient for H$_2$-dominated atmospheres, for which moist convection is inhibited \citep{leconte17}.}\citep{yang13}.
Its measured mass and radius \citep[8.6$\pm$1.4 M$_\text{Earth}$ and 2.71$\pm$0.07 R$_\text{Earth}$ from][]{cloutier19}  are quite similar to those of GJ 1214b, and they suggest it is surrounded by a low-molecular-weight (H$_2$/He-dominated) atmosphere, potentially with a fraction of water vapour.
A recent analysis of HST/WFC3 transit observations revealed an absorption at 1.4 $\mu$m, which is interpreted as water vapour absorption \citep{tsiaras19, benneke19b} or methane absorption \citep{bezard20} in a low-molecular-weight (H$_2$/He-dominated) atmosphere. Recent observations by HST/STIS also confirm the interpretation that K2-18b has a H2-dominated atmosphere \citep{DosSantos20}.
\cite{benneke19b} also suggested that water clouds (most likely icy, but possibly liquid for a Bond albedo around 0.3) may form in the atmosphere of K2-18b. K2-18 b appears as a prime target to characterise the atmospheric composition of super-Earth/mini-Neptune and the climate dynamics of moderately irradiated worlds. Previous atmospheric modelling studies of K2-18b are based on 1D models \citep{benneke19b, madhusudhan20, scheucher20, bezard20, blain20}. Such models cannot predict the cloud cover, driven by atmospheric dynamics and thermal contrasts, and thus the real impact of clouds on spectra and on the planetary albedo. Previous 3D climate studies for terrestrial planets showed how the atmospheric circulation and cloud radiative effects strongly affect the cloud distribution and the habitability \citep{leconte13a,leconte13b,yang13,wolf14b}. In particular, 3D models predict a strong cloud formation on the day side of slow, synchronous, rotating terrestrial planets around M-stars. For such conditions, the associated high planetary albedo could place the inner edge of the habitable zone closer to the host star \citep{yang13, kopparapu16}. This intense cloud formation could also limit the detectability of water vapour and other species from transmission spectroscopy \citep{fauchez19,komacek20,suissa20}. These studies illustrate the importance of 3D modelling to assess the effect of clouds on observational spectra.

In this work, we used a 3D general circulation model (GCM) to simulate the atmosphere of K2-18b for H$_2$-dominated atmosphere with water clouds. In Section 2, we describe the 3D model and the conditions of this study. In Section 3, we present our results on the thermal structure, the atmospheric dynamics, the cloud distribution and variability. We analyse the impact of atmospheric metallicity (from 1$\times$solar to 1000$\times$solar), the concentration of cloud condensation nuclei, and the rotation rate. We discuss the impact of clouds on transit spectra in Section 4. We finish with a summary and conclusions in Section 5.

\section{Model}

\subsection{The LMD Generic GCM}
We simulated the atmosphere of K2-18b using the Laboratoire de M\'et\'eorologie Dynamique Generic GCM (LMDG). This model has been specifically developed for exoplanet and paleoclimate studies. In particular, it has been used for studying the atmospheres of moderately irradiated terrestrial exoplanets \citep{wordsworth11, leconte13b, leconte13a, turbet16, turbet18} and warm sub-Neptunes \citep{charnay15b, charnay15c}. 
The model is derived from the LMDZ Earth \citep{hourdin06} and Mars \citep{forget99} GCMs. It solves the primitive hydrostatic equations of meteorology using a finite difference dynamical core on an Arakawa C grid.
The model includes schemes for adiabatic convective adjustment, large-scale water cloud condensation, moist convection, precipitation, and evaporation (see the description of the water cycle in \cite{charnay13} and in the next sub-section). \cite{mayne19} showed that the full Navier Stokes equations can produce some differences compared to the primitive hydrostatic equations for simulations of the warm sub-Neptune GJ1214b. These differences are significant for cases with strong day-night thermal contrasts. However, the thermal contrast is very small on K2-18b (see Section 3.1.), and its atmospheric scale height is approximately half that of GJ1214b for a similar planetary radius. We expect the shallow fluid approximation to remain fully acceptable under K2-18b's regime.

In this paper, simulations were performed with a horizontal resolution of 64$\times$48 (corresponding to resolutions of 3.75$^\circ$ latitude by 5.625$^\circ$ longitude). We also did a test with a 128$\times$96 resolution, but we did not notice significant differences. For the vertical discretisation, the model uses pressure coordinates. In this work, we used 40 layers,  which were equally spaced in log pressure, with the first level at 80 bars and the top level at 0.2 mbar. 

The radiative scheme is based on the correlated-k method with the absorption data calculated directly from high-resolution spectra. We used the k-coefficients computed by \cite{blain20} with bins of 200 cm$^{-1}$ (spectral resolution R$\sim$50 at 1 $\mu$m). We included absorption by H$_2$O, CH$_4$, CO, CO$_2$, NH$_3$, PH$_3$, H$_2$S, HCN, K, Na, FeH, TiO, and VO. We used HITEMP 2010 for H$_2$O, CO, and CO$_2$  \citep{rothman10}; TheoReTS for CH$_4$ \citep{rey18}; and ExoMol for NH$_3$ and PH$_3$ \citep{coles19, yurchenko15, sousa-silva15}.

The H$_2$-H$_2$ and H$_2$-He collision-induced absorptions (CIA) from the HITRAN database \citep{karman19} were included, as well as a water vapour continuum from \cite{clough89}. Rayleigh scattering by H$_2$,  H$_2$O, and He was included based on the method described in \cite{hansen74}. Radiative transfer was computed with the \cite{toon89} scheme including water cloud optical properties. 

\subsection{Water clouds}
The model simulates water cloud formation including condensation, evaporation, coalescence, and sedimentation. We used the routines for the water cycle described in \cite{charnay13}. The mixing ratio of water vapour at the first pressure level is fixed (see values in Table 1). Water vapour is advected by the atmospheric circulation and condenses where its partial pressure exceeds the water vapour saturation pressure. We fixed the density number of cloud condensation nuclei (CCN), exploring values from $10^4$ to $10^7$ CCN/kg. We based this range on cloud particle concentrations for terrestrial water clouds, with typically $10^7-10^8$ droplets/kg for marine cumulus and $10^4-10^5$ ice particles/kg for cirrus \citep{wallace06}. The CCN concentration is a key parameter that controls the properties of the water clouds. Exploring the sensitivity of the results to this poorly known parameter allows us to account for most of the uncertainties related to the water cloud's microphysics and particle size distribution. For reference simulations, we chose a concentration of $10^5$ CCN/kg corresponding to conditions for cirrus formation. For a sub-Neptune like K2-18b, CCN sources for water clouds could be micrometeorites, photochemical hazes, or NH$_4$Cl cloud particles. The latter are salts and could be efficient CCN. NH$_4$Cl clouds should form at $\sim$0.1 bar for K2-18b’s conditions \citep{blain20}.

In all simulations, water clouds form for water vapour pressures that are always lower than 6.11 mbar (i.e. below the triple point of water). They are thus expected to be composed of ice particles only. We assumed spherical particles or crystal shapes of rimed dendrites (geometric parameters from \cite{heymsfield77}). For K2-18b's conditions, the air flow around falling cloud particles can have a high Reynolds number. We modified our sedimentation routine (based on the Stokes law) to take into account these regimes.
The terminal velocity of cloud particles is given by

\begin{equation} 
V_{fall} = \sqrt{\frac{8}{3}\frac{\rho_{\rm atm}g}{\rho_{\rm particle}C_{\rm D}r_c}}
\label{eq1}
,\end{equation} 
where $r_c$ is the particle radius, $\rho_{\rm atm}$ is the atmosphere density,  $\rho_{\rm particle}$ is the cloud particle density, and $C_D$ is the drag coefficient, expressed as follows \citep{clift71}:
\begin{equation} 
C_{\rm D} = \frac{24}{Re} \left(1+0.15Re^{0.687} \right) + \frac{0.42}{1+\frac{42500}{Re^{1.16}}}
\label{eq2}
,\end{equation} 
where $Re=\frac{2 \rho_{\rm atm} r_c V_{fall}}{\eta}$ is the Reynolds number and $\eta$ is the dynamic viscosity.

\begin{figure}[!] 
\centering
        \includegraphics[width=10cm]{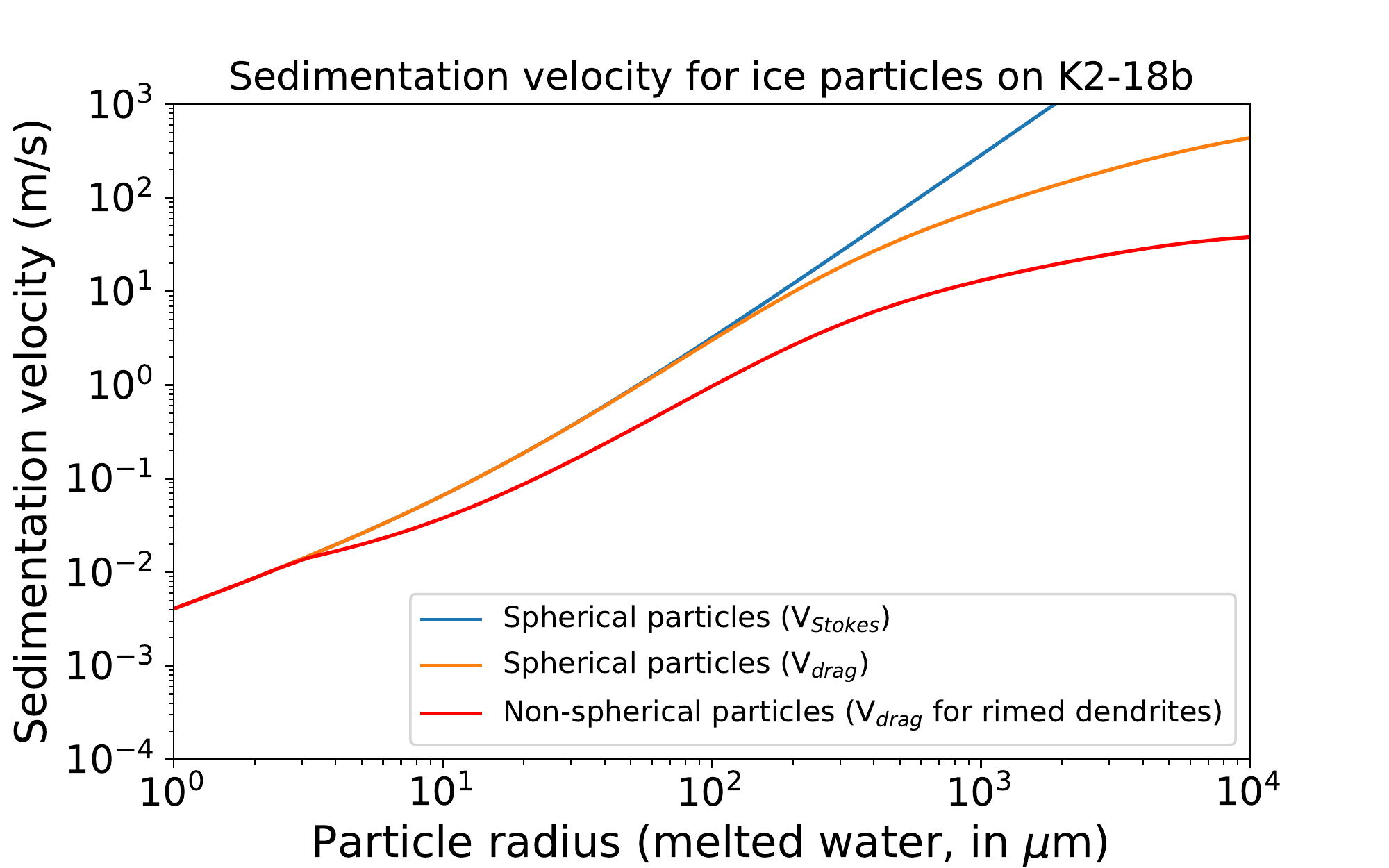} 
\caption{Sedimentation velocity as a function of particle radius (equivalent radius of melted water) computed for K2-18b at 10$^3$ Pa. The blue line follows the Stokes law for spherical particles. The yellow line shows the terminal velocity for the general drag coefficient. The red line is computed with the general drag coefficient and for rimed dentrites (parametrisation from \cite{heymsfield77}).}
\label{figure_1}
\end{figure}

Figure \ref{figure_1} shows the deviations between the Stokes law and the general terminal velocity for spherical particles (blue and yellow lines). Differences are significant for particles larger than 200~$\mu$m.
For rimed dendrites, the drag is increased compared to spherical particles because of a higher drag surface for the same mass. The terminal velocity is reduced by a factor of $\sim$4 for rimed dendrites compared to spherical particles with a 100~$\mu$m radius.

The model takes into account the conversion of cloud ice particles to snowflakes by coalescence with other particles following the parameterisation from \cite{boucher95}. For K2-18b, coalescence generally dominates for water ice precipitation.
Condensation and evaporation of water clouds also include latent heat release, which can potentially trigger moist convection. However, moist convection rarely occurs in our simulations of K2-18b because water condenses at high altitude, where the atmosphere is almost isothermal.
In addition, moist convection could be inhibited for sub-Neptunes with high fractions of water vapour due to the vertical gradient of mean molecular weight appearing with water condensation \citep{leconte17}. Table 1 indicates the critical water vapour mixing ratios above which moist condensation should be inhibited. The water vapour mixing ratios exceed these limits for all cases where water condenses in K2-18b's atmosphere (i.e. 100$\times$, 300$\times,$ and 1000$\times$solar). Moist convection should therefore not occur on K2-18b.

\begin{table}[!] 
\begin{tabular}{lccc}
\hline
\hline 
   Composition& r$_{vap}$ (mol/mol) &q$_{vap}$ (kg/kg) & q$_{crit}$ (kg/kg) \\ \hline
   100$\times$solar & 0.057 & 0.22 & 0.06  \\
   300$\times$solar & 0.088 & 0.2 & 0.08  \\  
   1000$\times$solar & 0.14 & 0.2 & 0.144  \\\hline
\end{tabular}
\label{table1}
\caption{Water vapour volume mixing ratio (r$_{vap}$), mass-mixing ratio (q$_{vap}$), and maximal mass-mixing ratio for moist convection (q$_{crit}$) for the different atmospheric compositions.}
\end{table}

\subsection{Model parameters and configurations}
Table 2 shows the orbital and physical parameters used for our simulations of K2-18b, mostly based on values from \cite{cloutier19} and \cite{benneke19b}.
An important and unknown parameter for 3D simulations is the rotation rate.
The time for tidal spin-down is given by \citep{guillot96}:
\begin{equation} 
\tau = Q \left(\frac{R_{\rm p}^3}{GM_{\rm p}} \right) \omega_{\rm p} \left(\frac{M_{\rm p}^3}{M_{\rm \star} }\right)^2 \left(\frac{D}{R_{\rm p}} \right)^6
\label{eq3}
,\end{equation} 
where $Q$ is the planet's tidal dissipation quality factor, $\omega_{\rm p}$ is the planet's primordial rotation rate ($\omega_{\rm p} \sim 1.7 \times 10^{-4}$ for Jupiter), and $D$ is the semi-major axis. From formula (\ref{eq3}), $\tau \sim$ 17 Ma for $Q=100$ (typical value for terrestrial planets) and $\tau \sim$ 17 Ga for $Q=10^{5}$ (typical value for gas-giant planets). We can expect that sub-Neptunes have intermediate values for $Q$, likely lower than $10^4$, giving $\tau \leqslant$ 1.7 Ga. With an estimated age of 2.4 $\pm 0.6$ Ga \citep{guinan19}, K2-18b would likely be tidally locked. We note that with its potential non-zero eccentricity ($e \sim 0.09$ in \cite{cloutier19}), K2-18b is as a good candidate for a spin-orbit resonance as Mercury. For most of our simulations, we assumed that the planet is tidally locked with a synchronous rotation around its host star. We also performed tests with different rotation rates (see Section 3.4).

\begin{table}[!] 
\begin{tabular}{lc}
\hline
\hline 
   Parameters &   \\ \hline
   $\Omega_p$ (planetary rotation rate, rad s$^{-1}$) & 2.2$\times$10$^{-6}$   \\
   $P_{rev}$ (revolution period, s) & 2.84$\times$10$^{6}$   \\
   $e$ (eccentricity) & 0   \\
   $R_p$ (planetary radius, m) & 1.73$\times$10$^{7}$   \\
   $g$ (gravitational acceleration, m s$^{-2}$) & 11.5   \\
   $F_{star}$ (stellar flux at the top of the atmosphere, W m$^{-2}$) & 1349  \\
   $F_{int}$ (internal thermal flux, W m$^{-2}$) & 3.7   \\ \hline
\end{tabular}
\label{table2}
\caption{Orbital and physical parameters used in the model.}
\end{table}

The simulations were performed for H$_2$-rich atmospheres with 1$\times$, 10$\times$, 100$\times$, 300$\times,$ and 1000$\times$solar metallicity. The atmospheric composition is based on calculations made with the 1D model Exo-REM, which has already been applied to K2-18b \citep{bezard20, blain20}. The 1D model was run with non-equilibrium chemistry and an eddy mixing coefficient K$_{zz}$=10$^6$cm$^2$/s. Figure \ref{figure_14} shows the atmospheric composition profiles used for the different cases. We assumed no longitudinal variation of the atmospheric composition except for water.
Table \ref{table3} shows the specific heat capacity (c$_p$), the atmospheric scale height at the 300 K level ($H$) and the mean molecular weight for the different compositions.
All simulations were initialised with a 1D thermal profile (computed with the 1D version of the model) and were run for more than one thousand K2-18b orbits ($\sim$100 Earth years). 
Simulations with 300$\times$solar metallicity provide the best fits to transit observations (discussed in Section 4). In this study, we used this atmospheric composition as a reference to explore sensitivity to cloud particle size and the rotation rate.

\begin{table*}[!] 
\begin{center} 
\begin{tabular}{lccc}
\hline
\hline 
   Atmospheric composition & c$_p$ (J kg$^{-1}$ K$^{-1}$) & $H$(km) & Mean molecular weight (g mol$^{-1}$) \\ \hline
   1$\times$solar & 11905 & 97 & 2.3  \\
   10$\times$solar & 10873 & 87 & 2.6  \\
   100$\times$solar & 6682 & 47& 4.8  \\
   300$\times$solar & 4764 & 29& 7.9  \\  
   1000$\times$solar & 3677 & 18 & 12.3  \\\hline
\end{tabular}
\label{table3}
\caption{Values of specific heat (c$_p$), scale height ($H$) at the 300 K level, and mean molecular weight used for the different atmospheric compositions.}
\end{center} 
\end{table*}

\section{Results}

\subsection{Thermal structure and atmospheric dynamics}

\begin{figure*}[!] 
\centering
        \includegraphics[width=8cm]{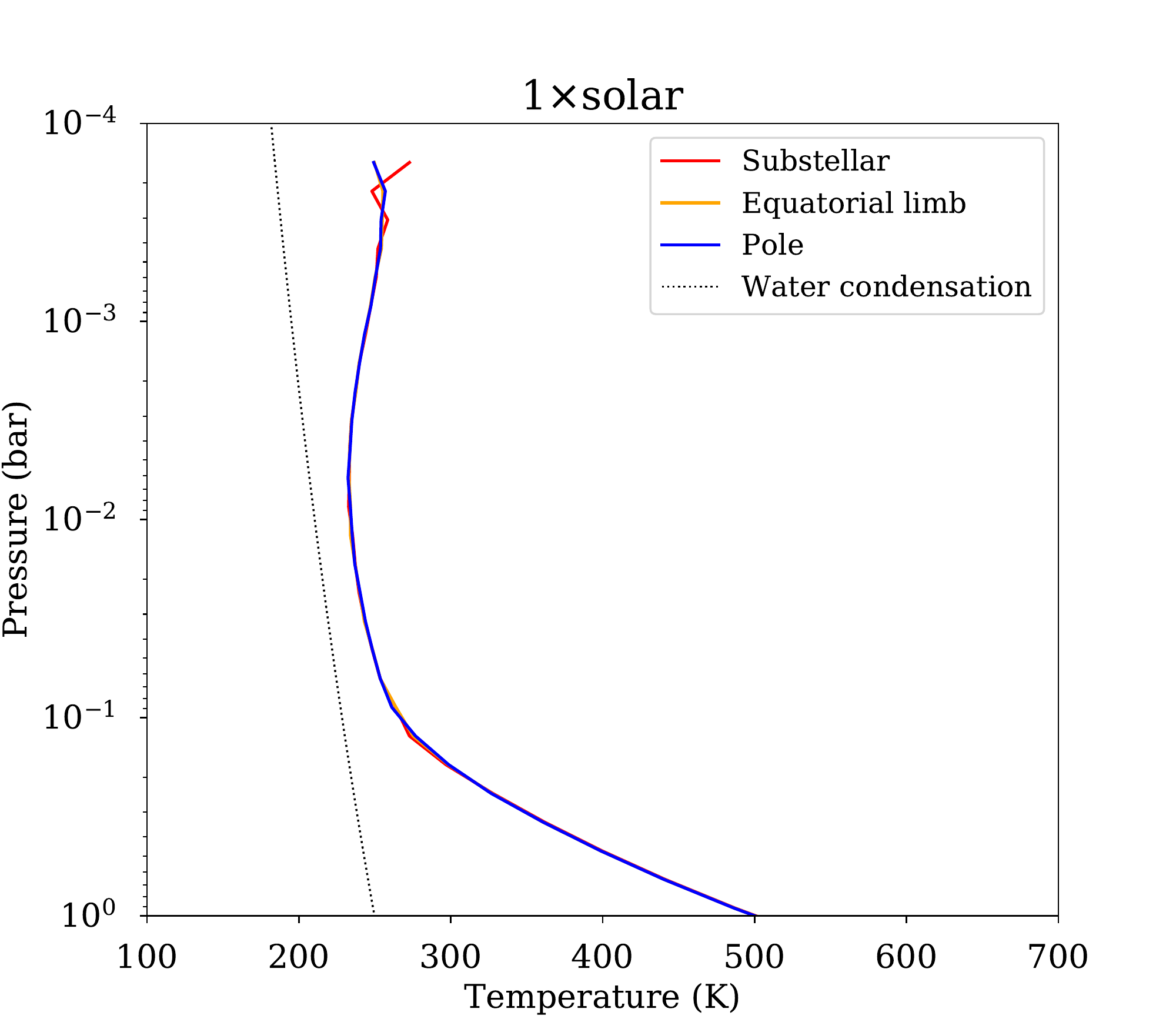}
        \includegraphics[width=8cm]{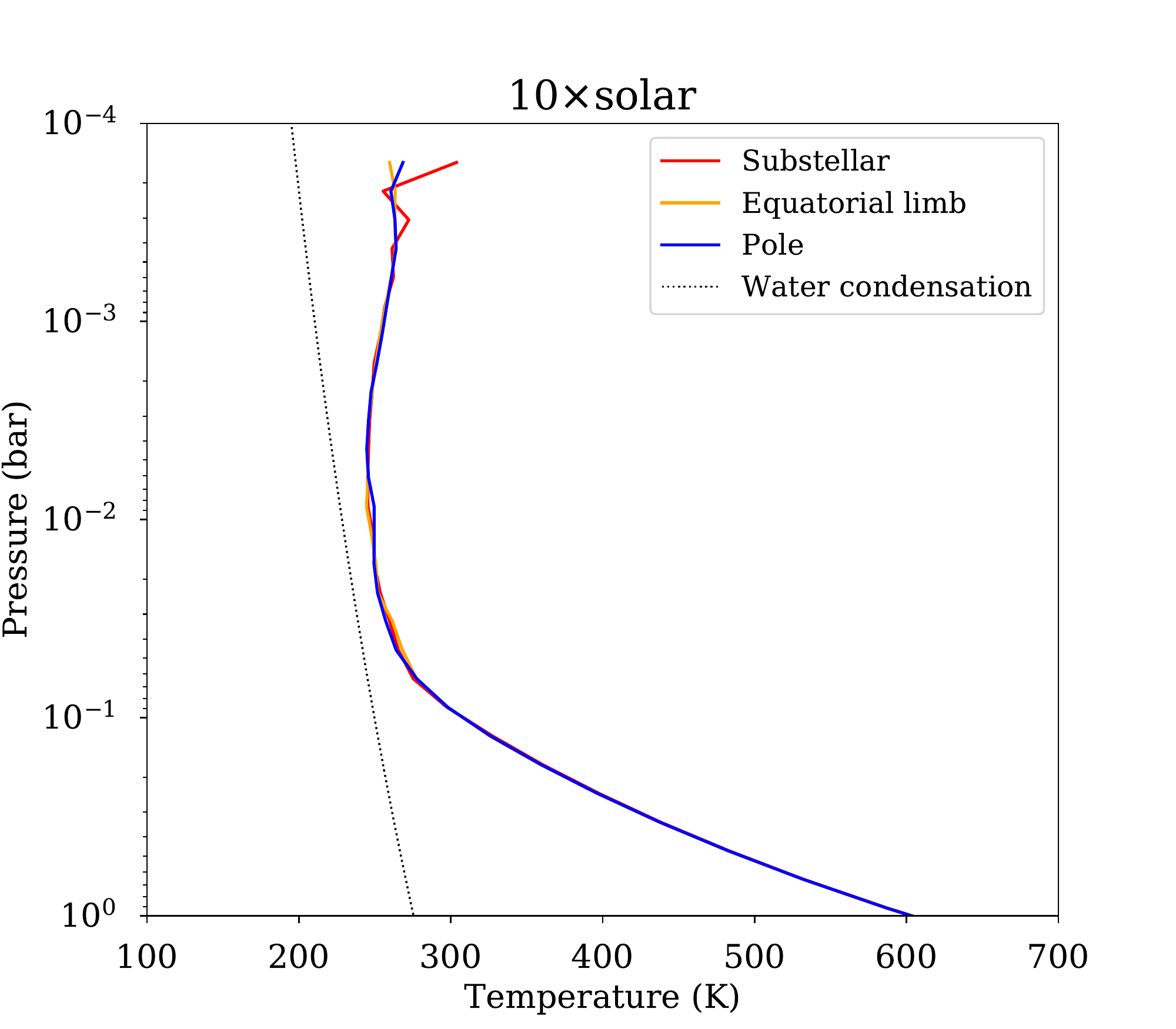}
        \includegraphics[width=8cm]{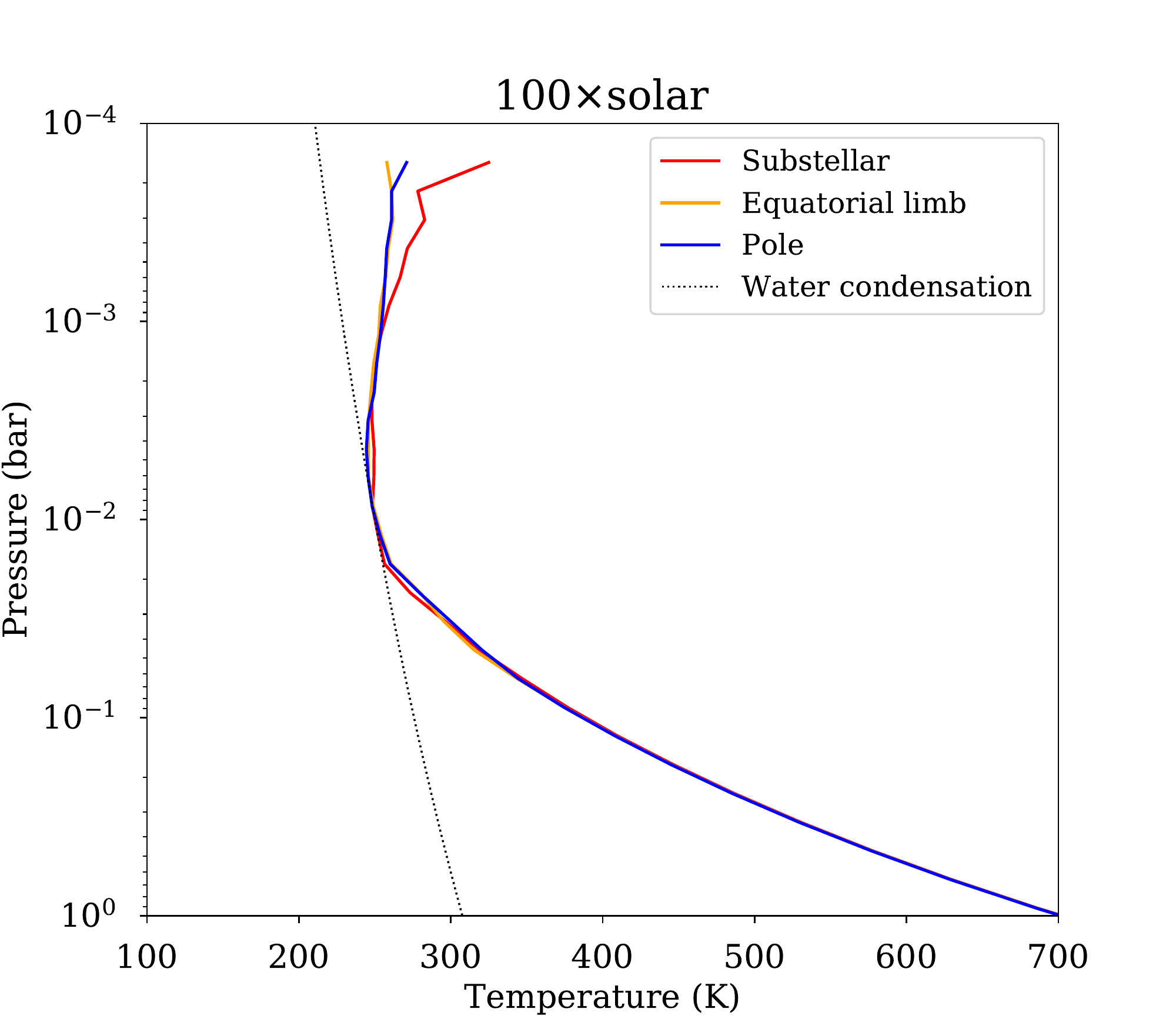}
        \includegraphics[width=8cm]{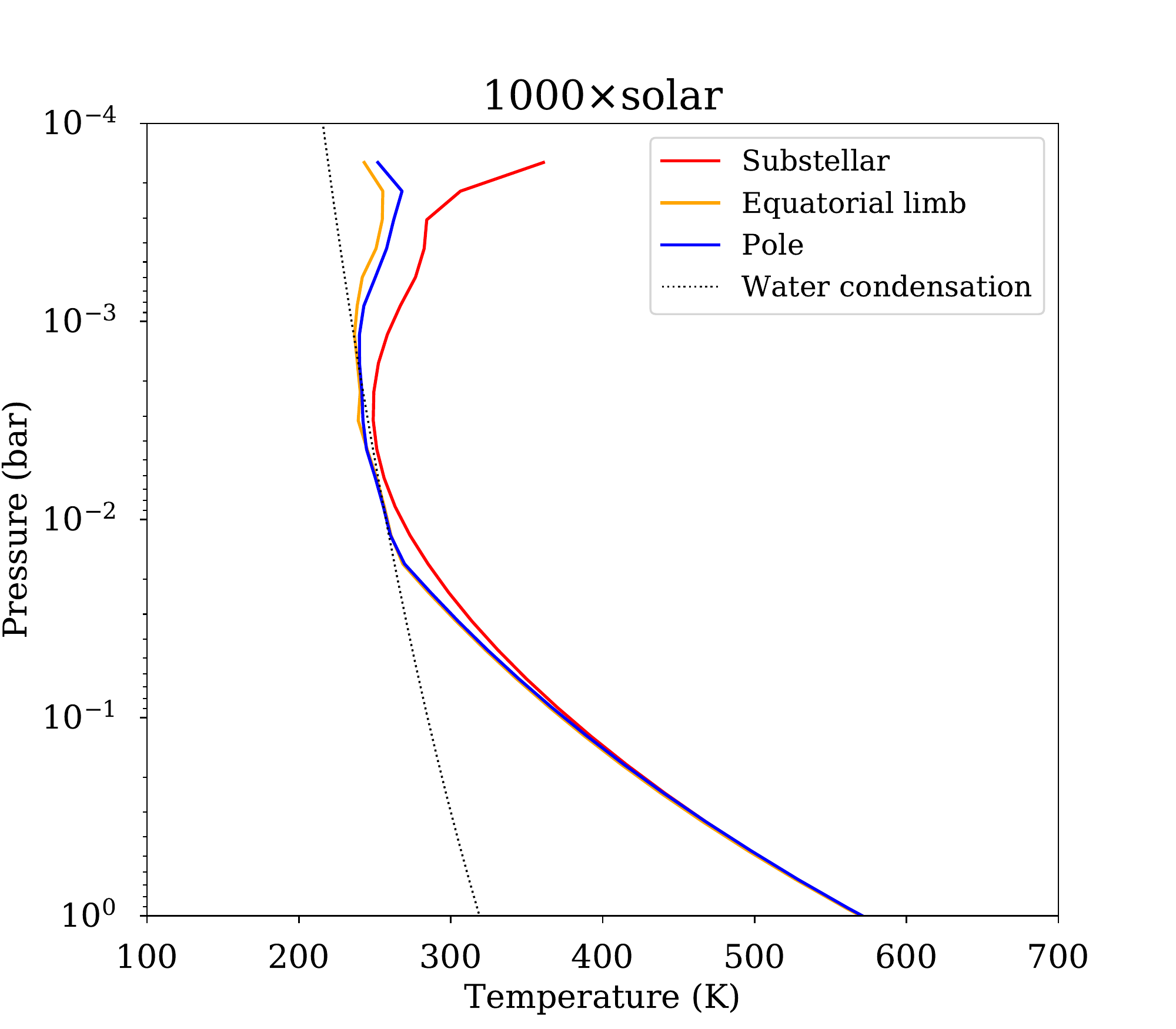}
\caption{Temperature profiles for 1$\times$, 10$\times$, 100$\times,$ and 1000$\times$solar metallicity at the sub-stellar point (red) at poles (blue) and at the equatorial morning terminator (yellow). The dashed lines are the condensation curves of water vapour.}
\label{figure_2}
\end{figure*}

Figure~\ref{figure_2} shows temperature profiles at the pole, at the sub-stellar point, and at the equatorial morning terminator for different atmospheric metallicities. 
We notice that the isothermal region appears at a lower pressure for the high-metallicity cases (e.g. $\sim$0.01 bar for 1000$\times$solar compared to $\sim$0.1 bar for 1$\times$solar). This is due to the enhanced infrared opacities and greenhouse effect for high metallicity cases. In addition, the thermal gradient and the temperature at 1 bar are reduced for the 1000$\times$solar case compared with the 100$\times$solar case. This is due to the specific heat capacity per molecule which increases with the metallicity and the abundances of H$_2$O, CO$_2$, CH$_4$, NH$_3$, which have more degrees of freedom (i.e. vibration modes) than H$_2$ and He.

There are few longitudinal/latitudinal temperature variations, apart from in the upper atmosphere (at pressures lower than 10~mbar) for high metallicity ($\geq$100$\times$solar). CH$_4$ is the main absorber of stellar flux in the upper atmosphere. It produces radiative heating and a stratospheric thermal inversion at the sub-stellar point for high-metallicity cases. The weak horizontal temperature variations are due to the long radiative timescale compared to the advection timescale. 1D modelling is therefore an excellent approach for computing the thermal structure of such a temperate planetary atmosphere.

\begin{figure*}[!] 
\centering
        \includegraphics[width=8cm]{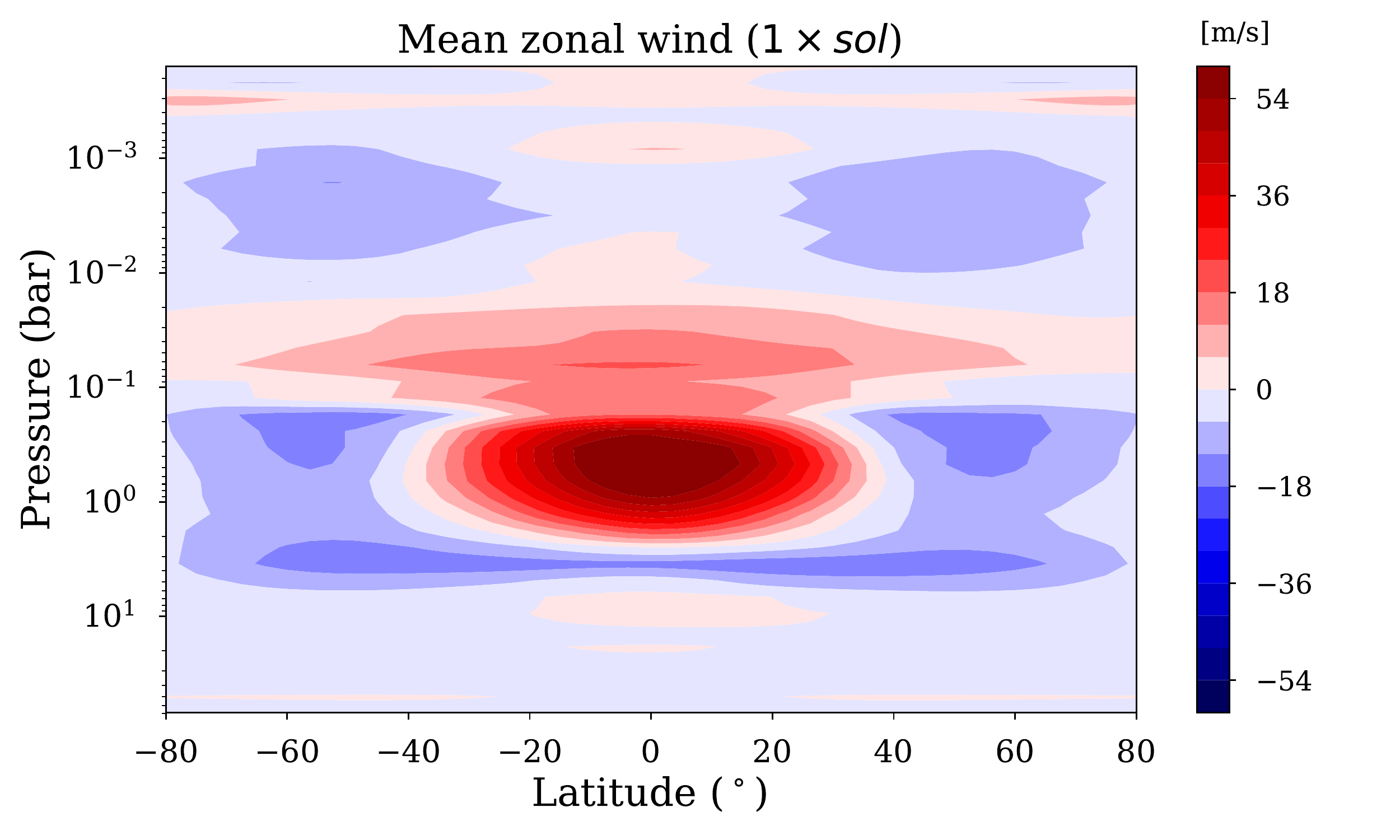}
        \includegraphics[width=8.06cm]{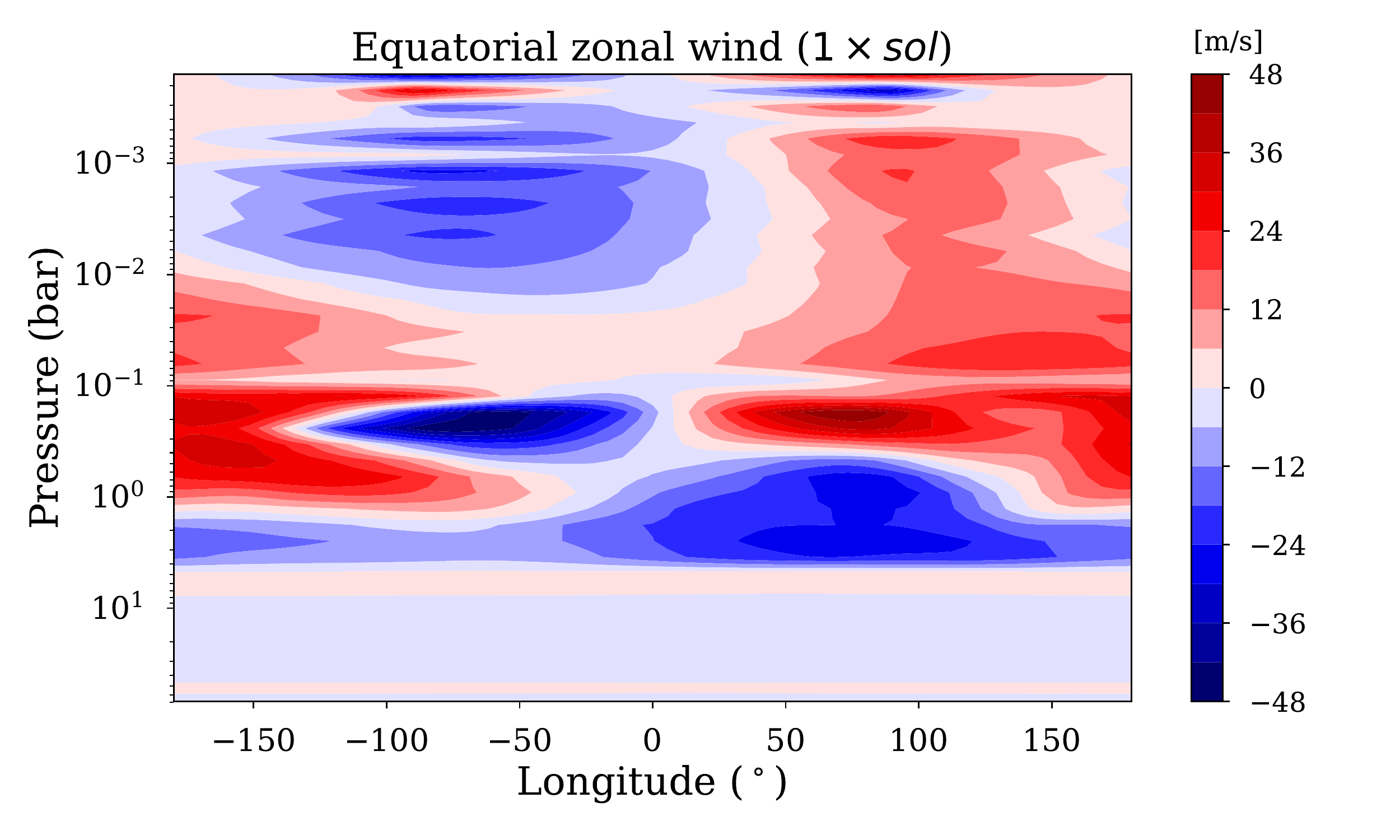}   
        \includegraphics[width=8cm]{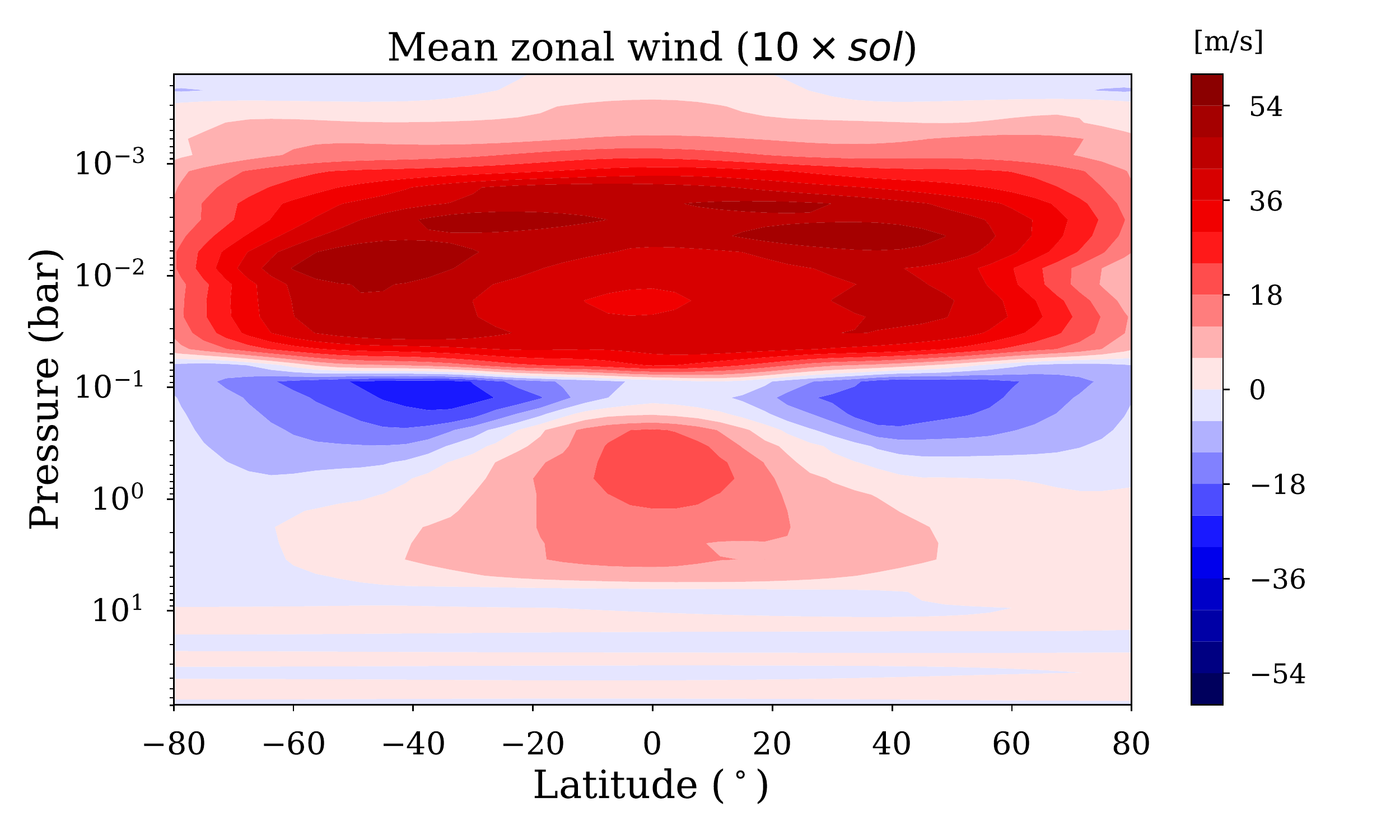}
        \includegraphics[width=8.06cm]{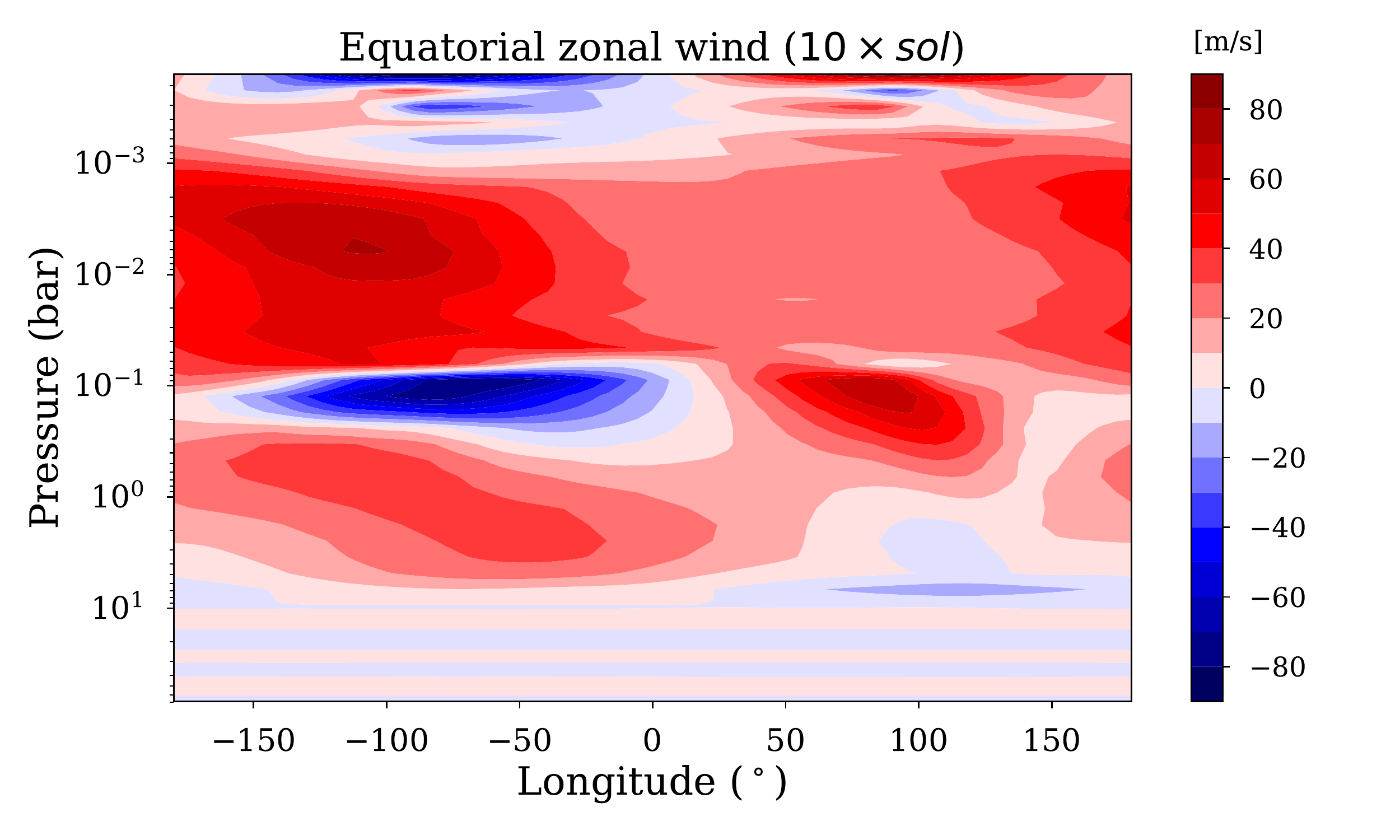}  
        \includegraphics[width=8cm]{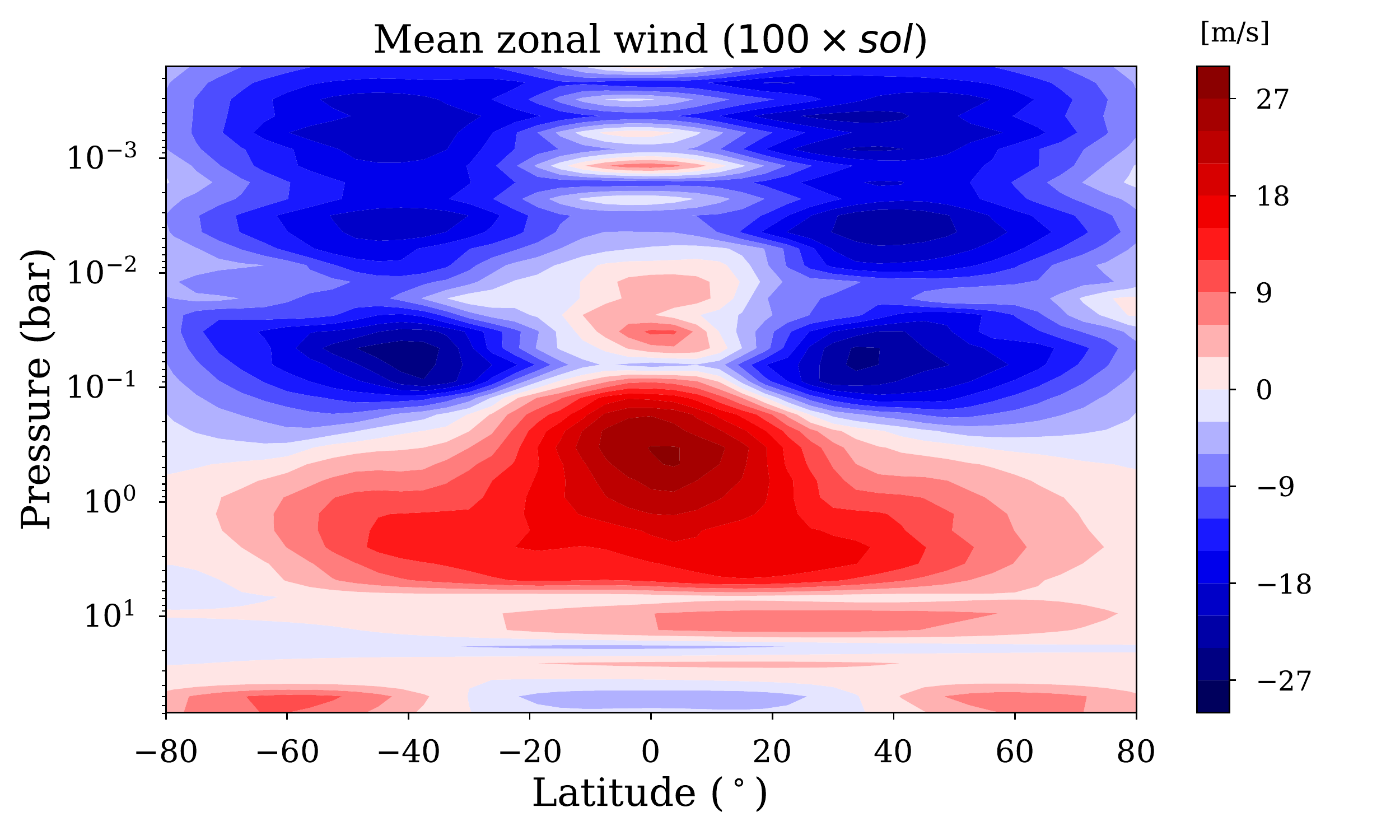}
        \includegraphics[width=8.06cm]{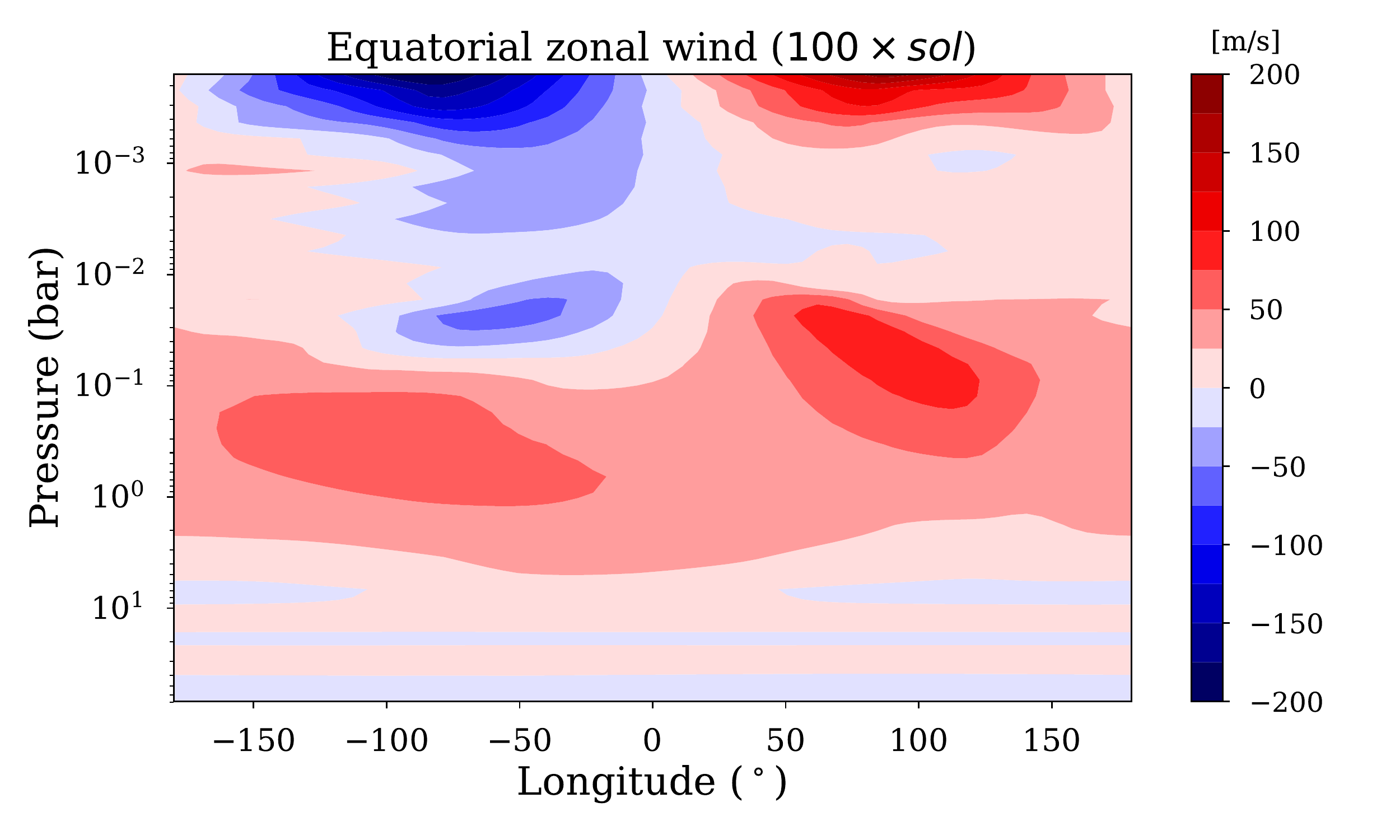} 
        \includegraphics[width=8cm]{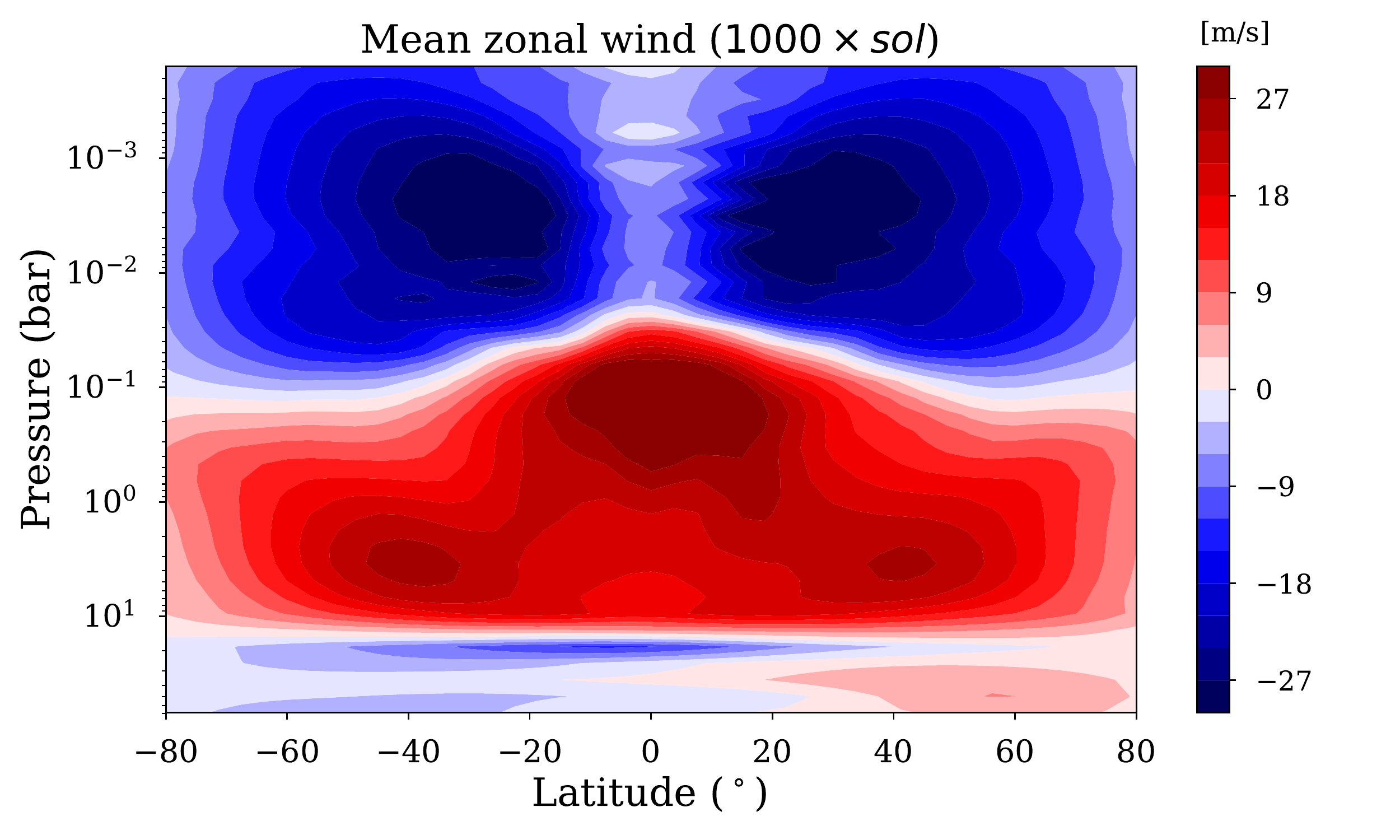}
        \includegraphics[width=8.06cm]{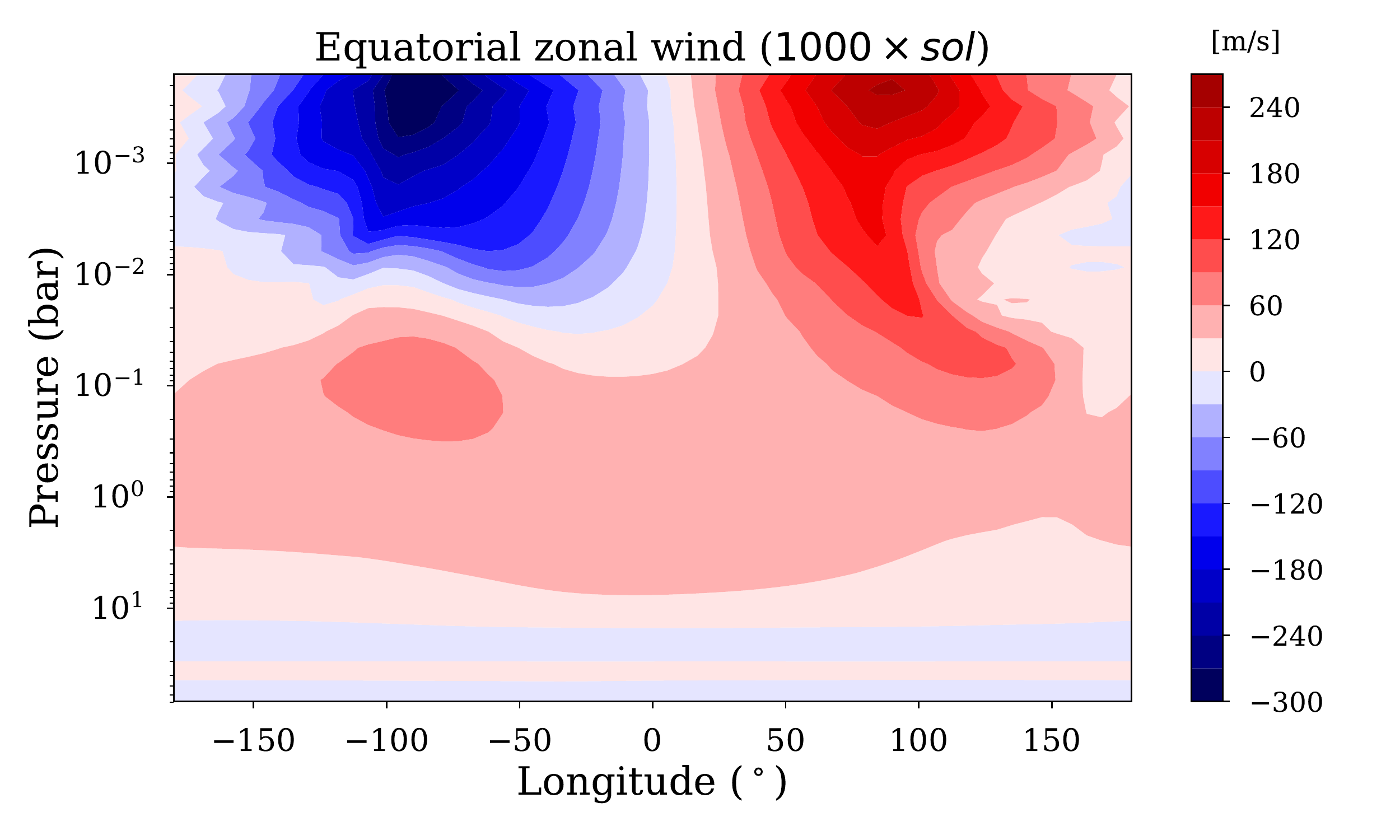}        
\caption{Zonally averaged zonal wind (left) and mean equatorial (averaged between latitude -30$^\circ$ and +30$^\circ$) zonal winds (right) for the different atmospheric compositions. Positive (negative) values correspond to westerly (easterly) winds.}
\label{figure_3}
\end{figure*}

\begin{figure}[!] 
\centering
        \includegraphics[width=8cm]{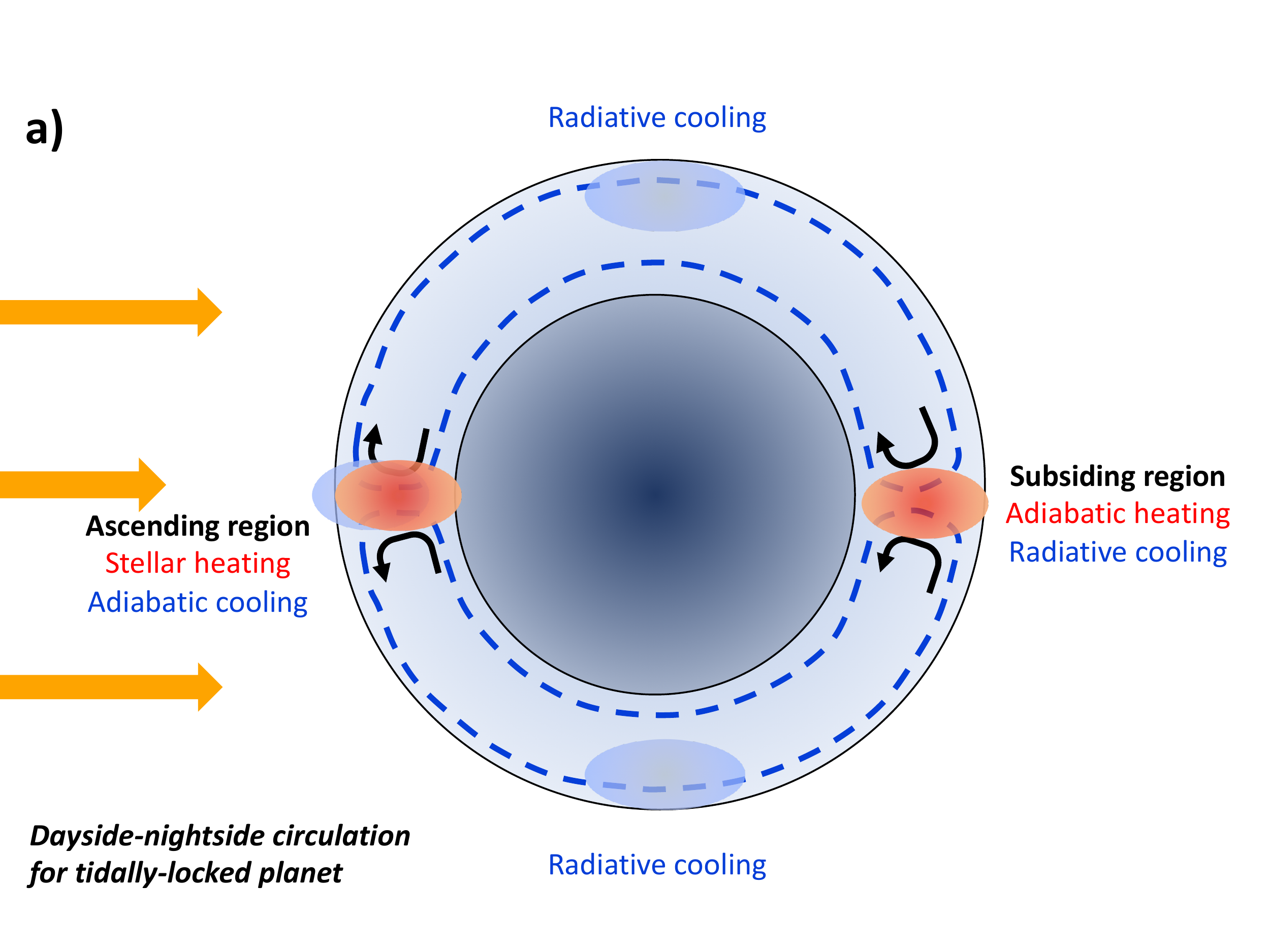}
        \includegraphics[width=8cm]{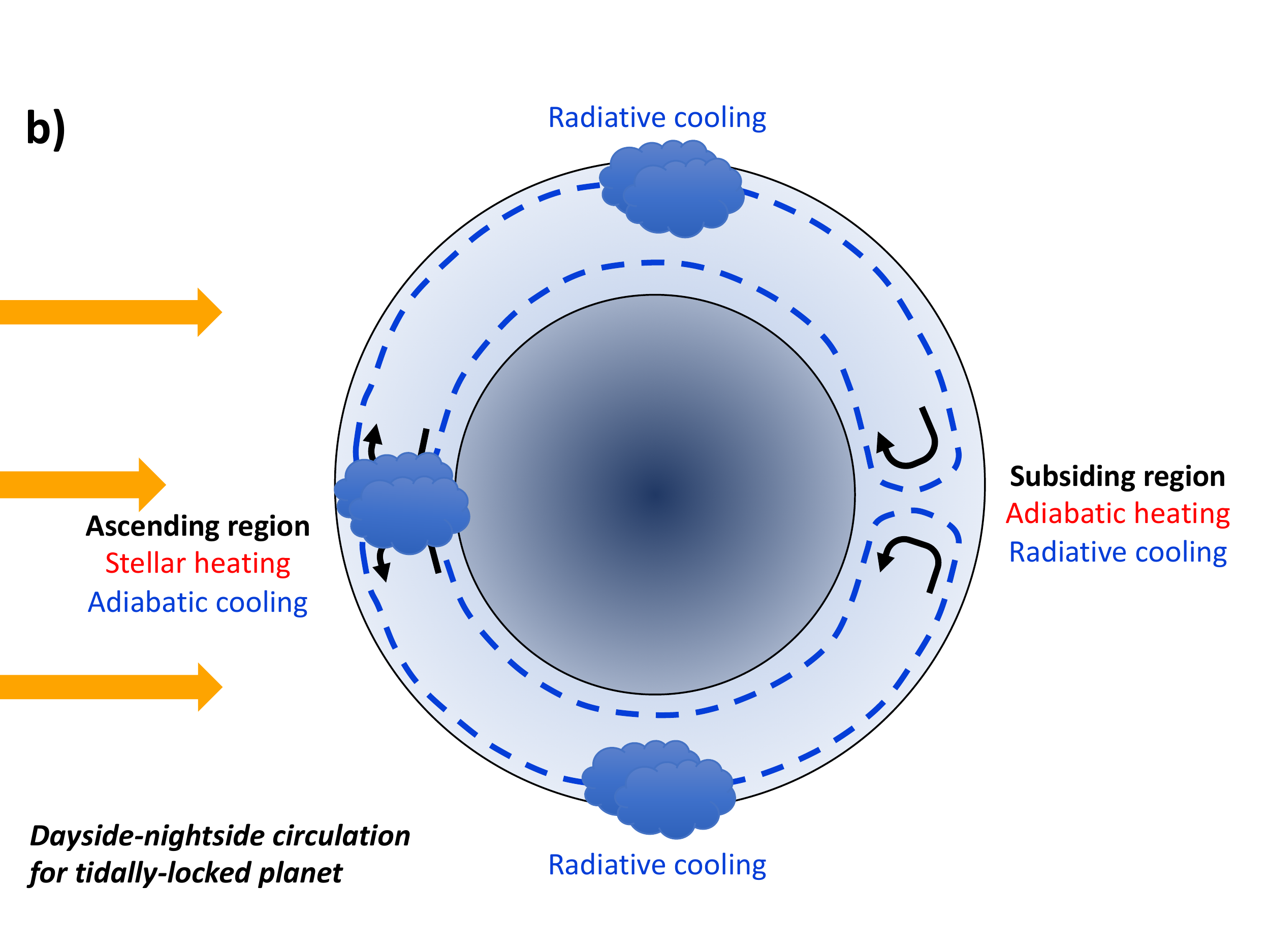} 
\caption{Illustration of day-night atmospheric circulation around K2-18b for a synchronous rotation. Panel a shows the circulation and warming/cooling zones. Panel b shows the preferential location of cloud formation.}
\label{figure_4}
\end{figure}

Pressure-latitude cross-sections of zonally averaged zonal wind and mean equatorial zonal winds (Fig.~\ref{figure_3}) show the presence of a weak tropospheric equatorial super-rotation jet (i.e. westerly winds) at pressures between 0.1 and 10~bars. \cite{showman11} showed that an equatorial super-rotation jet develops on tidally locked planets by the formation of standing planetary-scale equatorial and Rossby waves. A condition for this mechanism to occur is that the equatorial Rossby deformation radius be smaller than the planetary radius.
We can define the dimensionless equatorial Rossby deformation length \citep{leconte13a}:

\begin{equation} 
\mathcal{L} \equiv \frac{L_{\rm Ro}}{R_{\rm p}}=\sqrt{\frac{NH}{2\Omega R_{\rm p}}}
\label{eq4}
,\end{equation} 
where $H$ is the atmospheric scale height and $N=\sqrt{\frac{g}{T}\left(\frac{g}{c_{\rm p}}+\frac{dT}{dz}\right)}$ is the Brunt-Vaisala frequency.
In the troposphere, the vertical temperature gradient deviates from the dry adiabatic gradient ($\Gamma_{\rm dry}=-\frac{g}{c_{\rm p}}$) by around 1-10$\%$ and $\mathcal{L} \sim$ 0.7. An equatorial super-rotation jet forced by stationary planetary waves can develop there. In contrast, the stratosphere is almost isothermal with $\mathcal{L}\sim$ 2. An equatorial jet cannot develop from stationary planetary waves at pressures lower than than $\sim$0.1~bar. We note that for the case with 10$\times$solar metallicity, super-rotation develops at pressures lower than 0.1~bar and at all latitudes. Super-rotation is likely triggered by a different process here (e.g. by barotropic waves as on Titan). In any case, the atmospheric circulation for pressures lower than 0.1~bar is dominated by a day-night circulation, with up-welling air on the day side and down-welling air on the night side (see Fig.~\ref{figure_4}a for illustration). Winds are relatively axisymmetric around the sub-stellar-anti-stellar axis. For 300$\times$solar metallicity, horizontal winds are at their maximum at the terminator and reach 200~m/s at top model level (at 0.2~mbar). The vertical wind is around 0.2~m/s at the sub-stellar point for pressures lower than 0.1~bar.

\begin{figure*}[!] 
\centering
        \includegraphics[width=8cm]{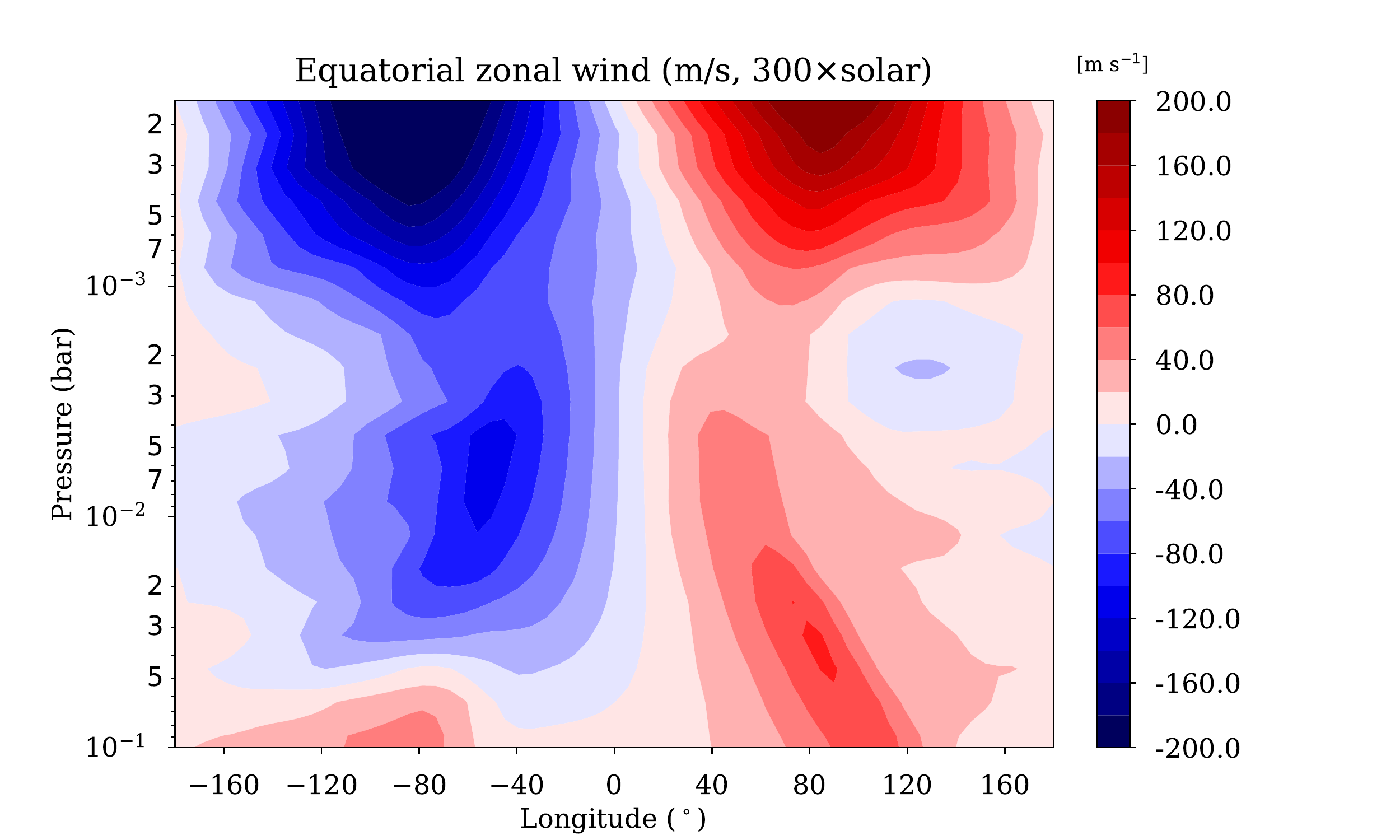}
        \includegraphics[width=8cm]{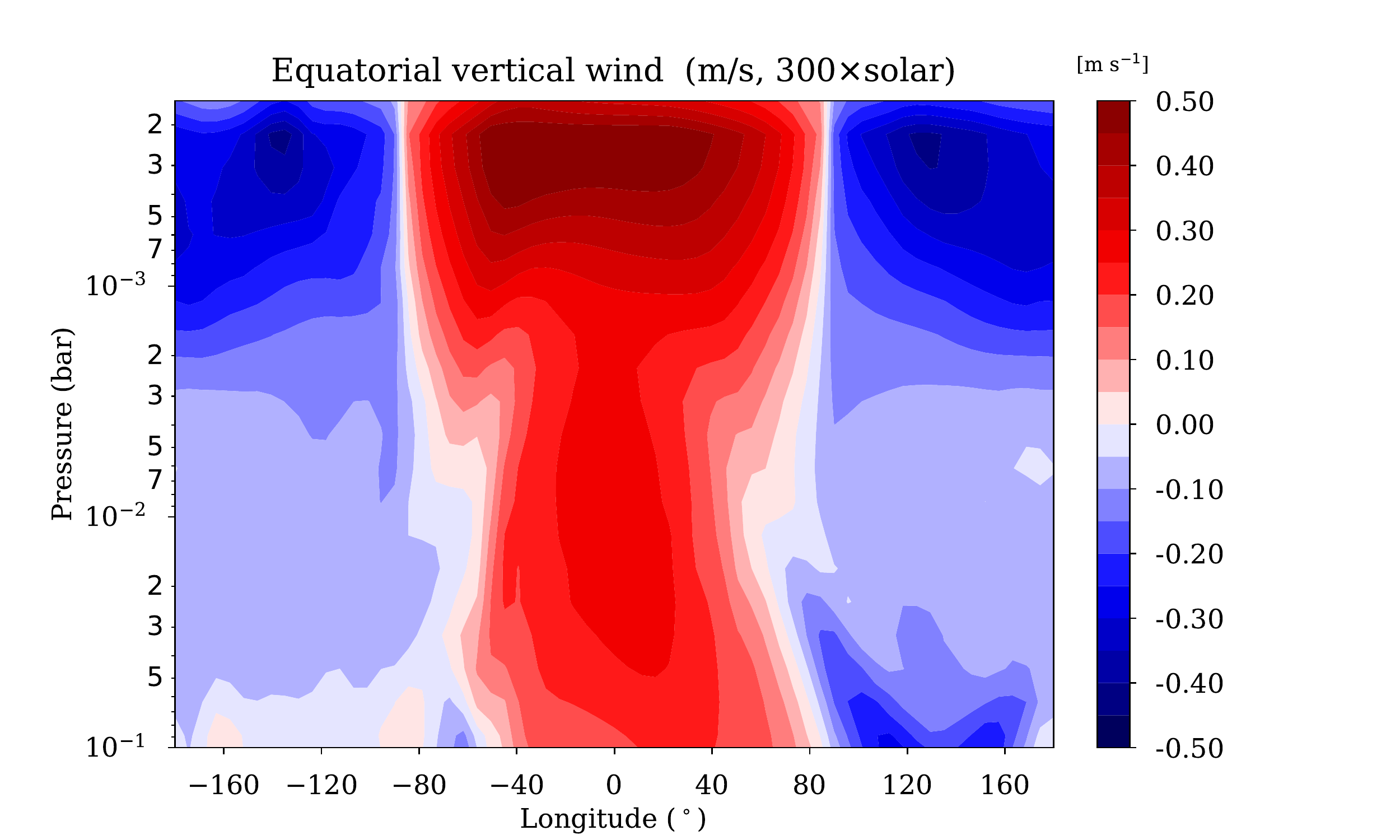}
        \includegraphics[width=8cm]{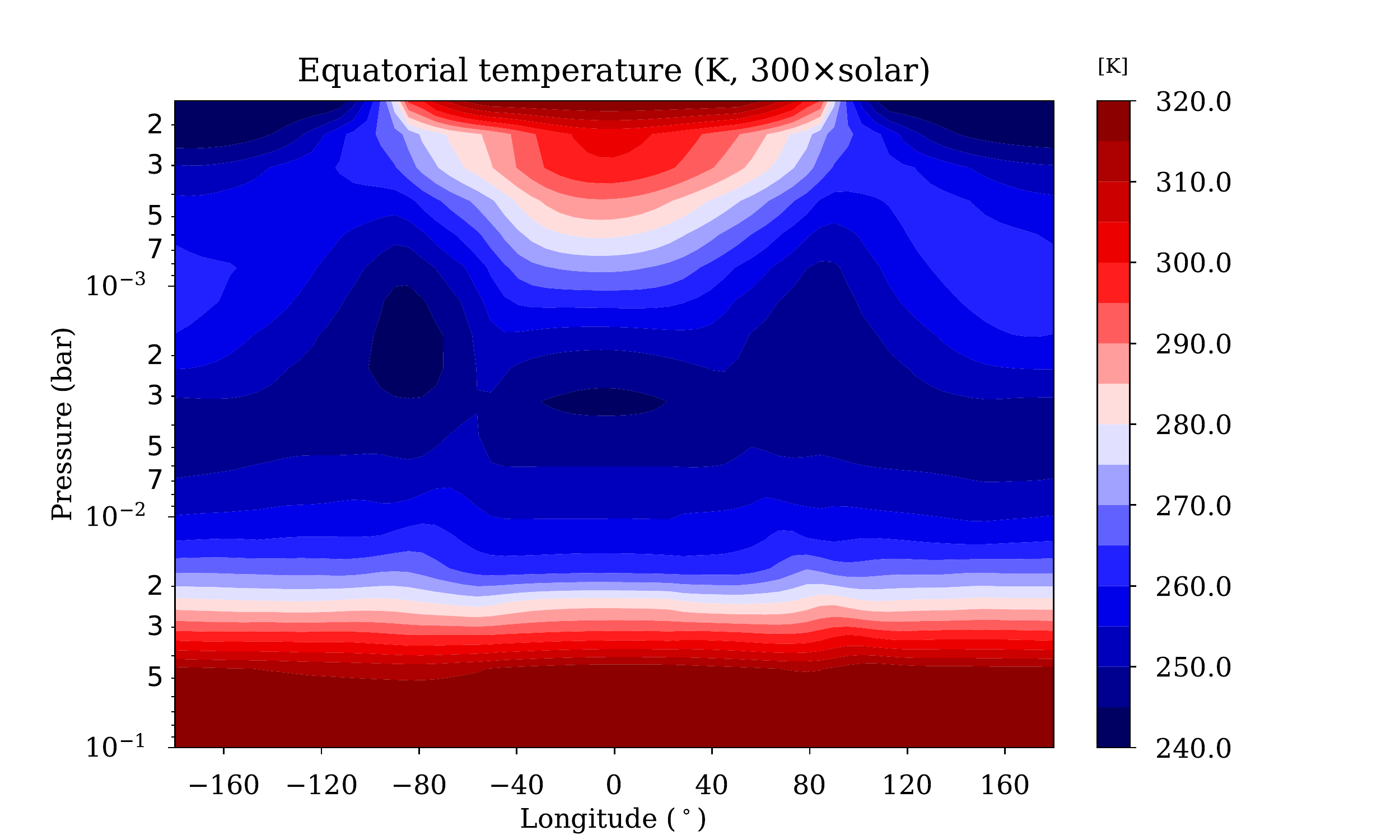}         \includegraphics[width=8cm]{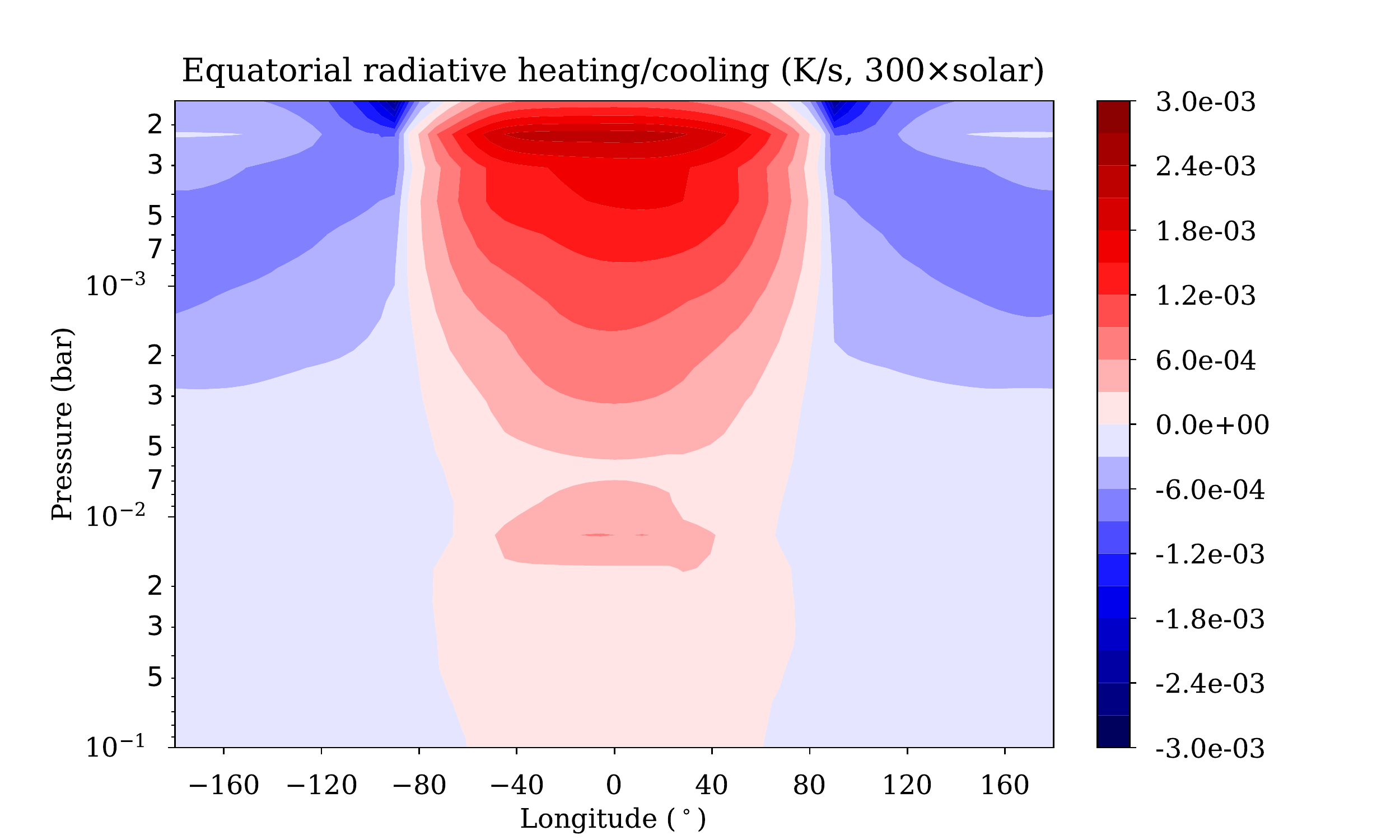}               
         \includegraphics[width=8cm]{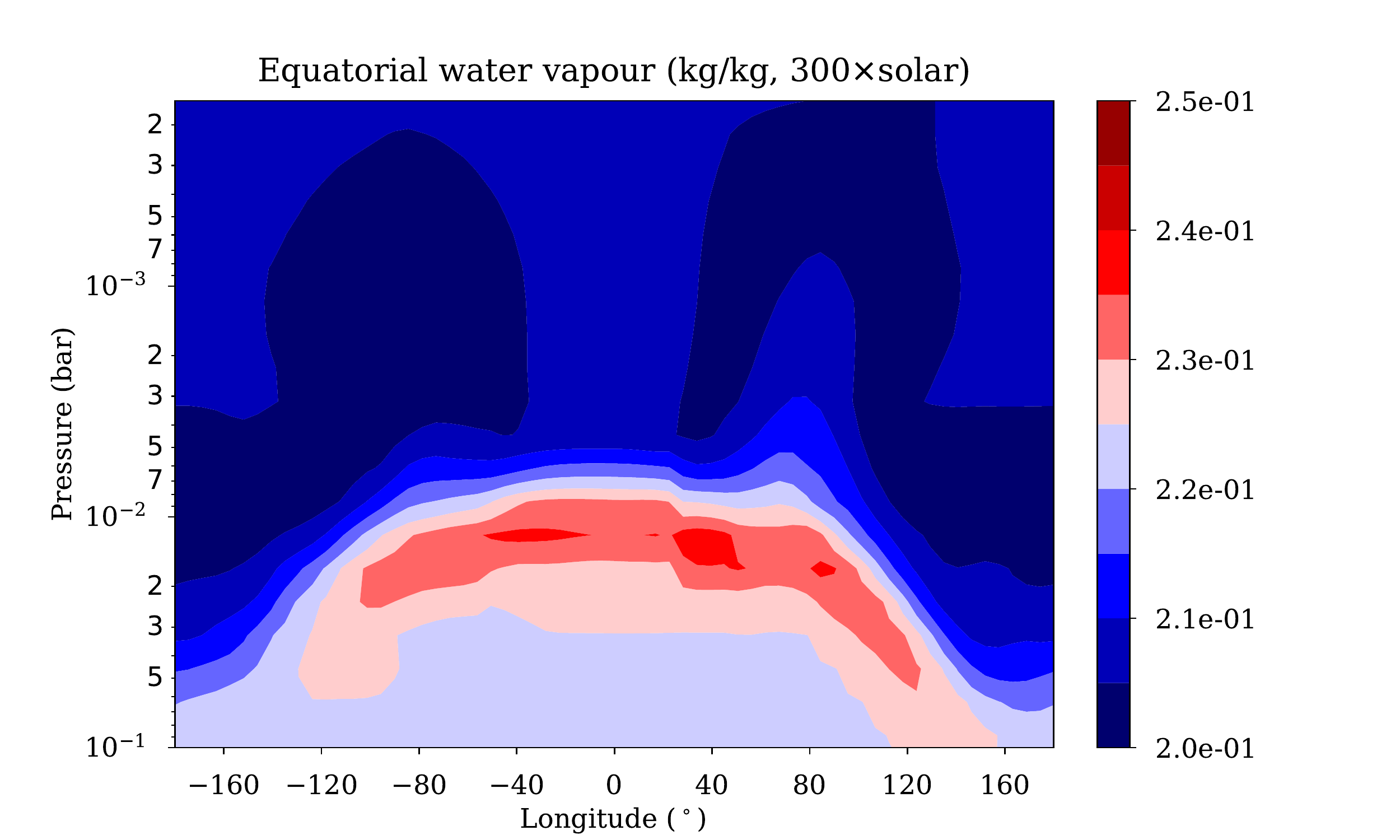}
         \includegraphics[width=8cm]{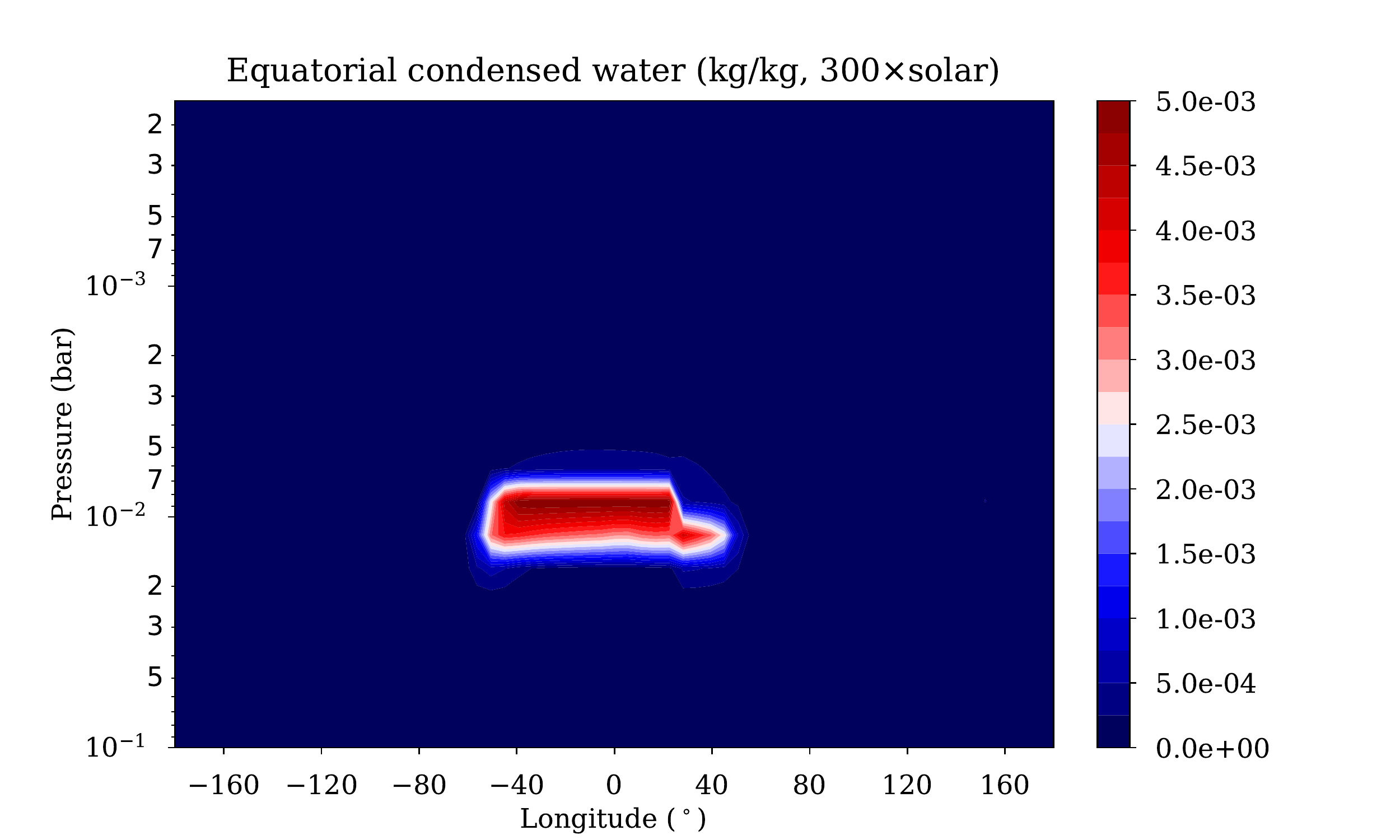}
\caption{Mean equatorial (averaged between latitude -30$^\circ$ and +30$^\circ$) zonal winds, vertical wind, temperature, radiative warming/cooling, water vapour mixing ratio, and water cloud mixing ratio as a function of longitude and pressure for 300$\times$solar metallicity.}
\label{figure_5}
\end{figure*}

The day side is heated by stellar radiation and cooled by ascending air producing adiabatic cooling. The temperature of the night side is controlled by radiative cooling and by adiabatic warming produced by down-welling air. Counterintuitively and because of this night-side adiabatic warming, the coldest point in the atmosphere between 0.5 and 5 mbar is not at the anti-sub-stellar point but at the terminator (see Fig.~\ref{figure_5}). For the 300$\times$solar metallicity, the temperature at longitude $\pm 90^{\circ}$ and at 1~mbar is $\sim$20~K lower than at the sub-stellar point (see Fig.~\ref{figure_5}).

\subsection{Cloud formation}

\begin{figure*}[!] 
\centering
        \includegraphics[width=6cm]{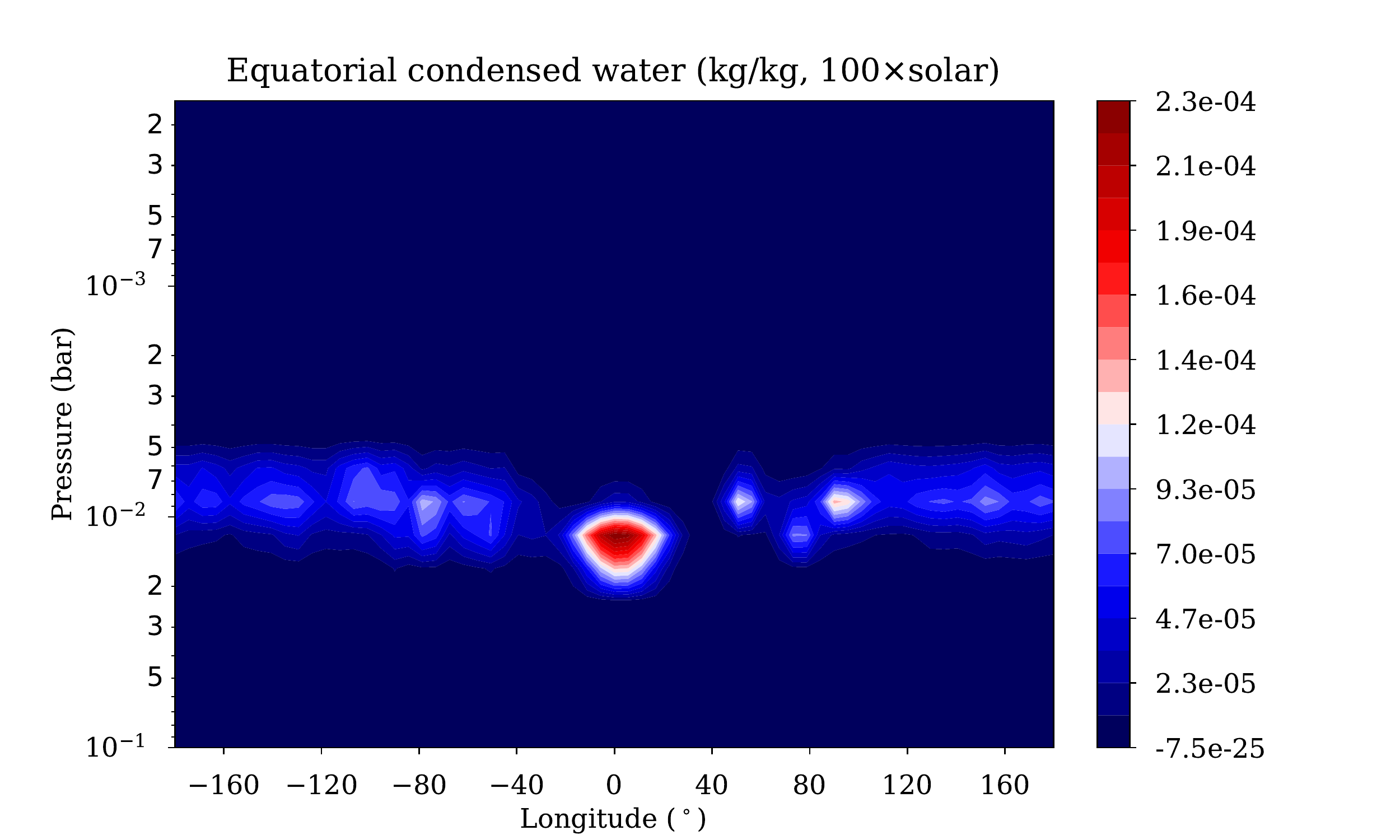}
        \includegraphics[width=6cm]{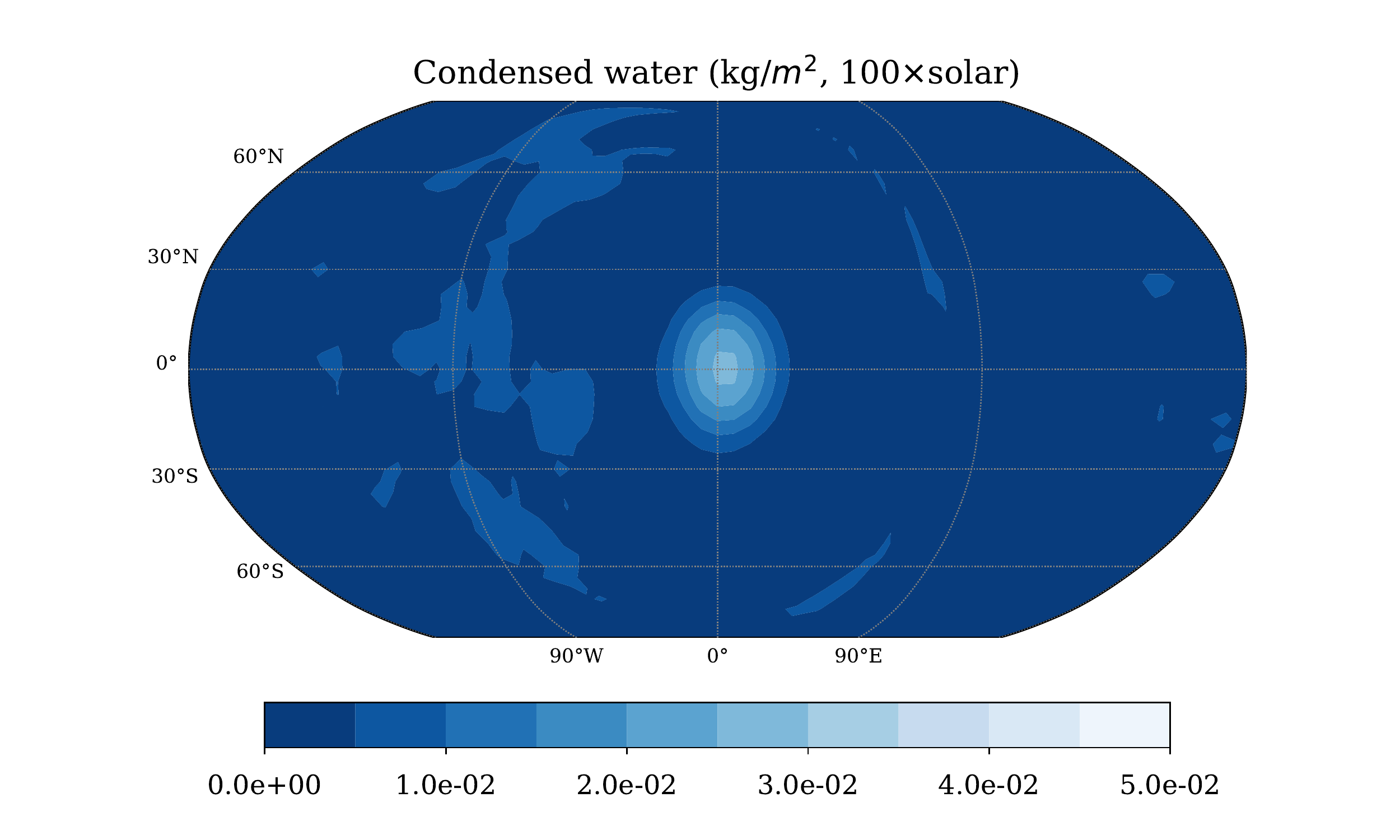}
         \includegraphics[width=6cm]{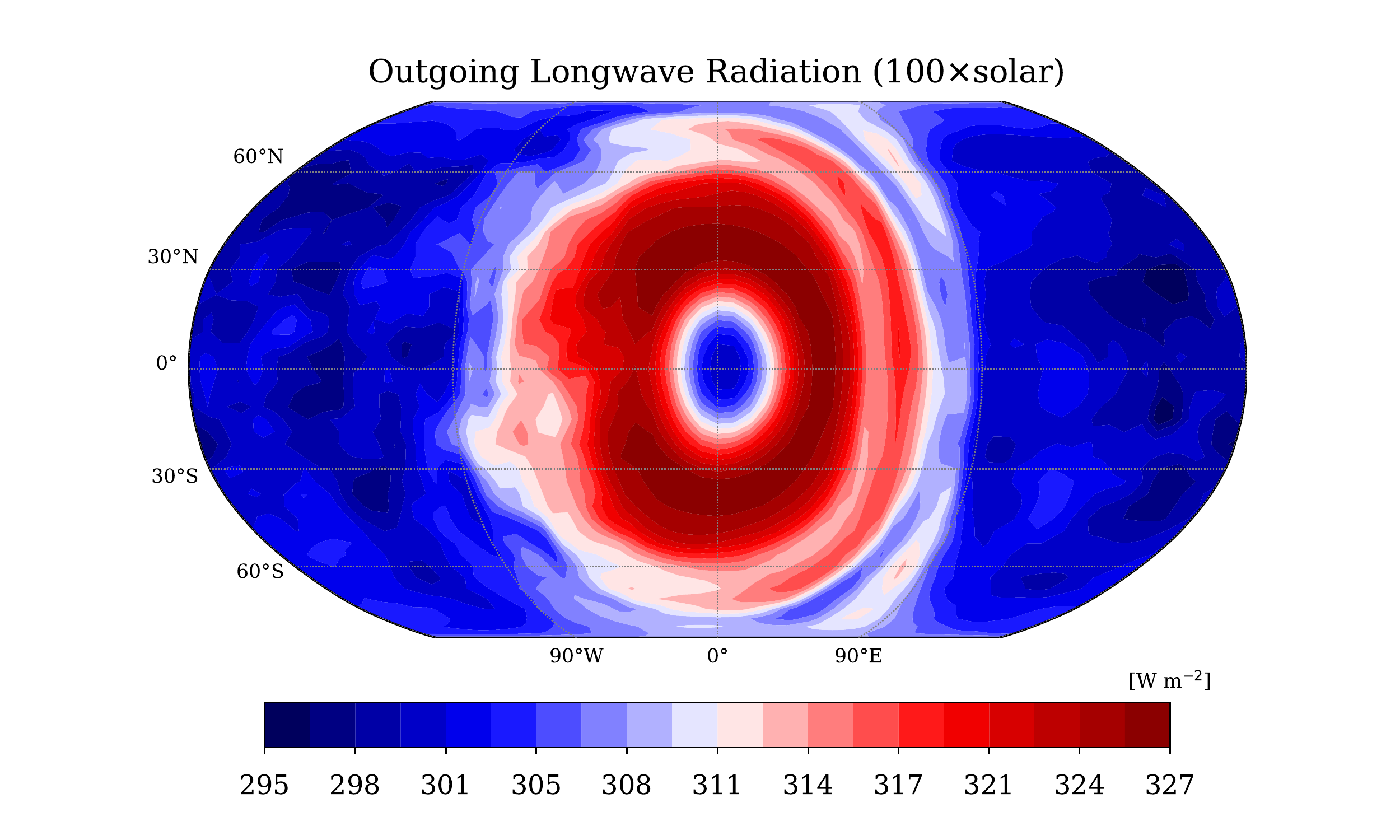}
        \includegraphics[width=6cm]{eq_h2oice_300xsol.pdf}
        \includegraphics[width=6cm]{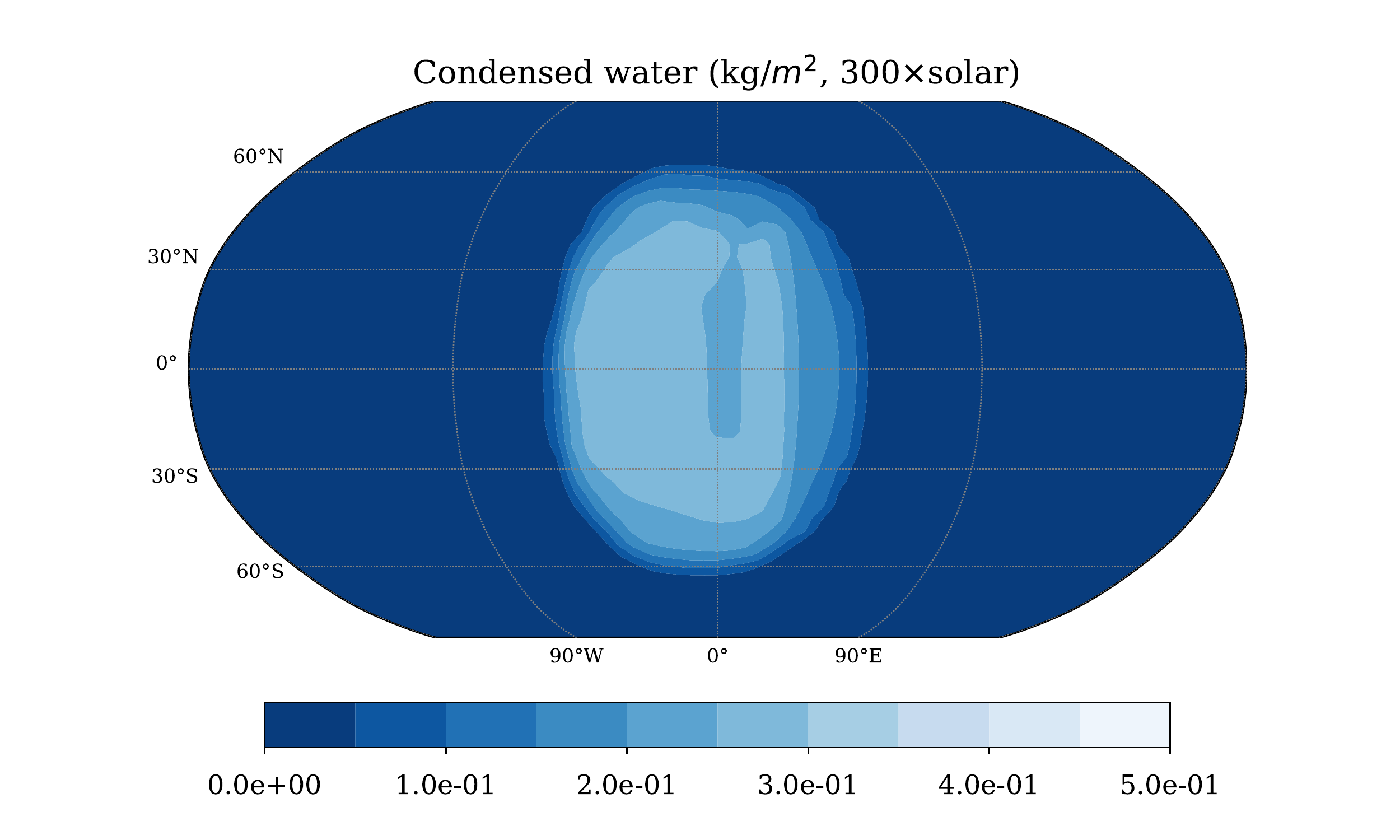}
         \includegraphics[width=6cm]{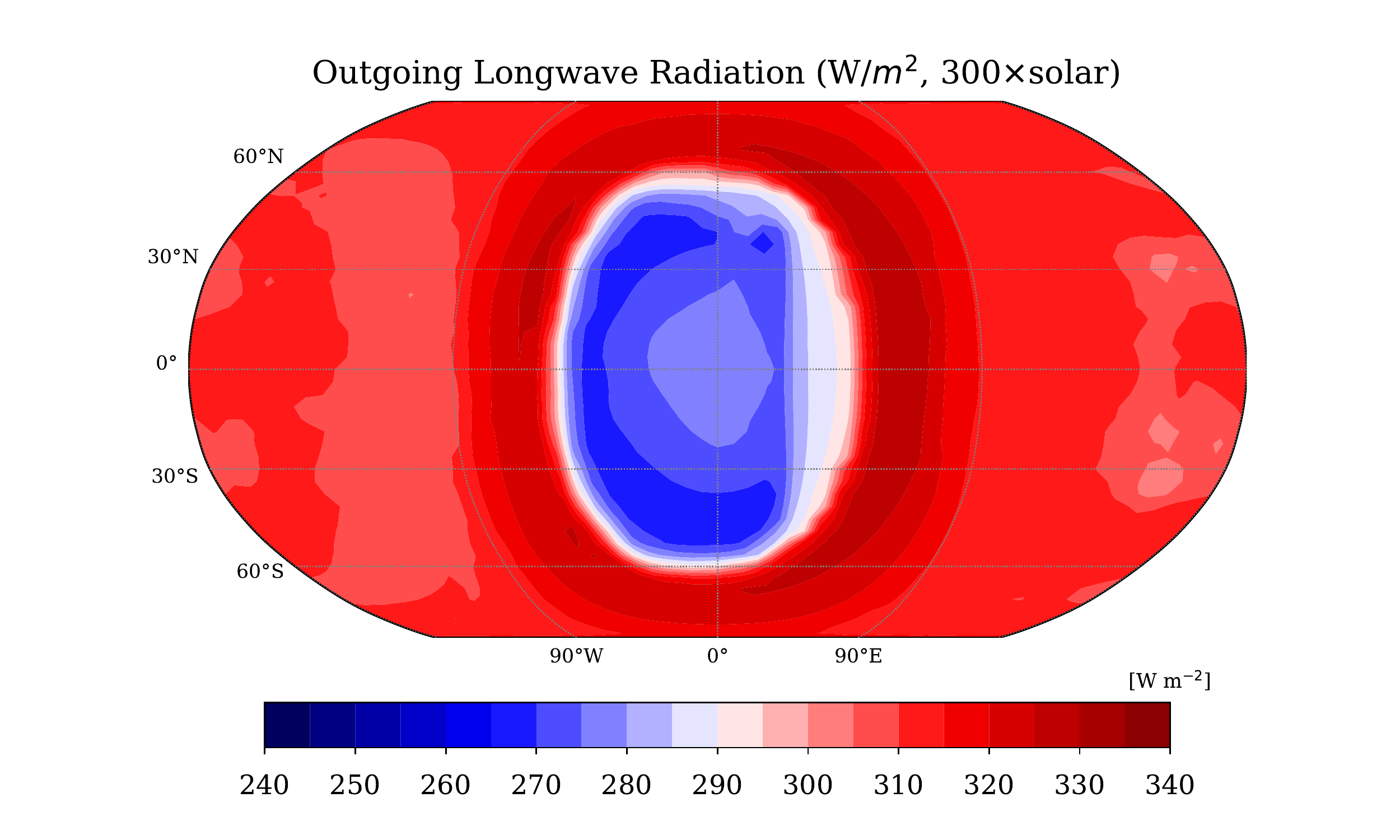}
        \includegraphics[width=6cm]{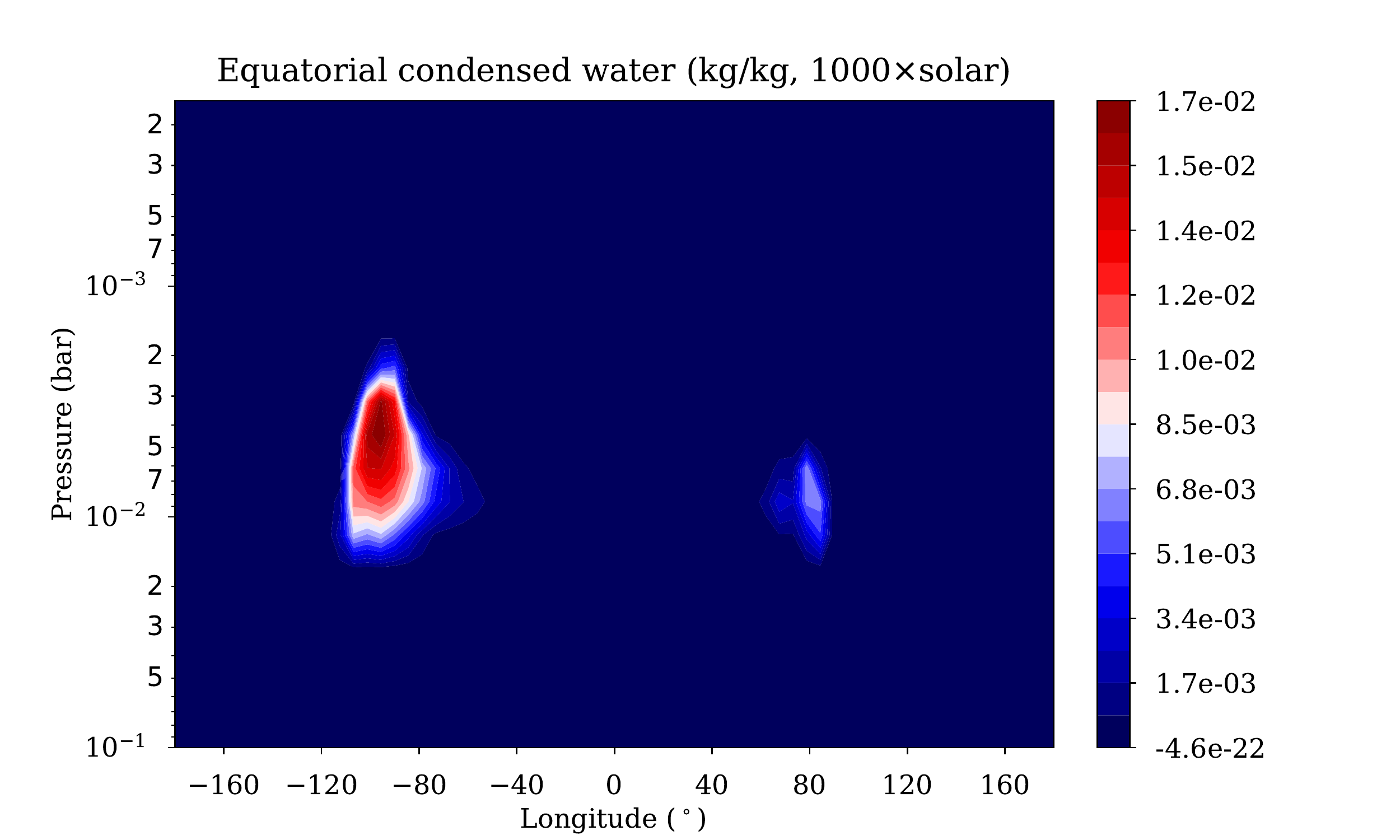}
        \includegraphics[width=6cm]{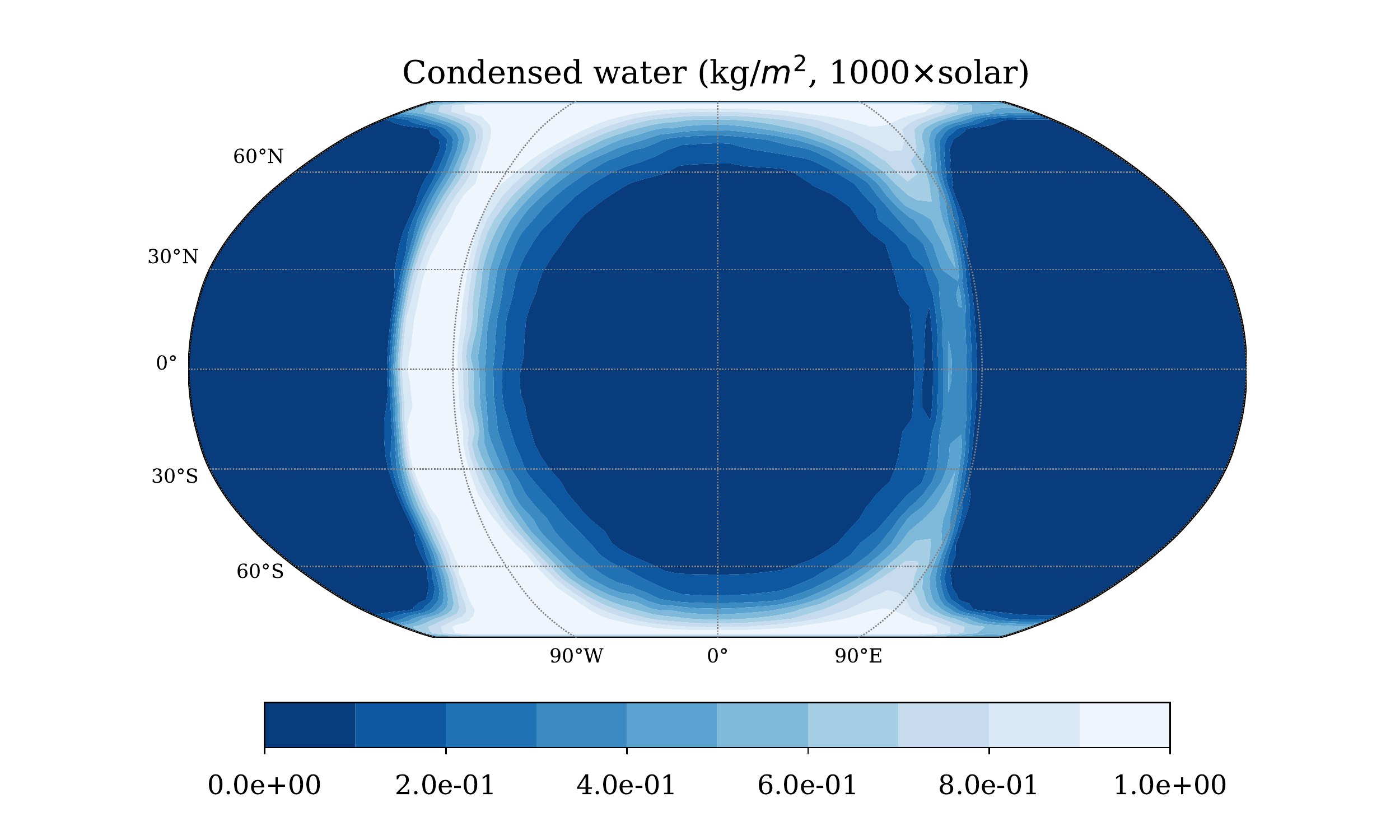}
         \includegraphics[width=6cm]{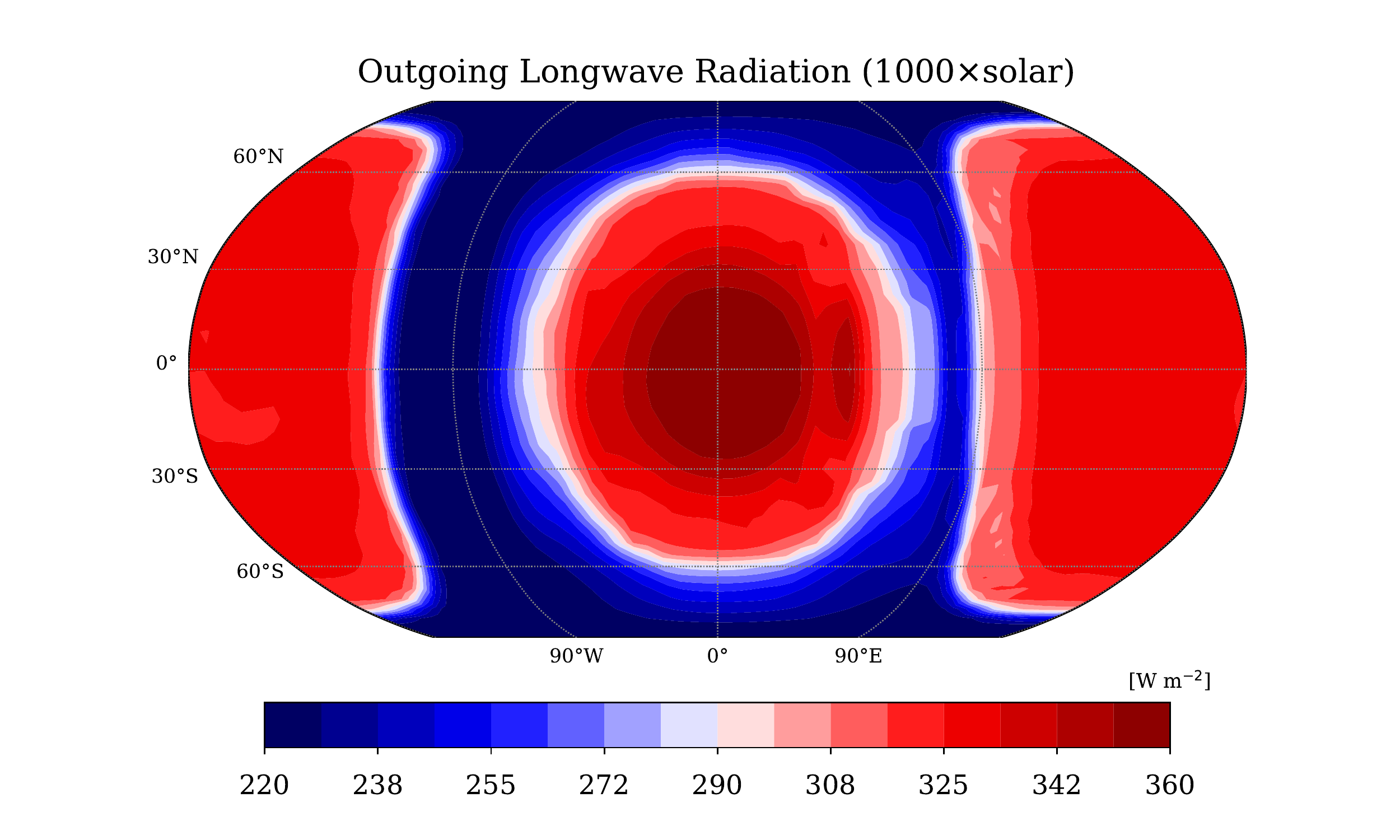}
\caption{Mean equatorial water cloud mixing ratio (left), map of vertical integrated mass of condensed water (middle) and map of outgoing long-wave radiation for 100$\times$, 300$\times,$ and 1000$\times$solar metallicity from top to bottom. We note that the colour bars are significantly different in each of the three cases.}
\label{figure_6}
\end{figure*}

As illustrated in Fig.~\ref{figure_2}, water clouds can only form on K2-18b for high atmospheric metallicity (100-1000$\times$solar) if the atmosphere is H$_2$-dominated with a solar C/O. In this case, water condensation occurs between 2 and 10~mbar, where the atmosphere is almost isothermal. Cloud formation is ruled by the day-night atmospheric circulation described in the previous subsection. For such low pressures, clouds should be composed of water ice. 
They preferentially form  at the sub-stellar point at the tropopause by the vertical advection of water vapour, or at the terminator due to the strong radiative cooling occurring when the air moves towards the night side (see Fig.~\ref{figure_4}b). Cloud precipitations evaporate below the cloud layer in undersaturated air (typically in one atmospheric scale height), as virgae on Earth. This increases the water vapour mixing ratio below the cloud layer as illustrated in the bottom-left panel of Fig.~\ref{figure_5}. Figure~\ref{figure_6} shows the equatorial cloud mass-mixing ratio and the vertical integrated mass of condensed water for 100$\times$, 300$\times$ and 1000$\times$solar metallicity. For 100$\times$ and 300$\times$solar, clouds essentially form at the sub-stellar point. They appear at similar pressures at other locations, but intermittently and with much smaller mixing ratios. 
For 1000$\times$solar metallicity, the thermal contrast between the day side and the night side is enhanced. This is due to the shorter atmospheric radiative timescale for a higher metallicity. The latter implies a higher mean molecular weight, a lower specific heat capacity per mass, and a photosphere at a lower pressure, reducing the radiative timescale (see \cite{menou12}). The day side is warmer and less favourable to cloud formation. The shorter radiative timescale makes the terminator colder and very favourable to cloud formation. Optically thick clouds form there, circling the whole planet. Despite the strong condensation, the abundance of water vapour above the cloud layer at the terminator is almost unchanged because it is advected by horizontal winds. The adiabatic warming caused by subsiding air combined with the humidity reduced by precipitation at the terminator limits cloud formation on the night side.

\begin{figure*}[!] 
\centering
        \includegraphics[width=8cm]{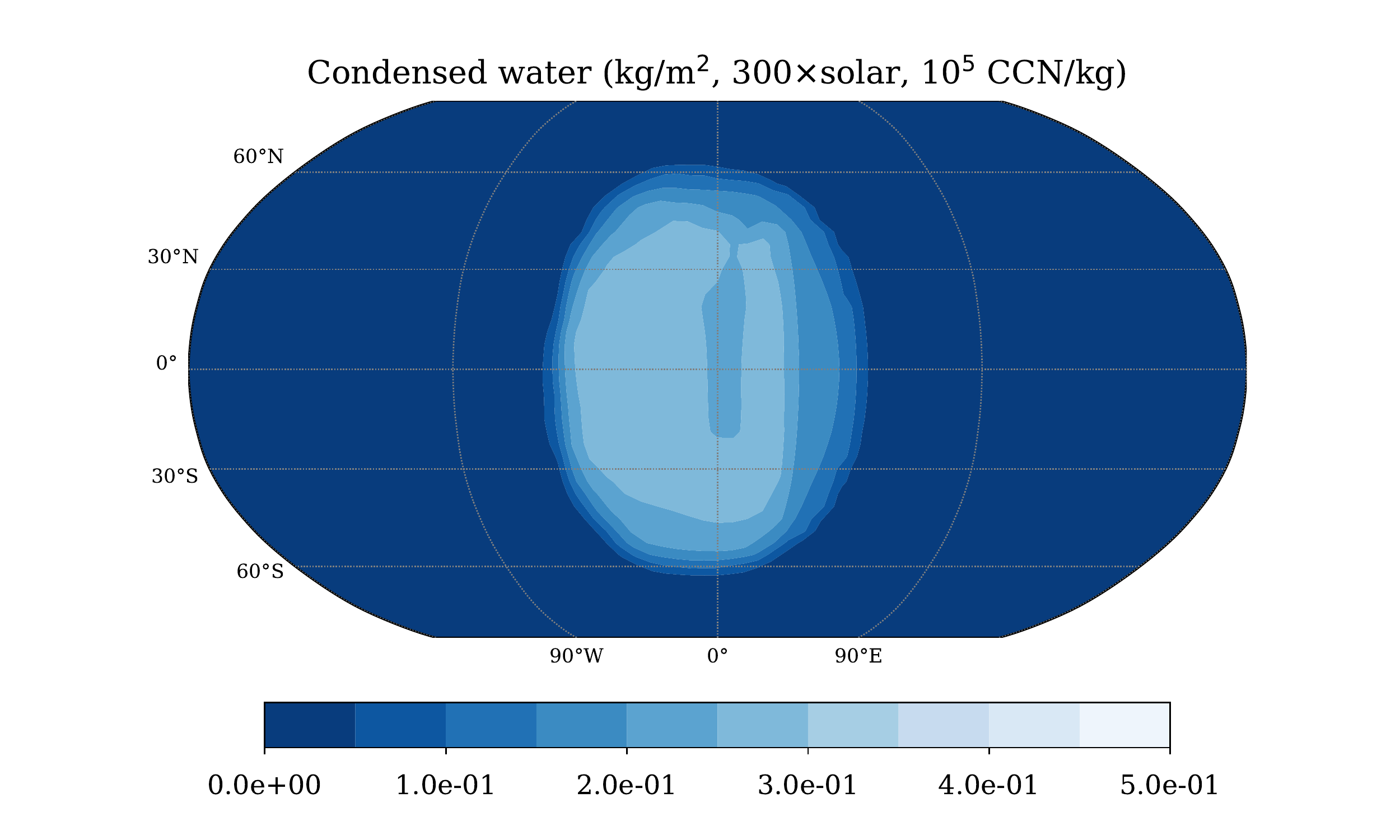}  \includegraphics[width=8cm]{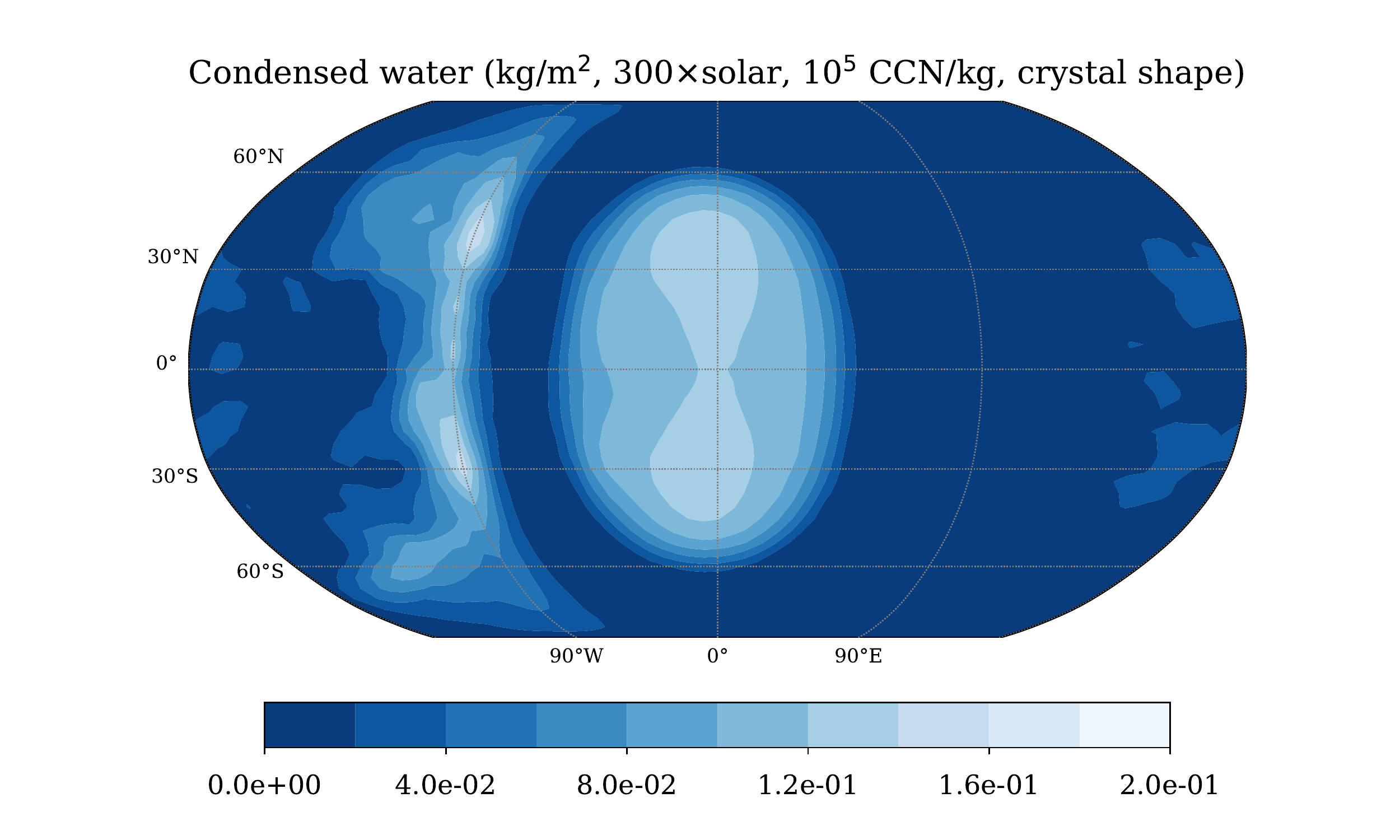}
        \includegraphics[width=8cm]{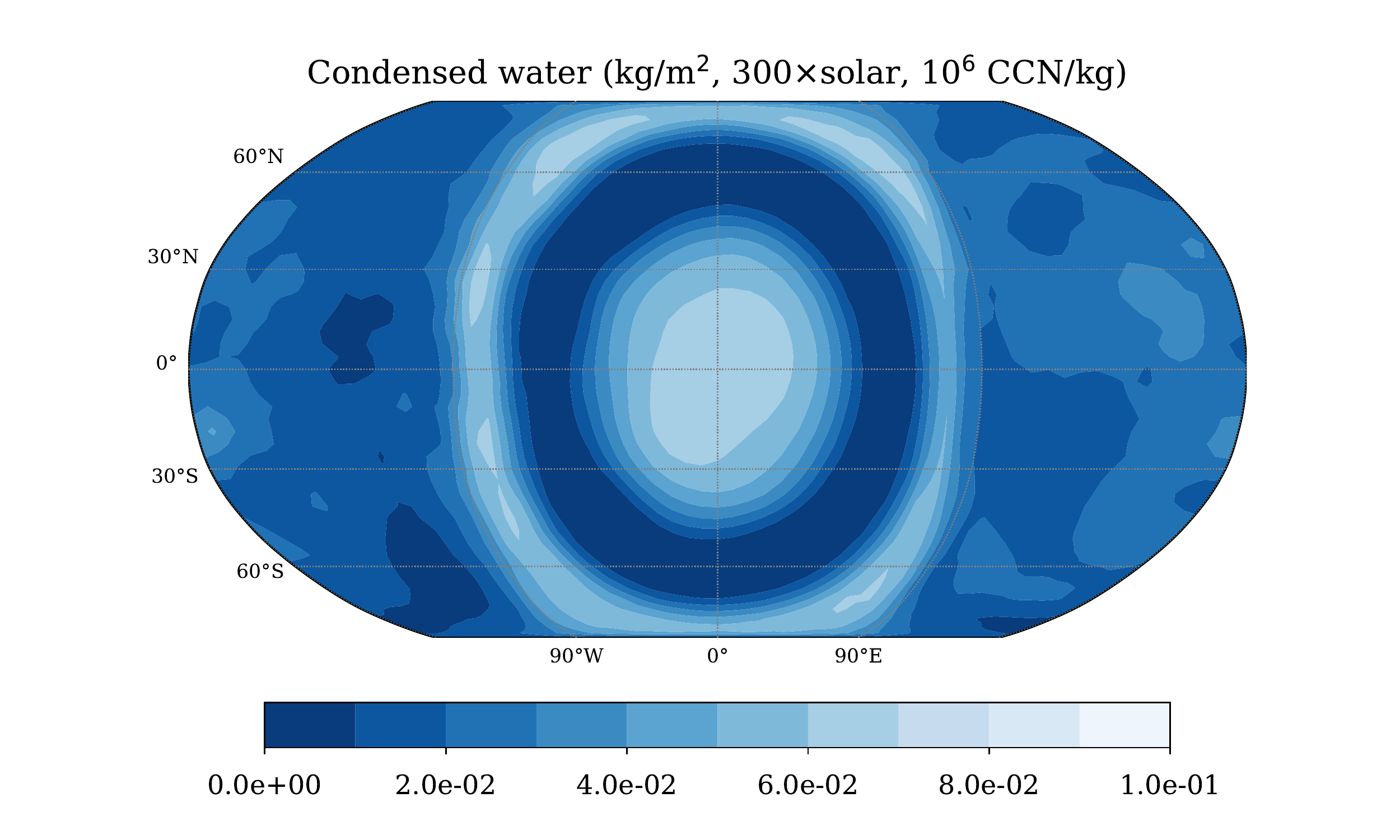}  \includegraphics[width=8cm]{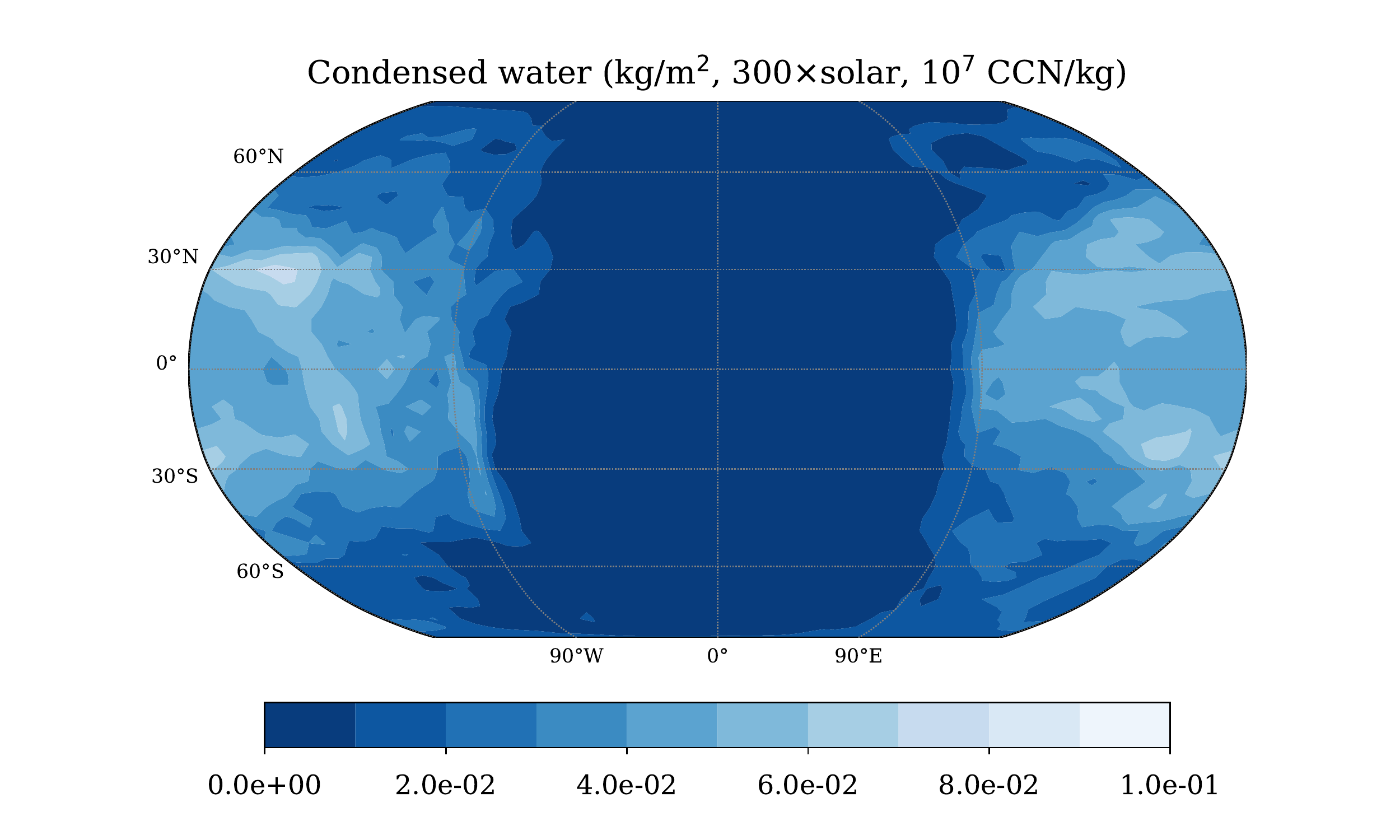} 
\caption{Map of vertical integrated mass of condensed water for CCN concentrations of 10$^5$, 10$^6,$ and 10$^7$ nuclei/kg for spherical and non-spherical (upper right) cloud particles.}
\label{figure_7}
\end{figure*}

We explored the sensitivity to cloud particle size by changing the CCN concentration for the 300$\times$solar metallicity. Figure~\ref{figure_7} shows the equatorial cloud mass-mixing ratio for spherical cloud particles with
CCN concentrations of 10$^5$, 10$^6,$ and 10$^7$ nuclei/kg of air. The cloud distribution is significantly changed with a regime of preferential cloud formation at the sub-stellar point for low CCN concentrations (10$^4$-10$^5$/kg of air) and a regime of preferential cloud formation at the terminator and on the night side for high CCN concentrations (10$^6$-10$^7$/kg of air).
By fixing the CCN concentration $n_{\rm CCN}$, the mean cloud particle size is $r_c=\left( \frac{3 q_{\rm cloud}}{4 \pi n_{\rm CCN} \rho_{\rm ice}}\right)^{1/3}$, where  $q_{\rm cloud}$ is the mass-mixing ratio of condensed water. The particle size is typically  $r_c$ = 450~$\mu$m, 200~$\mu$m, 60~$\mu$m, and 30~$\mu$m for $n_{\rm CCN}$ = 10$^4$, 10$^5$, 10$^6,$ and 10$^7$~kg$^{-1}$, respectively. For $n_{\rm CCN}$ = 10$^4$ and 10$^5$~kg$^{-1}$, the sedimentation velocity of spherical particles is $\sim$10-30~m/s at 10~mbar  (see Fig.~\ref{figure_1}), much higher than the vertical wind speed ($w\sim$ 0.2~m/s). The sedimentation timescale ($\tau_{sed}=H/v_{sed}$) is also much shorter than the advection timescale ($\tau_{adv}=R_p/v$). Clouds are not efficiently transported and are limited to their initial formation location (essentially the sub-stellar point). For $n_{\rm CCN}$ = 10$^6$ and 10$^7$, the sedimentation velocity of spherical is $\sim$0.3-1~m/s, and the sedimentation timescale is relatively close to the advection timescale ($\tau_{sed}/\tau_{adv}=0.1-0.3$). This ratio is enhanced by upward vertical winds on the day side that reach values close to 1 for $n_{\rm CCN}$ = 10$^7$~kg$^{-1}$ taking into account vertical winds. Clouds are then quite efficiently horizontally advected to the terminator, reducing the efficiency of the day-side cold trapping. This enhanced global water transport favours cloud formation at terminator and on the night side. Surprisingly, cloud formation is reduced at the day side. This is due to a cloud radiative feedback. Clouds formed at the terminator or on the night side warm the atmosphere below the cloud deck by a greenhouse effect. The greenhouse effect is illustrated in Fig. \ref{figure_6} with a reduction of the outgoing long-wave radiation where clouds are present. Because of the efficient heat redistribution, the day side becomes warmer and potentially too warm for cloud formation, suppressing the associated cold trapping and the horizontal/vertical gradient of humidity. This reinforces cloud formation at the terminator and on the night side. In contrast, we found that the cloud albedo effect is limited. For K2-18's spectrum (M2.8 stellar spectral type) and for K2-18b's CH$_4$- and H$_2$O-rich atmosphere, a large part of the stellar radiation is absorbed by CH$_4$ and H$_2$O above clouds. The planetary albedo is always lower than 0.15, and the cloud distribution only marginally changes this value.

We also performed a simulation with $n_{\rm CCN}$ = 10$^5$ and with non-spherical ice particles (rimed dendrites, see Section 2). Sedimentation speed is reduced by a factor of $\sim$4, and the results are intermediate between the case with $n_{\rm CCN}$ = 10$^5$ and the case with $n_{\rm CCN}$ = 10$^6$. The morning (west) terminator appears more cloudy than the evening (east) terminator because it is slightly colder.
Exploring a large range of CCN concentration is a simple way to cover the possible impact of microphysics and particle shape. 

From these simulations, we conclude that the cloud distribution on K2-18b is controlled by the global day-side-to-night-side circulation, particle size, and cloud radiative effects. Radiative feedback can profoundly alter cloud distribution, making it very sensitive to cloud microphysics. In any case, the terminator is at least partially cloudy for metallicity $\geqslant$100$\times$solar with clouds confined between 2 and 10 mbar.

\subsection{Cloud variability}

\begin{figure}[!] 
\centering
        \includegraphics[width=8cm]{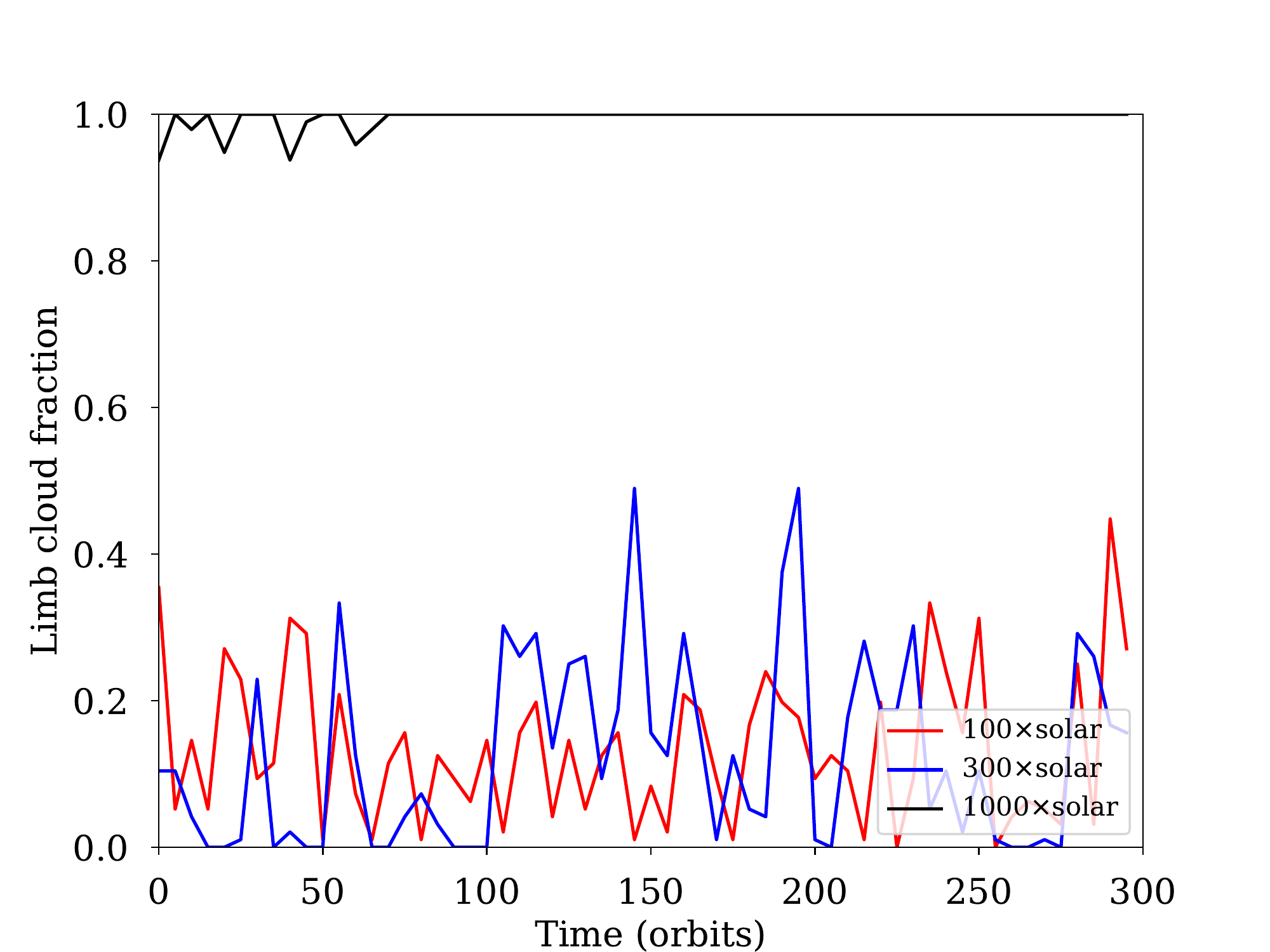}
\caption{Time variability of cloud fraction at terminator for 100$\times$, 300$\times,$ and 1000$\times$solar metallicity and a CCN concentration of 10$^5$nuclei/kg.}
\label{figure_8}
\end{figure}

\begin{figure}[!] 
\centering
        \includegraphics[width=8cm]{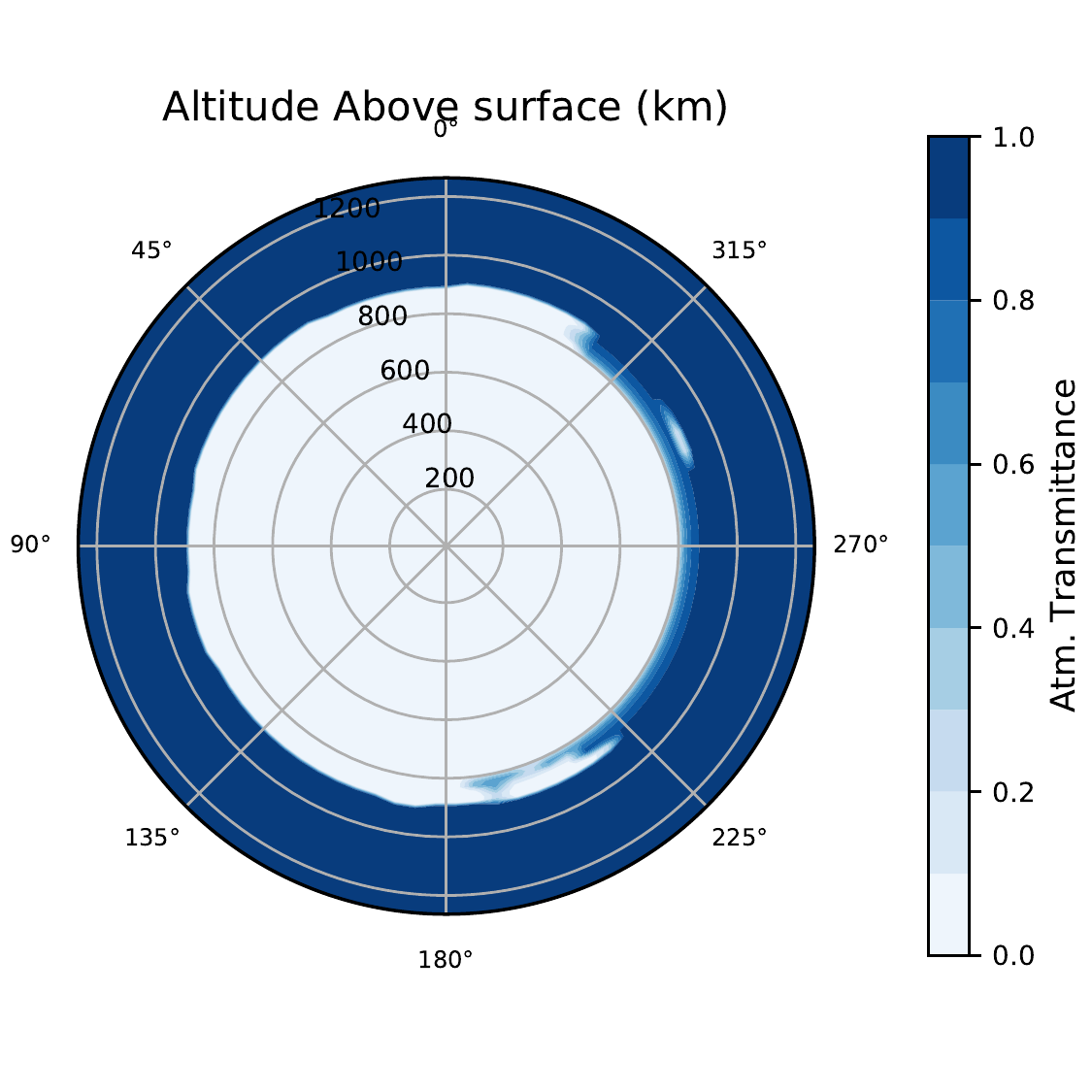}
        \includegraphics[width=8cm]{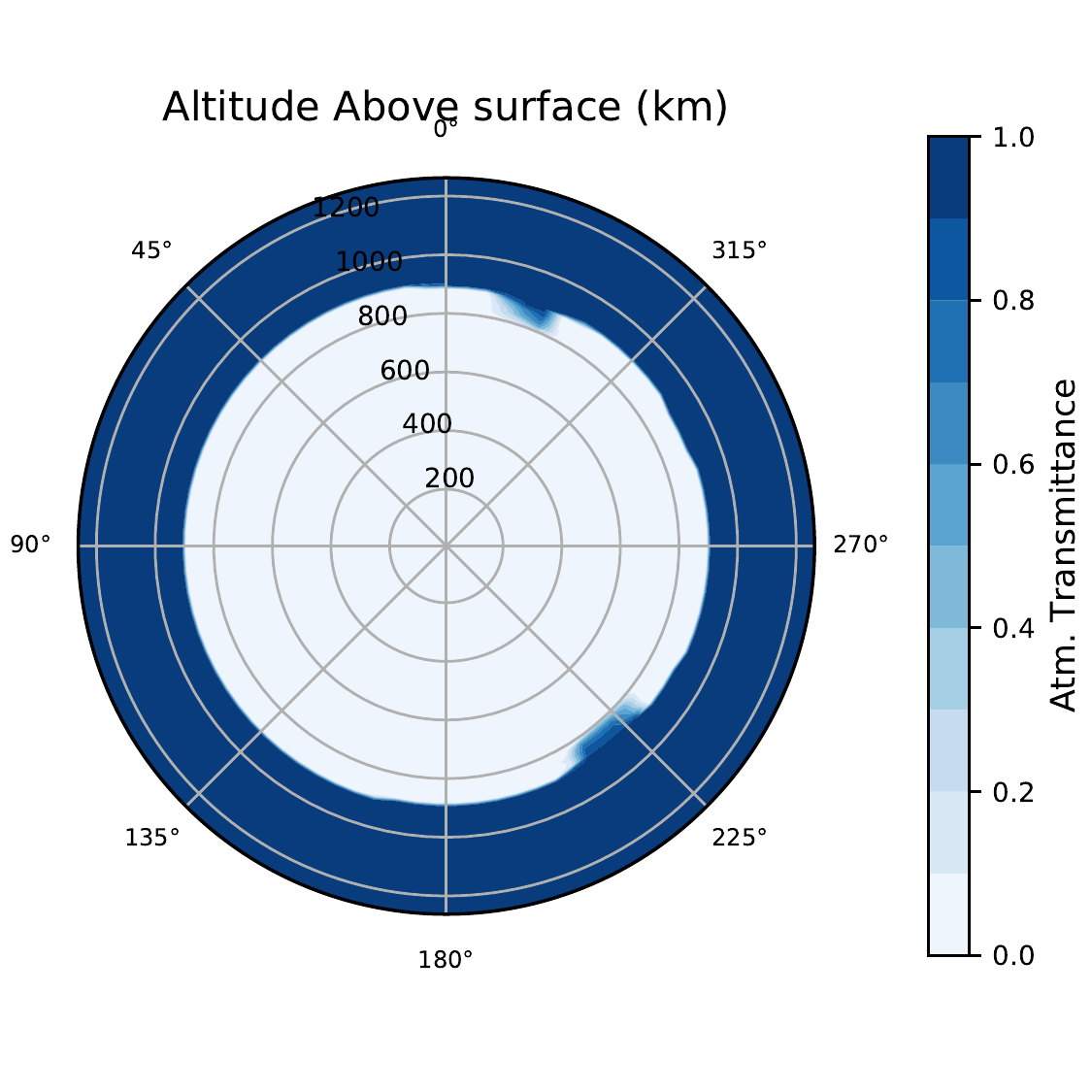}
\caption{Maps of transmittance of the atmosphere at two different times for the 300$\times$solar metallicity case with 10$^6$nuclei/kg computed with Pytmosph3R \citep{caldas19}. The colour shading shows the transmittance that goes from zero where the atmosphere is opaque to one where it completely transmits the light. The surface is defined here as the first model level (80 bar). These maps are computed at 0.4 micron where the contribution of water clouds is particularly visible. The atmosphere is seen from the night side so that the north pole corresponds to 0$^\circ$ and the equatorial east limb (longitude of +90$^\circ$) corresponds to 90$^\circ$ in the plots.}
\label{figure_8b}
\end{figure}

Cloud infrared opacity produces a significant greenhouse effect, warming the atmosphere below the cloud layer. As discussed in the previous paragraph, this cloud radiative effect has a strong impact on the spatial cloud distribution. We also found that it produces time variability in the cloud fraction. We computed the global cloud fraction and the cloud fraction at the terminator for the different atmospheric compositions assuming that the atmosphere is cloudy when the cloud optical depth in the visible range ($\tau^{\rm cloud}$) is greater than 1. The cloud optical depth in transit ($\tau^{\rm cloud}_{\rm H}$) is related to the normal cloud optical depth  ($\tau^{\rm cloud}_{\rm V}$) by \cite{fortney05}:
\begin{equation} 
\eta =\frac{\tau^{\rm cloud}_{\rm H}}{ \tau^{\rm cloud}_{\rm V}}= \sqrt{\frac{2 \pi H}{R_p}}
\label{eq5}
,\end{equation}
with $\eta$$\sim$65 for 300$\times$solar metallicity. In the optical regime ($r\gg\lambda$), the particle extinction cross-section is equal to $2\pi r^2,$ and the normal cloud optical depth can be expressed as follows: 
\begin{equation} 
\tau^{\rm cloud}_{\rm V} = \frac{3 m_{\rm cloud}}{2 r_c \rho_{ice}}
\label{eq6}
,\end{equation} 
where $m_{\rm cloud}$ is the vertical integrated mass of condensed water. Clouds are optically thick ($\tau^{\rm cloud}_{\rm H} \geqslant 1$) for $m_{\rm cloud}\geqslant \frac{2 r_c \rho_{ice}}{3 \eta}$.  For $r_c$ = 200 $\mu$m, this corresponds to $m_{\rm cloud}\geqslant$ 1.9$\times$10$^{-3}$ kg/m$^2$. 
From Fig. \ref{figure_6}, the vertical integrated mass of condensed water is generally much larger than 0.01 kg/m$^2$ when clouds form, meaning that water clouds on K2-18b are optically thick at the limbs. We note that what matters is the cloud optical depth in the visible part of the atmosphere. Our simplified calculation is valid because clouds form high enough in the atmosphere, with few latitudinal variations of the vertical cloud extent at the terminator.
Using the time-averaged GCM outputs, clouds would be optically thick at the terminator at all latitudes for all cases. We note that atmospheric columns close to the terminator can significantly contribute to the limb's optical depth. We computed the mean optical depth over an opening angle of 15 to 30$^\circ$ longitude, depending on the atmospheric scale height and based on the work by \cite{caldas19} (see Fig. 2 in their paper).
If we use the instantaneous GCM outputs, the mean cloud fraction is around 99$\%$ for 1000$\times$solar, and around  13$\%$ for 100$\times$solar and 300$\times$solar metallicity (see Fig. \ref{figure_8}). For the last two cases, the cloud fraction varies between 0$\%$ and 50$\%$. This difference between the instantaneous and time-averaged atmospheric state can have important implications for the observations, as discussed in Section 4. Figure \ref{figure_8b} shows two maps of the transmittance of the atmosphere at two different times for a relatively cloudy case. These maps were computed with Pytmosph3R, which computes the optical depth of light rays passing through the 3D atmospheric structure of the GCM on a grid of altitudes and  azimuths \citep{caldas19}. The transmittance shows where the atmosphere becomes opaque and does not transmit the light from the star to the observer (transmittance goes to zero). Such maps can be computed at every wavelength and integrated spatially to yield the effective transit radius. The maps shown in Figure \ref{figure_8b} are computed at 0.4 micron, where the contribution of water clouds is particularly visible because there are few strong molecular bands. At this wavelength, Rayleigh scattering creates an opaque floor at an altitude of about 800 km above the first model level (80 bar). The altitude of this opaque floor can rise to the altitude of the cloud deck ($\sim$900 km), where these clouds are present at the limb (the whole limb in the upper panel, and the east limb in the lower panel).

\subsection{Asynchronous rotation}
\begin{figure*}[!] 
\centering
        \includegraphics[width=8cm]{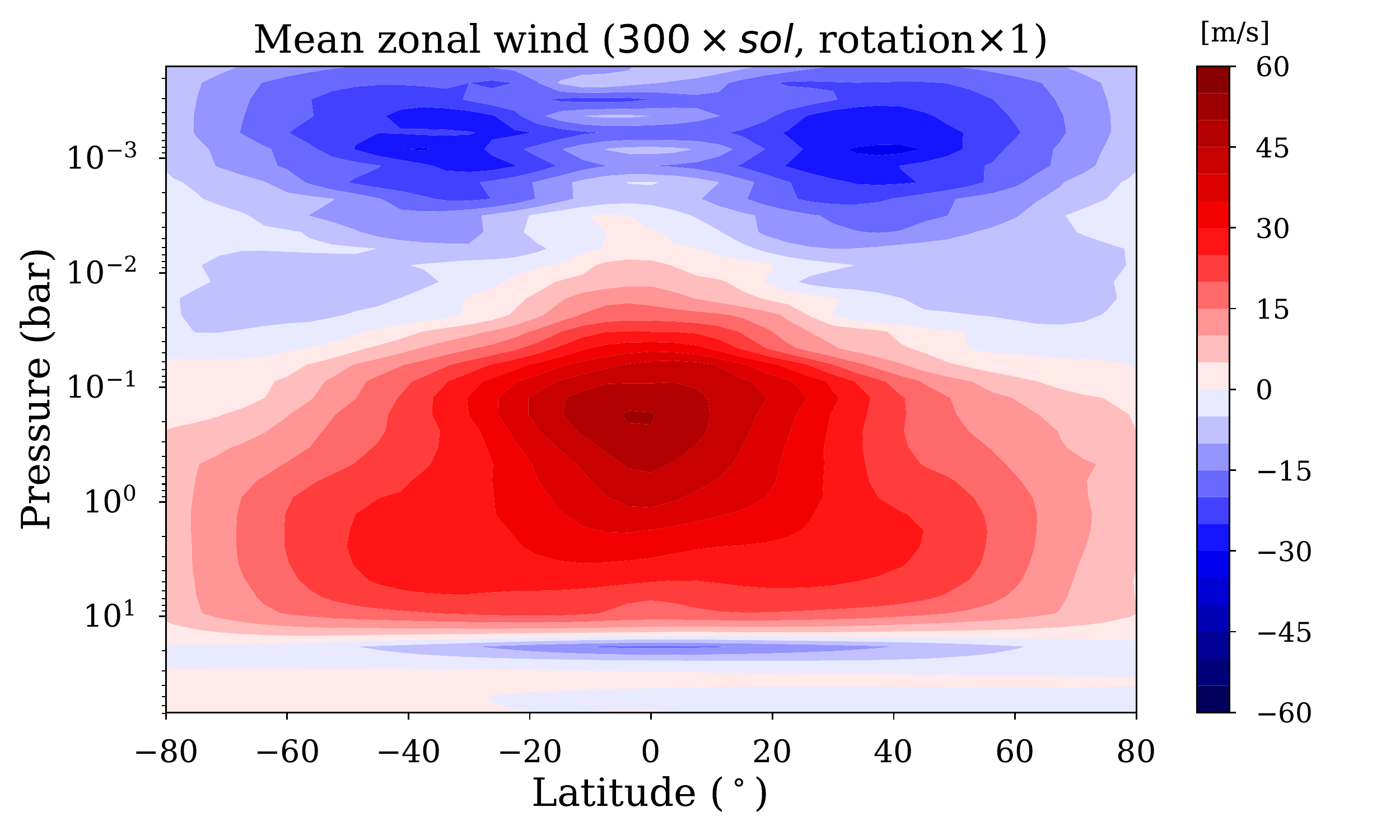} 
        \includegraphics[width=8cm]{h2oice_col_300xsol.pdf}
        \includegraphics[width=8cm]{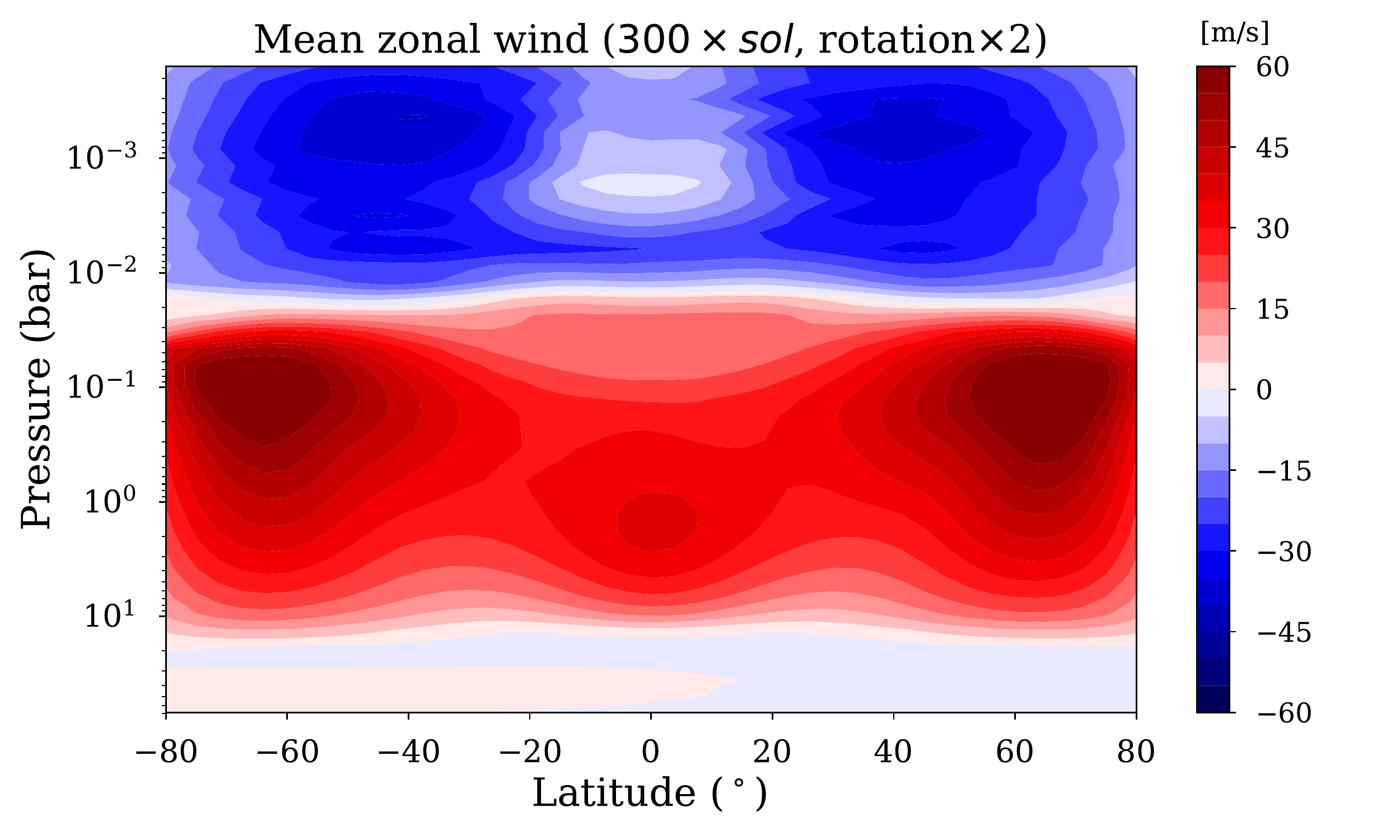}
        \includegraphics[width=8cm]{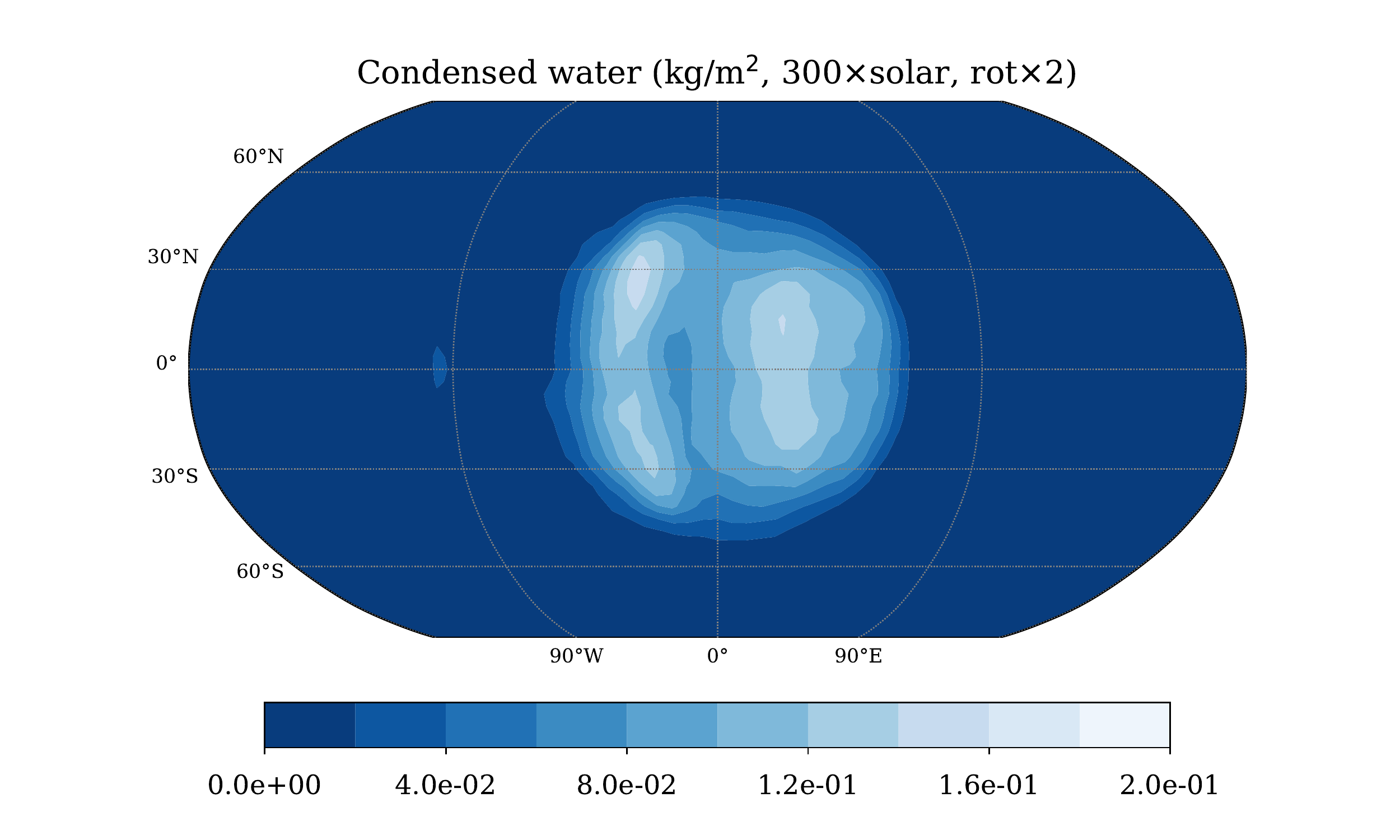}                
        \includegraphics[width=8cm]{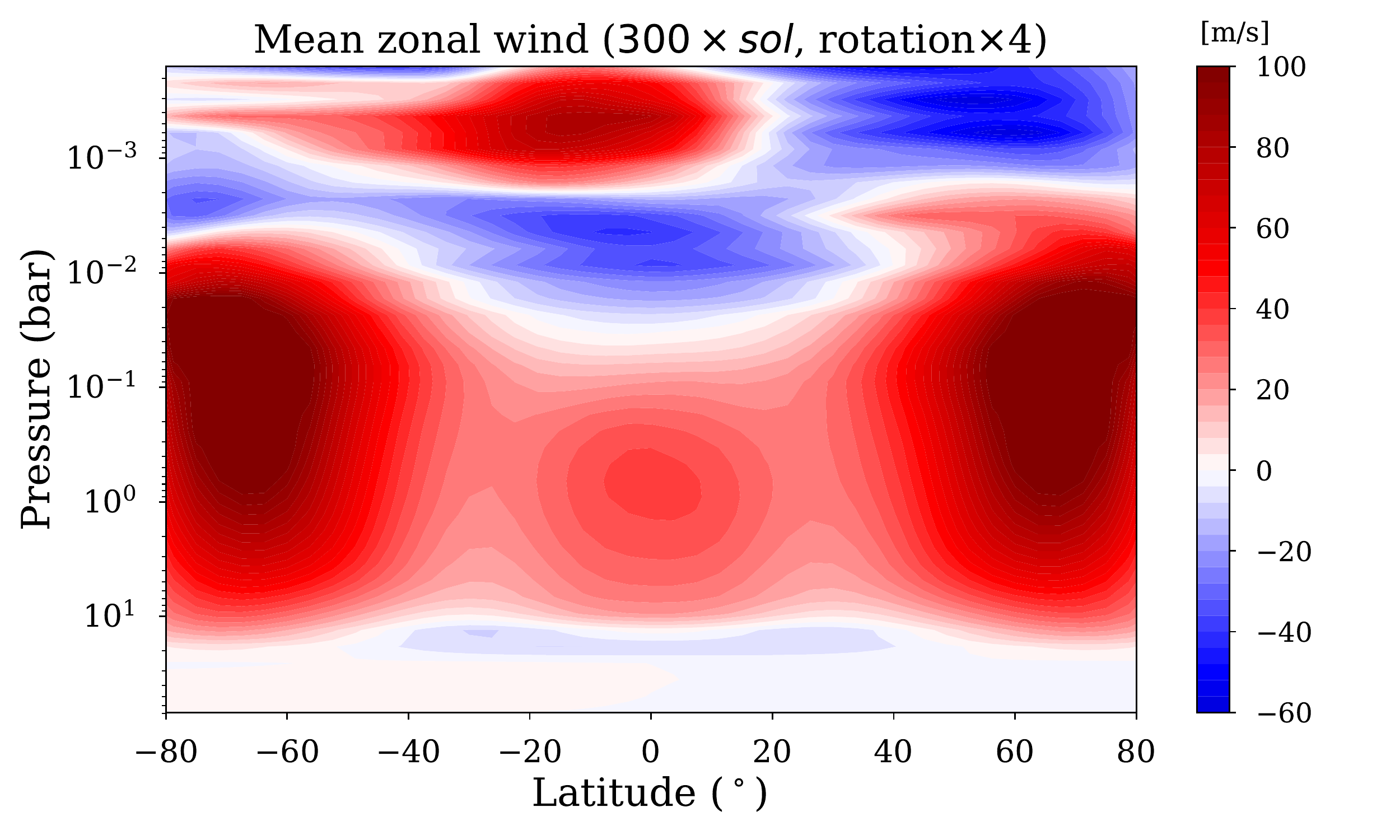}
        \includegraphics[width=8cm]{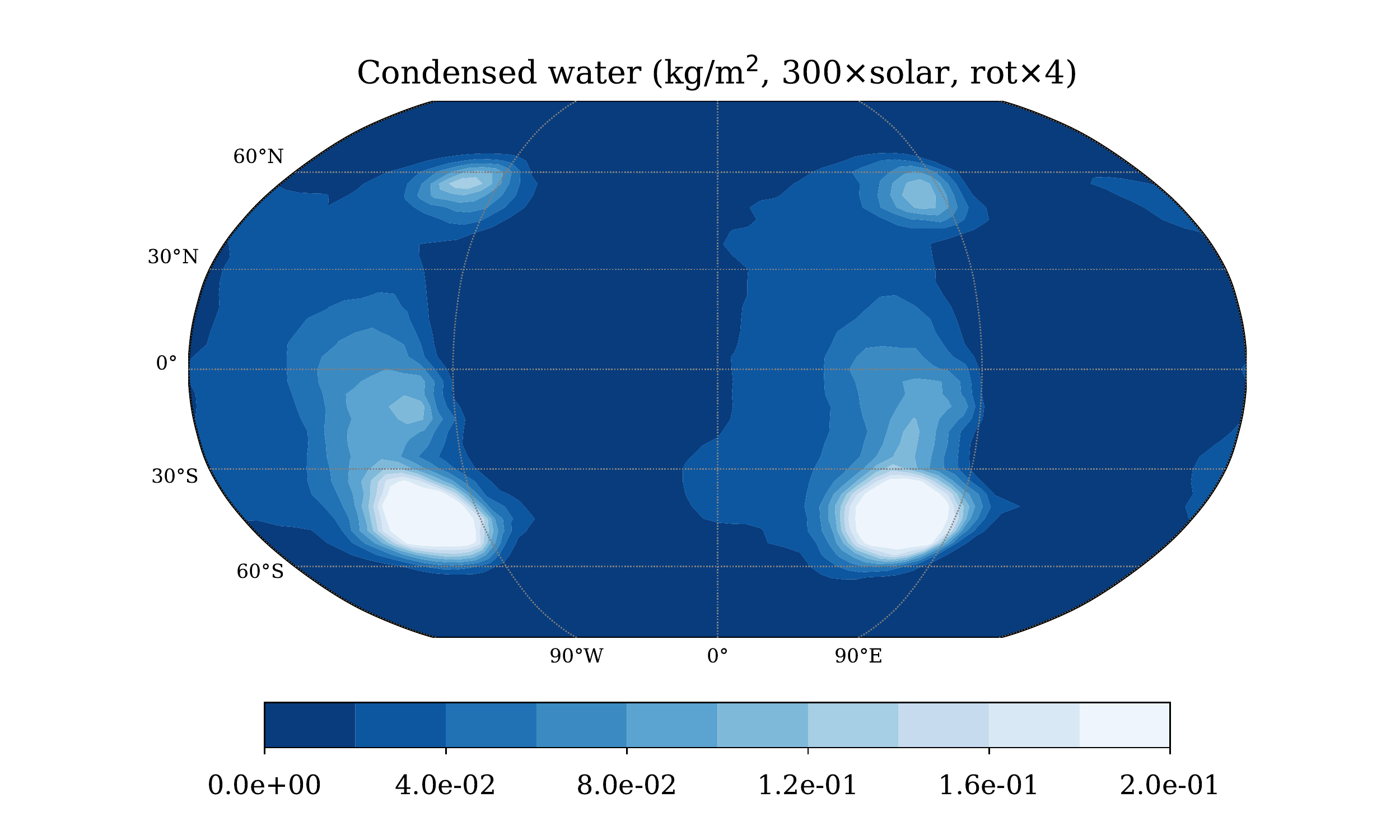}                
        \includegraphics[width=8cm]{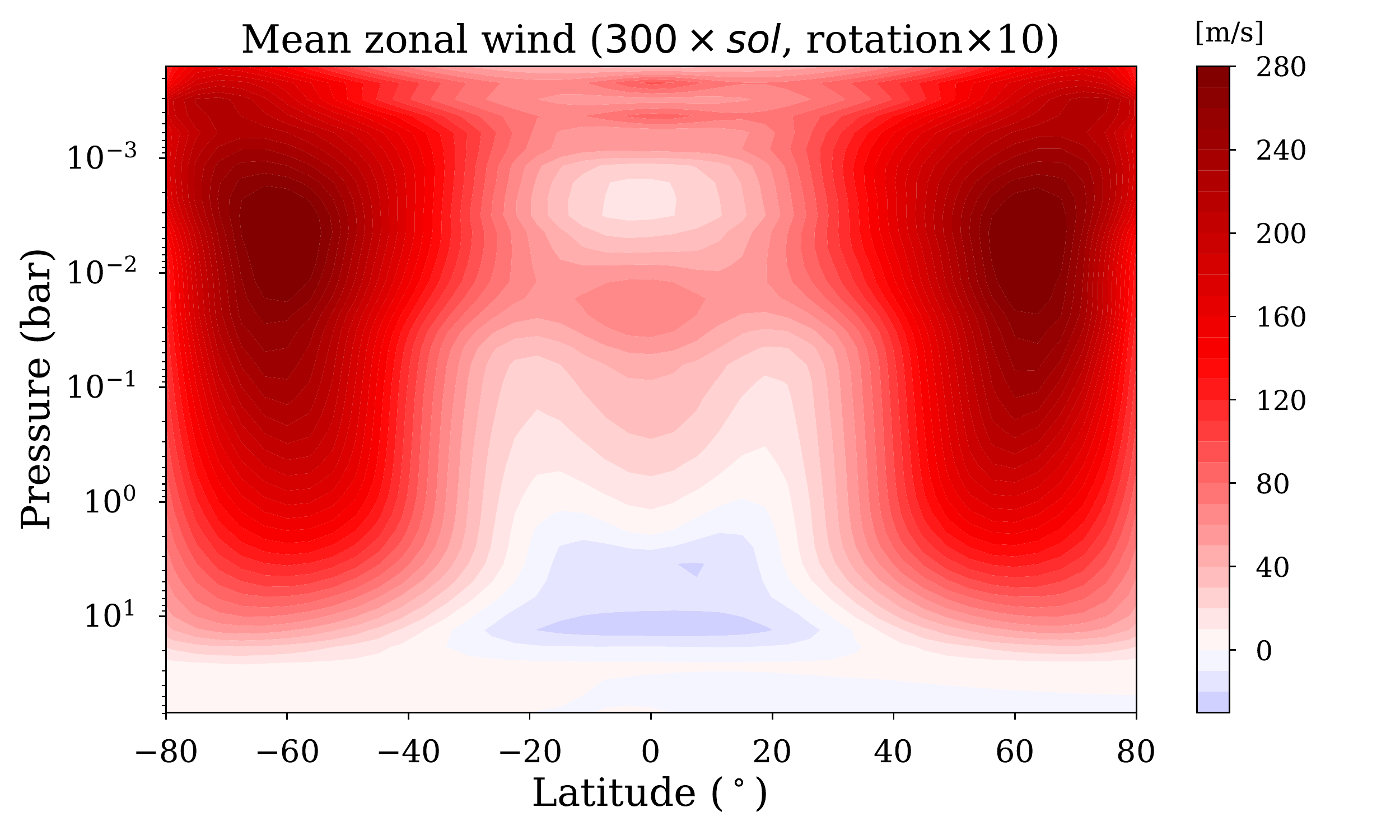}
        \includegraphics[width=8cm]{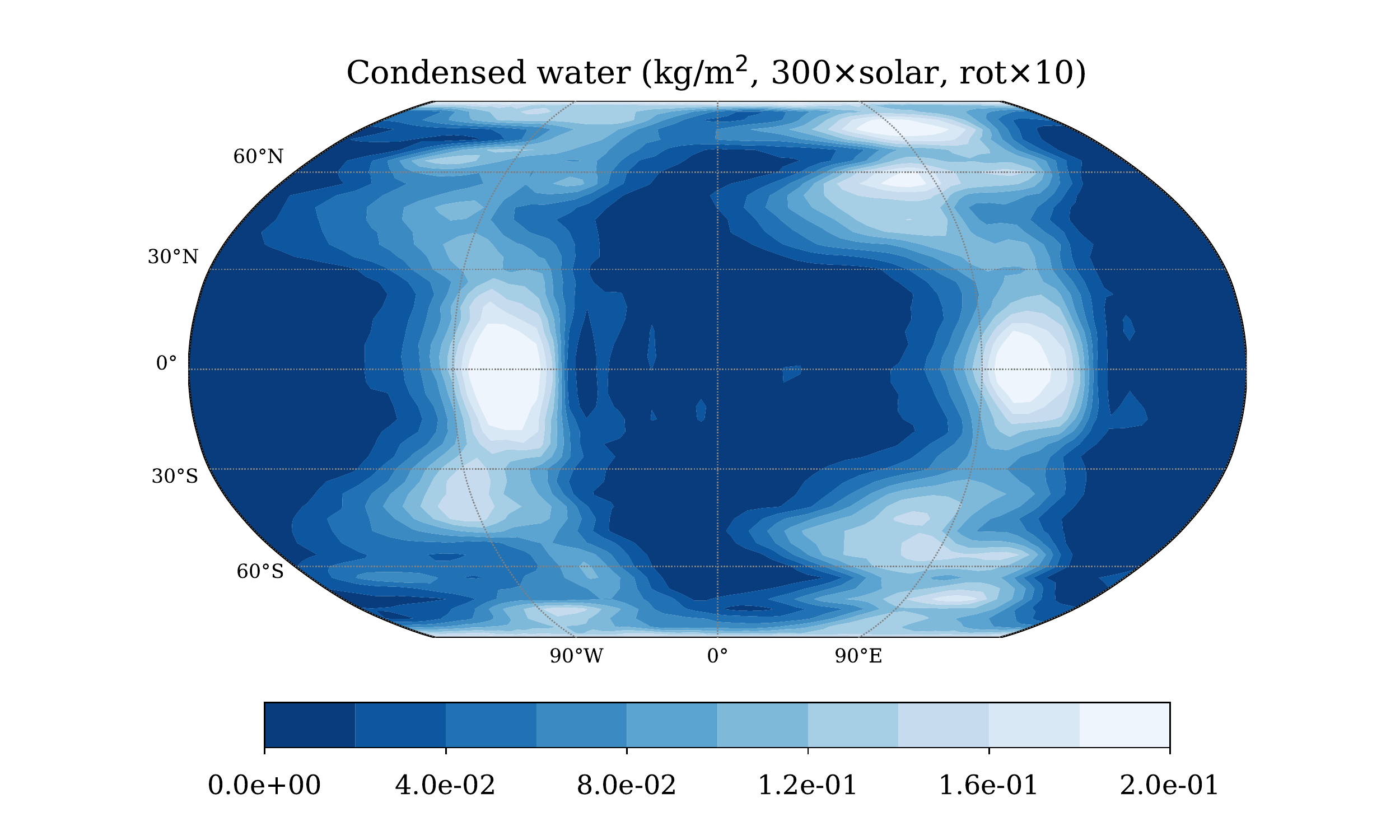}                
\caption{Zonally averaged zonal wind as a function of latitude and pressure, and a map of vertical integrated mass of condensed water for 1:1, 2:1, 4:1, and 10:1 spin-orbit resonance (from top to bottom).}
\label{figure_9}
\end{figure*}

As discussed in Section 2.3, K2-18b may have a non-synchronous rotation, since it is close to the limit for tidal locking. We tested the impact of the rotation rate on the atmospheric dynamics and cloud distribution for 2:1, 4:1, and 10:1 spin-orbit resonance. Figure \ref{figure_9} shows zonally averaged zonal wind and
maps of the vertical integrated mass of condensed water. For the 2:1 resonance, two zonal jets appear at high latitudes. The cloud distribution is mostly unchanged, with a preferential formation at the sub-stellar point, but the amount of cloud at the sub-stellar point is reduced by a factor of $\sim$2 due to a weaker day-night circulation.
The cloud distribution for a 3:2 spin-orbit resonance would therefore be similar to our reference 1:1 case, with a reduction of the amount of cloud at the sub-stellar point. For a four-times faster rotation, the high-latitude jet is reinforced. The cloud distribution is significantly changed, with preferential formation at the terminator. The cloud maximum is shifted eastward due to the easterlies at pressures lower than 10 mbar, where clouds form. For a ten-times faster rotation, the super-rotation is well developed at mid and high latitudes, and easterlies appear at the equator in the troposphere. The circulation is closer to an Earth-like circulation. Again, clouds preferentially form at the terminators, but their maximum density is shifted eastward due to the super-rotation, which is developed everywhere in the stratosphere.

\section{Observational transit spectra}

\begin{figure}[!] 
\begin{center} 
        \includegraphics[width=8cm]{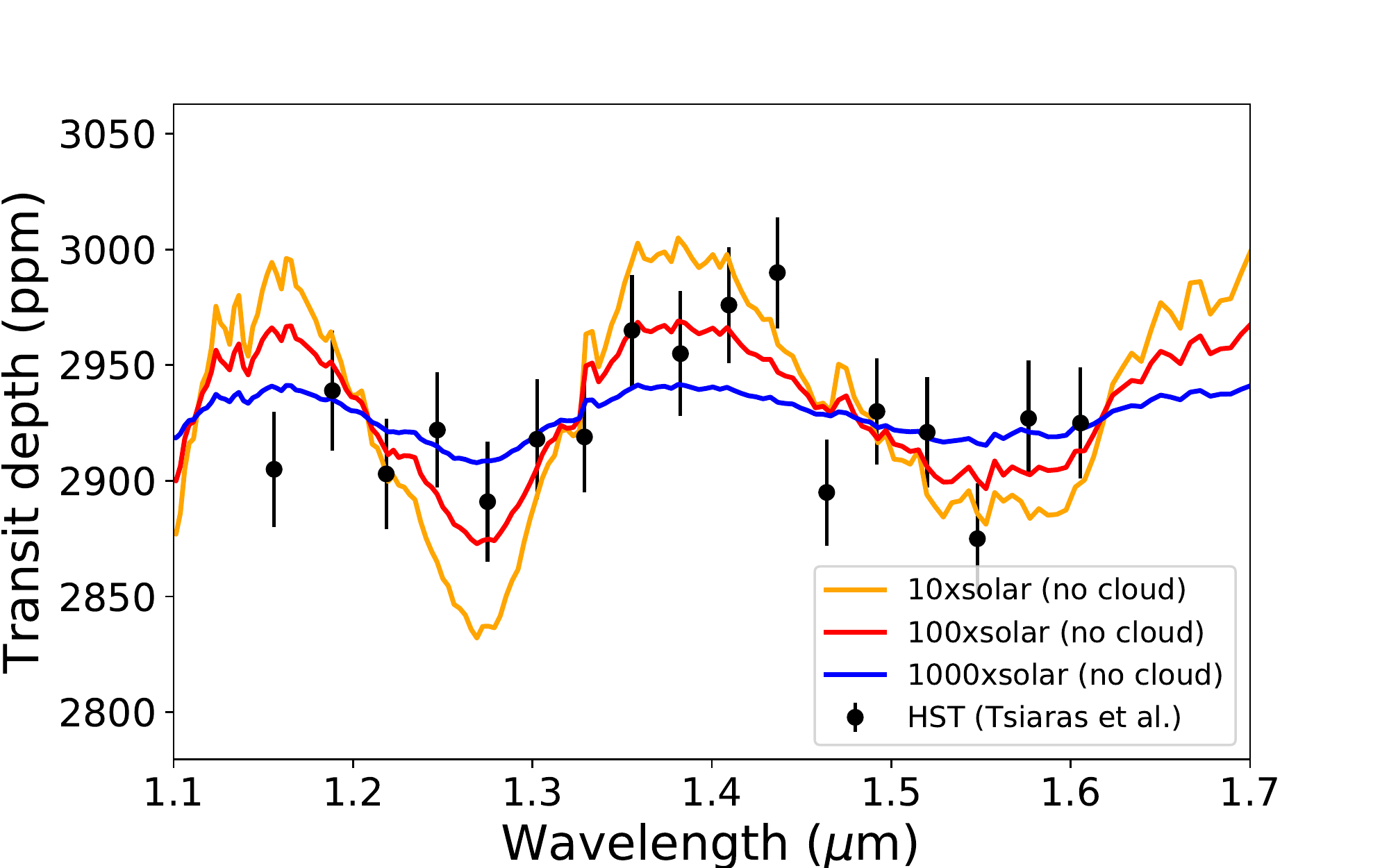}
\end{center}  
\caption{Transit spectra computed with Exo-REM from the outputs of the GCM for 10$\times$, 100$\times$ and 1000$\times$solar metallicity without cloud. HST data from \cite{tsiaras19} are indicated with black dots.}
\label{figure_10}
\end{figure}

Using the outputs of the 3D GCM, we computed transit spectra using Exo-REM \citep{baudino15, baudino17, charnay18, blain20}, and we compared them to HST data from \cite{tsiaras19} and \cite{benneke19b} in order to constrain the atmospheric metallicity and to highlight implications for future observations.
We computed spectra with k-coefficient bands of 20 cm$^{-1}$ (R$\sim$500 at 1 $\mu$m) using the temperature and cloud profiles from the GCM at the terminators. Figure \ref{figure_10} shows the transit spectra with no cloud for 10$\times$, 100$\times,$ and 1000$\times$solar metallicity compared to HST/WFC3 data from \cite{tsiaras19}. As shown by \cite{bezard20} and also discussed in \cite{blain20}, the transit spectrum in the HST/WFC3 band, and in particular the 1.4 $\mu$m band, is dominated by CH$_4$ absorption for a H$_2$-dominated atmosphere with solar C/O. Spectral features are too deep for the 1$\times$solar and 10$\times$solar metallicities compared to observations. These two cases are ruled out if they are cloud-free, as predicted by a 3D model.

\begin{figure*}[!] 
\centering
        \includegraphics[width=8cm]{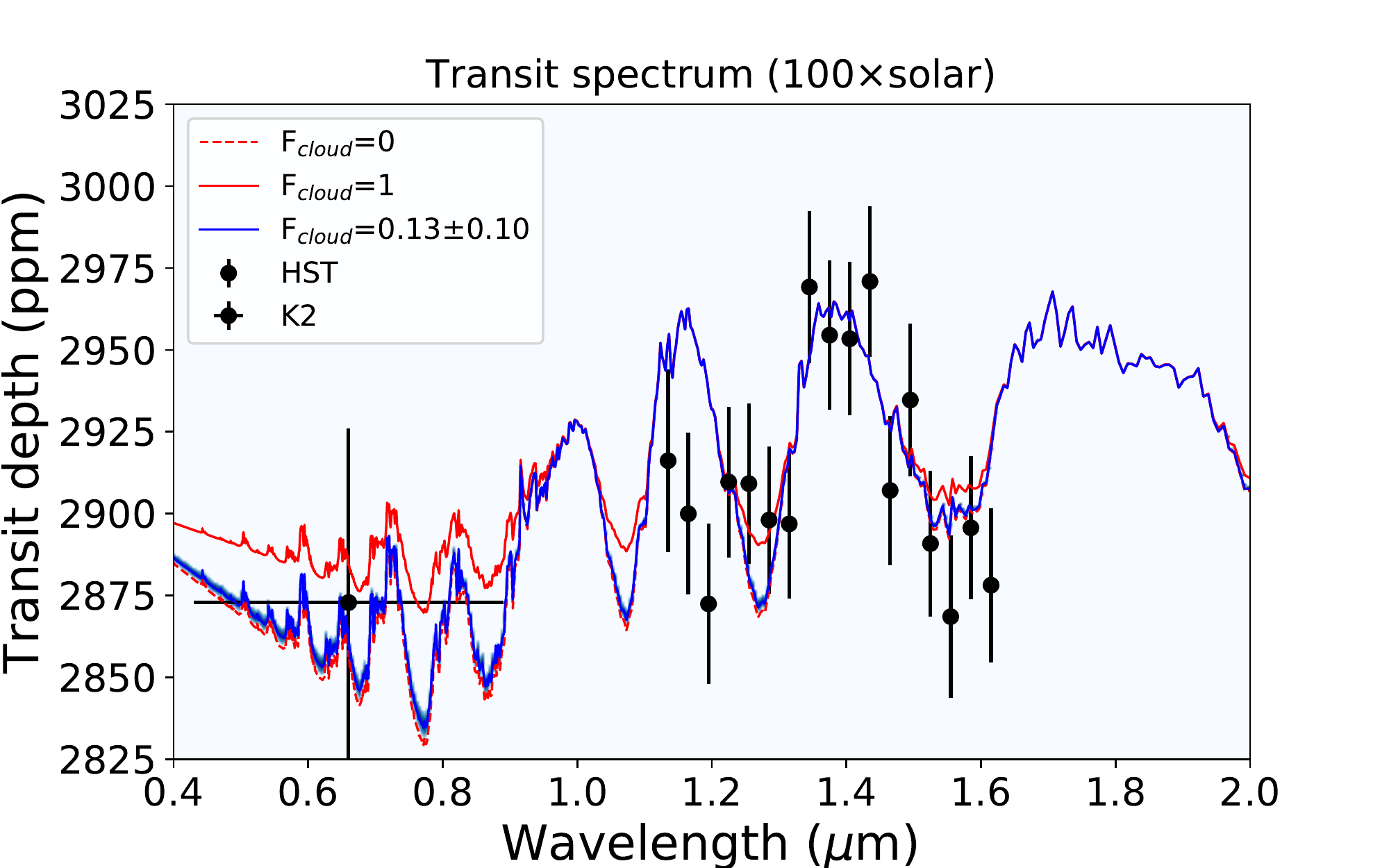}   
        \includegraphics[width=8cm]{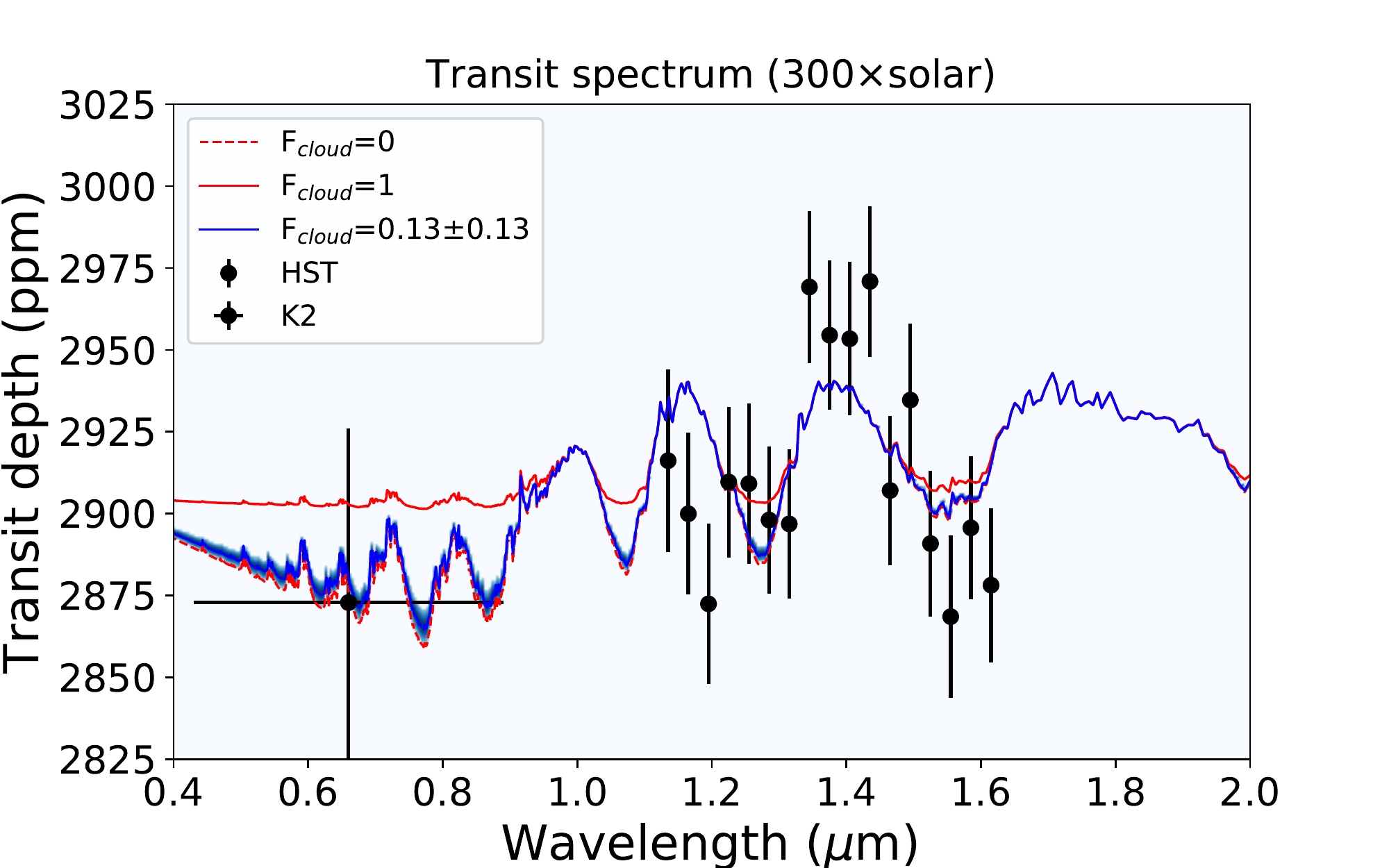}   
        \includegraphics[width=8cm]{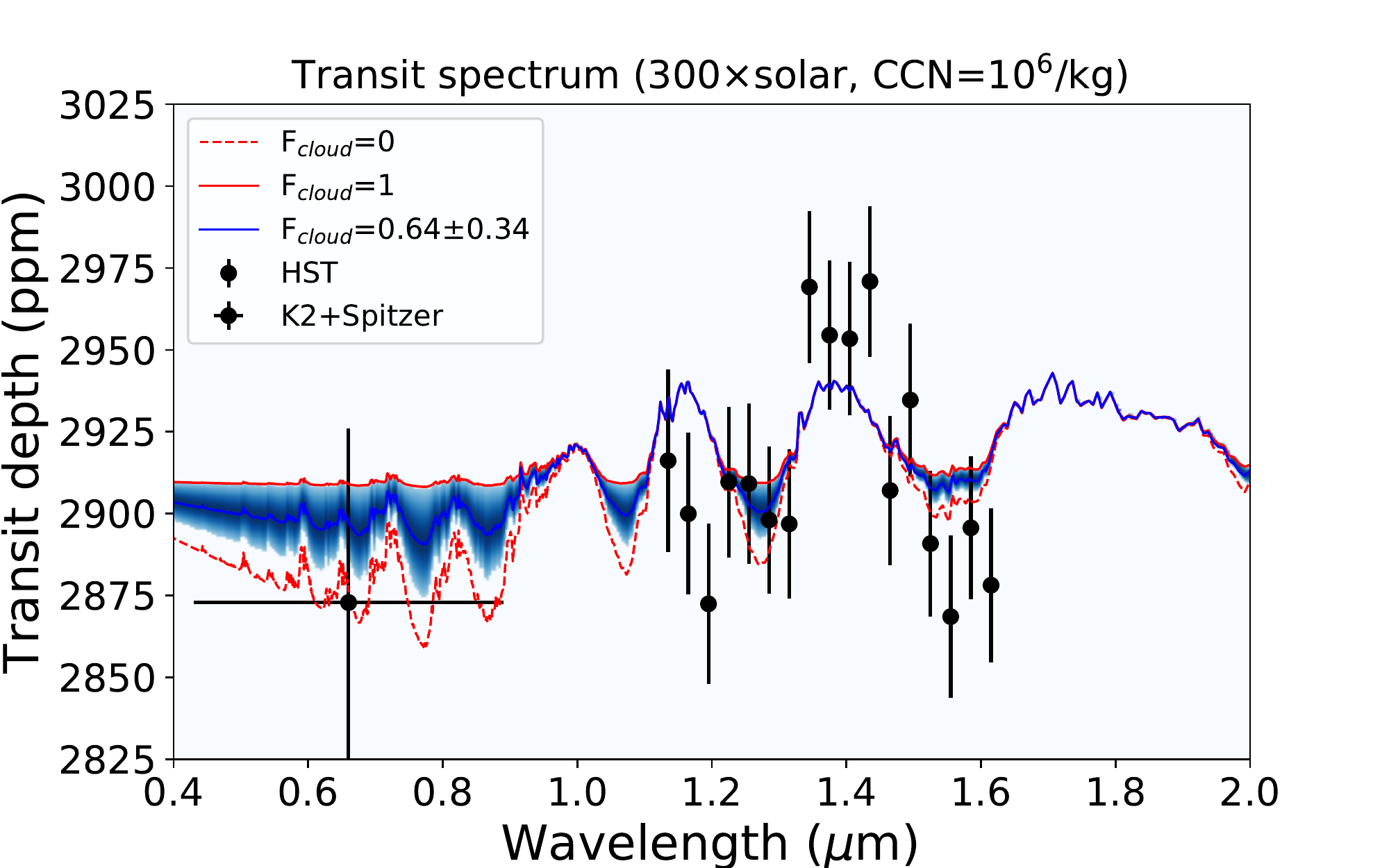}       
        \includegraphics[width=8cm]{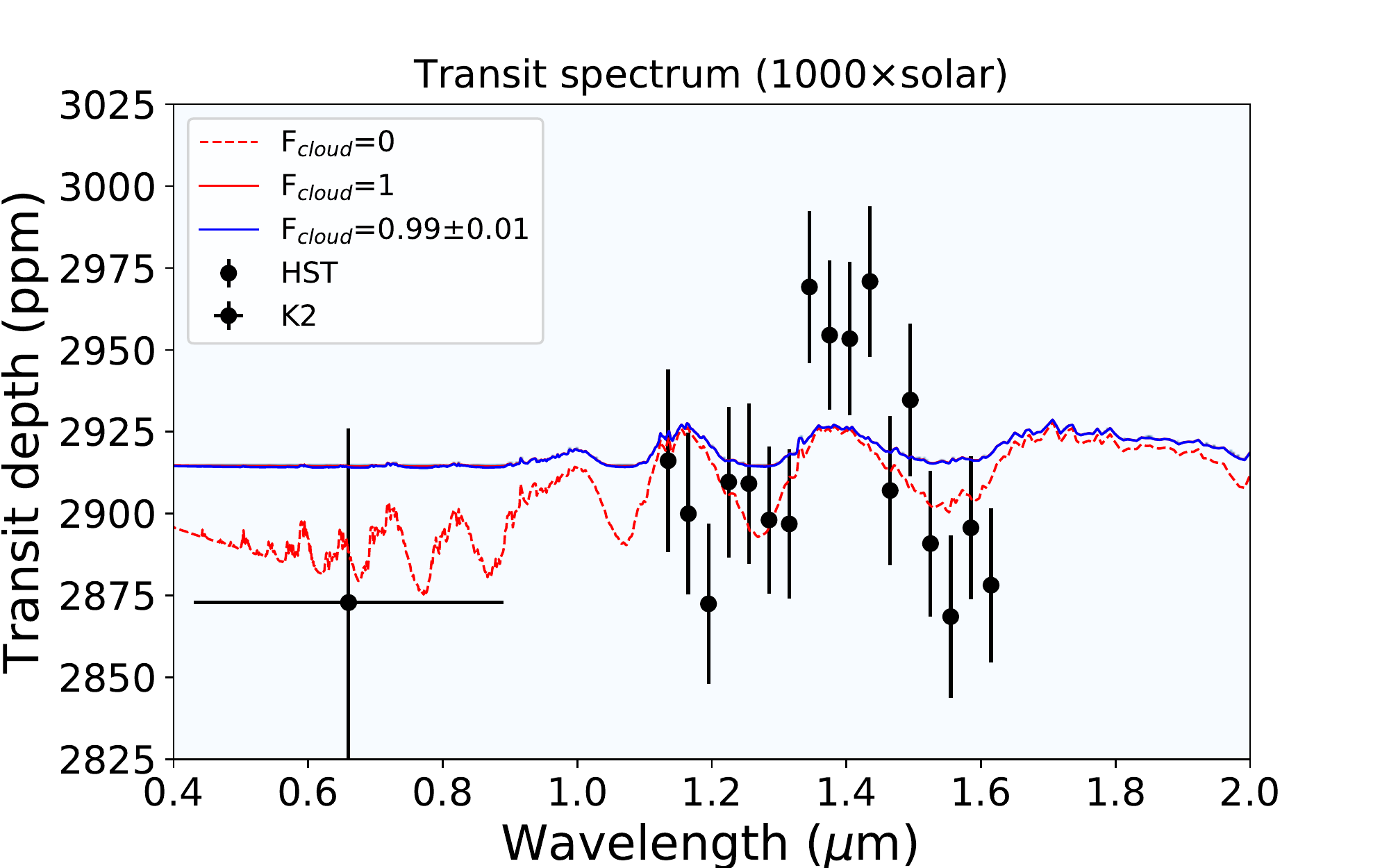}          
\caption{Transit spectra for 100$\times$, 300$\times$ and 1000$\times$solar metallicity, for cloud-free cases (solid red lines), fully cloudy cases (dotted red lines), and with partial cloud cover (blue regions). The simulations were performed with 10$^6$ CCN/kg for the bottom left panel and with 10$^5$ CCN/kg for the other panels. K2 and HST data from \cite{benneke19b} are indicated with black dots.}
\label{figure_11}
\end{figure*}

We computed the transit spectrum with partial cloud cover as the sum of a cloudy spectrum and a clear spectrum pondered by the cloud fraction. This approximation is acceptable because the temperature profile and the altitude of clouds do not  significantly change with latitude. The transit depth is given by (see also \cite{line16})
\begin{equation} 
D = (1 - f_{\rm c}) D_{\rm clear} + f_{\rm c} D_{\rm cloudy}
\label{eq7}
,\end{equation} 
where $D_{clear}$ and $D_{cloudy}$ are the clear and fully cloudy transit depth, and $f_{\rm c}$ is the cloud fraction at the terminator.
Figure \ref{figure_11} shows simulated transit spectra from 0.4 to 2 $\mu$m with $f_{\rm c}=0$, $f_{\rm c}=1$ and with the cloud fraction computed in Section 3.3, including the 1 sigma error (blue areas). The cases with 100$\times$solar and 300$\times$solar metallicity are weakly impacted by clouds. We note that for 100$\times$solar, clouds are not optically very thick even for $f_{\rm c}=1$. For the case with 300$\times$solar metallicity and 10$^6$ CCN/kg, the cloud top is slightly higher and the cloud fraction is much larger, flattening the transit spectrum in the visible range and reducing the depth in spectral windows. Absorption bands at 1.15, 1.4, and 1.8 $\mu$m are still strong. For the case with 1000$\times$solar, the almost complete cloud cover at the terminator strongly flattens the transit spectrum masking almost completely the molecular absorptions.

For each case, we computed the reduced chi-squared for HST data from \cite{tsiaras19}, HST data from \cite{benneke19b}, and HST+K2+Spitzer data for the HST data from \cite{benneke19b}. Table \ref{table4} summarises these results with the planetary bond albedo, the global cloud fraction, and the cloud fraction at the terminator for the different atmospheric compositions. For all datasets, a minimum amount of the chi-squared is obtained for the 300$\times$solar metallicity. A higher concentration of CCN weakly changes the chi-squared. Using HST from \cite{tsiaras19}, the cases with 300$\times$solar and 100$\times$solar metallicity are within 1 sigma. The case with 1000$\times$solar metallicity is within 2 sigma, while the cases with 1$\times$solar and 10$\times$solar metallicity are ruled out. The chi-squared is increased for all cases using HST data from \cite{benneke19b}. Cases with 100$\times$solar and 300$\times$solar metallicity are the only ones acceptable with observations (close to 1 sigma).
We conclude that the atmospheric metallicity of K2-18 b is likely $\sim$100-300$\times$solar if it has a solar C/O ratio.

As discussed in Section 3.3, the cloud fraction at the terminator of K2-18b could be highly variable. This would produce variability in the transit depth, correlated to spectral windows. It would be at its maximum at 0.77 $\mu$m. For the most variable case from our simulations (i.e. 300$\times$solar with CCN=10$^6$/kg), the transit depth varies with a standard deviation of 14 ppm at 0.77 $\mu$m and 7 ppm at 1.07 $\mu$m. 
The transit depth uncertainty of individual HST-WFC3 transits is $\sim$80 ppm, too large to search for spectral variability even with nine transits. However, this variability could be detectable with multiple transits observed by JWST-NIRISS. The variability of transit depth due to a variable cloud fraction can be expressed as
\begin{equation} 
\delta D = \delta f_{\rm c} \left( D_{\rm cloudy}-D_{\rm clear} \right)\label{eq8}
.\end{equation} 
$ D_{\rm cloudy}$ = $D_{\rm clear}$ at wavelengths probing above the cloud layer, and $D_{\rm cloudy}$ is constant in spectral windows (in the optical regime with $\lambda \ll r_c$).
$\delta D$=0 when the stellar light is absorbed above the cloud layer. Measurements with high signal-to-noise ratio (S/N) of transit variability could thus provide $D_{\rm cloudy}$ and a constraint on the altitude/pressure of the cloud top. If the atmospheric composition can be retrieved from transit depth above the cloud layer, then the cloud fraction at the terminator, its variability, and the whole clear transit spectrum could also be derived.

\begin{figure}[!] 
\centering
        \includegraphics[width=8cm]{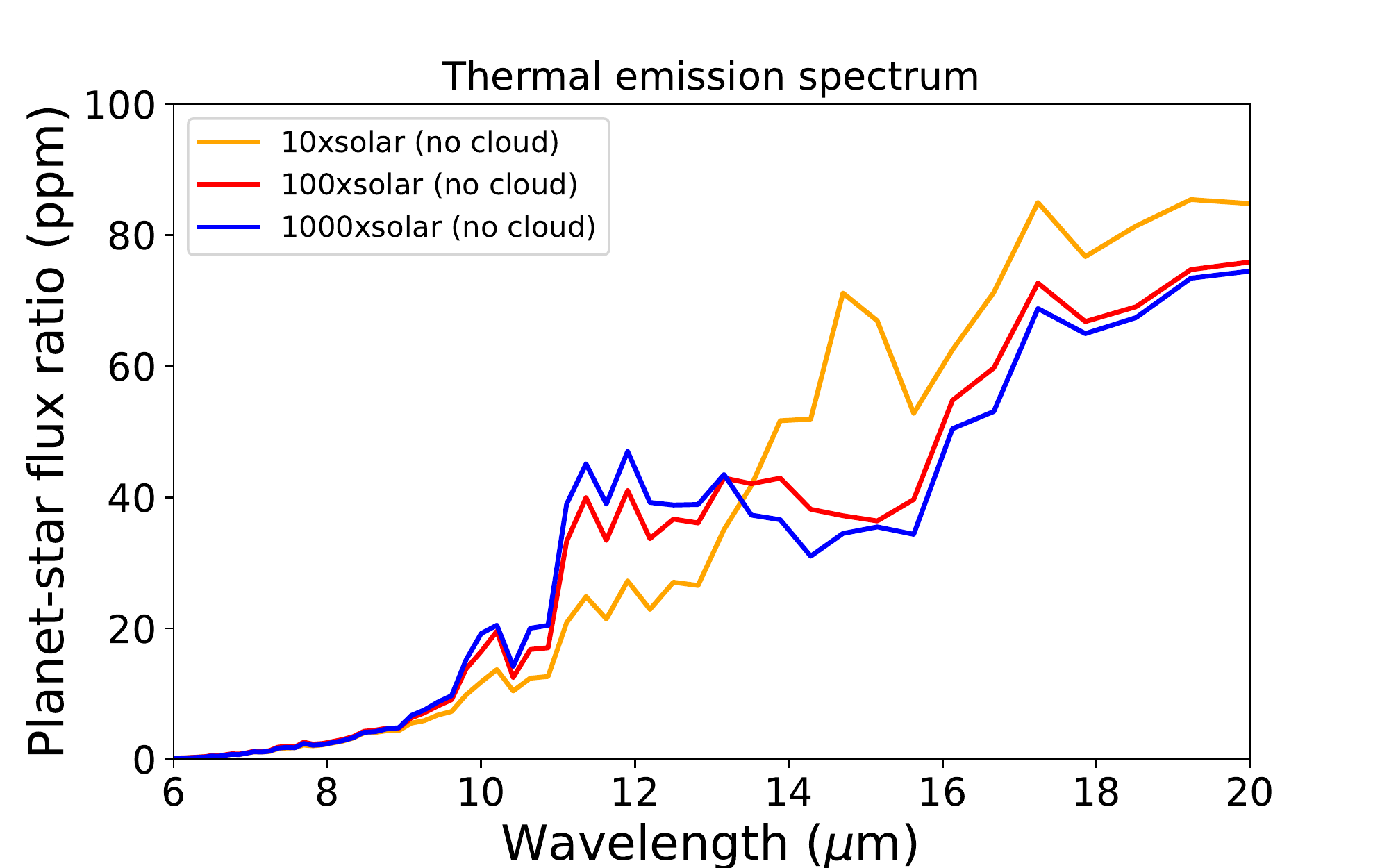} 
\caption{Secondary eclipse depth computed with the thermal emission for 10$\times$, 100$\times,$ and 1000$\times$solar metallicity without cloud. }\label{figure_12}
\end{figure}

Figure \ref{figure_12} shows the thermal emission flux from 1D simulations with Exo-REM with no cloud. We can notice a reduction of the thermal flux at 15 $\mu$m for high atmospheric metallicity due to the absorption by CO$_2$. At 15 $\mu$m, with high metallicity, the secondary eclipse depth is around 40 ppm. Using the JWST Exposure Time Calculator, we found that the photometric uncertainty for one secondary eclipse with MIRI-Imaging is around 70 ppm with the F1280W filter (centred at 12.8 $ \mu$m), and 80 ppm with the F1500W filter (centred at 15 $ \mu$m). Several eclipses would be required to detect K2-18b, which is thus too cold for observations in thermal emission.

Finally, complementary information on K2-18b's atmosphere could be provided by high-resolution Doppler spectroscopy from ground-based instruments. Such observations can be efficient to probe cloudy atmospheres and to constrain abundances and cloud-top pressure \citep{gandhi20}. As shown by \cite{blain20}, one full transit of K2-18b observed with VLT-CRIRES+ might be sufficient to detect CH$_4$.

\begin{table*}[!] 
\begin{tabular}{|l|c|c|c|c|c|c|}
\hline 
   Atmospheric & Albedo & Cloud fraction  & Cloud fraction  & $\chi^2_{\rm red}$ & $\chi^2_{\rm red}$ & $\chi^2_{\rm red}$\\ 
   composition & (Bond)         & (global)      & (terminator)  & \tiny{(Tsiaras et al.)} & \tiny{(Benneke et al.)} & \tiny{(HST+K2+Spitzer)} \\ \hline
   1$\times$solar & 0.12        &  0.0  & 0.0   &2.19 &2.44& 2.34       \\
   10$\times$solar & 0.08       &   0.0 & 0.0   &2.56 &2.91& 2.73        \\
   100$\times$solar & 0.07 &  0.26$\pm$0.03     & 0.13$\pm$0.10&1.11 & 1.48& 1.39    \\
   300$\times$solar & 0.08 &  0.32$\pm$0.06     & 0.13$\pm$0.13 &0.99 & 1.35&1.26       \\  
   300$\times$solar (CCN=10$^6$/kg) & 0.08 &  0.74$\pm$0.06     & 0.64$\pm$0.34 &1.04 & 1.37& 1.28      \\     
   1000$\times$solar & 0.09 & 0.57$\pm$0.01     & 0.99$\pm$0.01 &1.32 &1.64& 1.53    \\ \hline
\end{tabular}
\label{table4}
\caption{Table of Bond albedo, cloud fraction (global and at terminator), and reduced chi-squared from the 3D simulations for different atmospheric compositions with clouds. Concerning the comparison to transit observations, $\chi^2_{\rm red} $(1$\sigma$)=1.13 and $\chi^2_{\rm red}$(2$\sigma$)=1.67 for HST data points from \cite{tsiaras19} and \cite{benneke19b} (17 data points and 16 free parameters). $\chi^2_{\rm red} $(1$\sigma$)=1.12 for HST+K2+Spitzer data points from \cite{benneke19b} (20 data points and 19 free parameters).}
\end{table*}

\section{Summary and conclusions}
In this study, we analysed the atmospheric dynamics, cloud formation, and observational implications for K2-18b with a H$_2$-dominated atmosphere and solar C/O. We explored the effects of atmospheric metallicity, CCN concentration, and the rotation rate. Assuming a synchronous rotation, we found that the atmospheric circulation in the upper atmosphere (above 0.1 bar) corresponds essentially to a symmetric day-to-night circulation. The heat transport is very efficient and there are only modest horizontal temperature changes. This simple regime leads to preferential cloud formation between 2 and 10 mbar at the sub-stellar point or at the terminator, which is the coldest region.
We found that water clouds never form for low atmospheric metallicities (i.e. 1$\times$solar and 10$\times$solar). Clouds always form for $\geqslant$100$\times$solar metallicity, but the cloud cover is never total.
For 100-300$\times$solar metallicity, they preferentially form at the sub-stellar point and the cloud fraction at the terminators is small. For 1000$\times$solar metallicity, the weaker heat redistribution leads to an absence of cloud on the day side, but very thick clouds at the terminator. 
Due to the high fraction of water vapour, we predict large ice cloud particles with radii of 30-450 $\mu$m for a realistic range of CCN concentration.
We found that the cloud particle size can strongly impact the cloud distribution because of cloud radiative feedbacks. Increasing the rotation rate to that for a 2:1 resonance does not significantly impact the cloud distribution, although it tends to decrease cloud formation at the sub-stellar point and enhance it at the terminators.
Because of the inhomogeneous cloud cover and the absorption of stellar radiation by CH$_4$ and H$_2$O being high in the atmosphere, the planetary albedo is very low (lower than 0.1).

\begin{figure}[!] 
\centering
        \includegraphics[width=9cm]{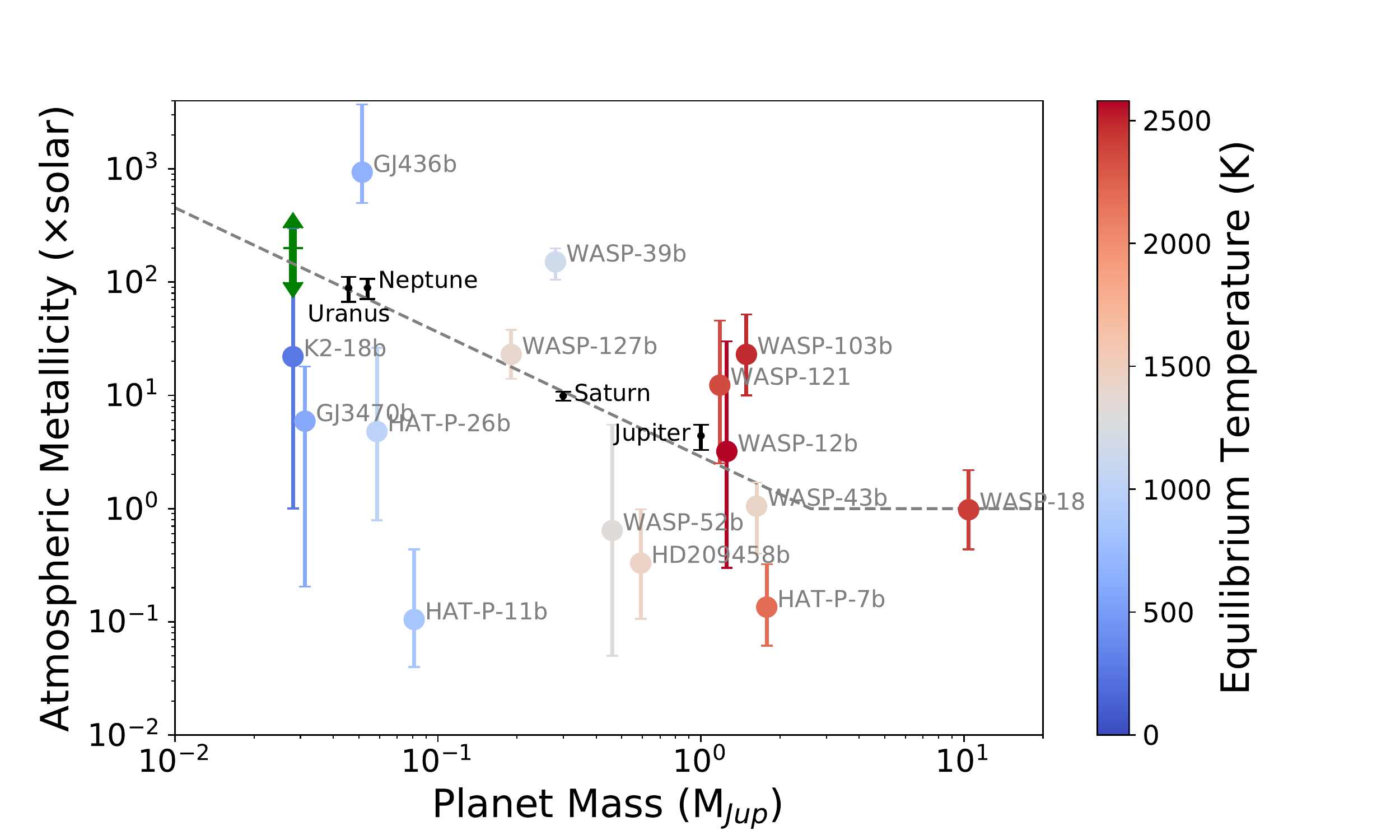} 
\caption{Mass-metallicity diagram for Solar System giant planets and exoplanets. Our estimation for K2-18b with solar C/O ratio is indicated with a green double arrow. Metallicity for Solar System giant planets is based on CH4 abundance. The grey dashed line corresponds to the linear fit (in log log diagram) for Solar System planets and assuming that the metallicity cannot be lower than 1. Figure adapted from \cite{wakeford17} and \cite{wakeford20}.}
\label{figure_13}
\end{figure}

Comparing transit spectra simulated from the outputs of the 3D model to HST observations, we found that data are compatible with a 100-300$\times$solar metallicity, similarly to the conclusions from \cite{bezard20} and \cite{blain20}. A 100-300$\times$solar metallicity would be consistent with the mass-metallicity trend of the Solar System \citep{kreidberg14b, blain20}, as shown in Fig. \ref{figure_13}.
For such a composition, clouds have a relatively small impact on transit spectra in the near-infrared, even with high CCN concentrations. An important implication of the day-night circulation is that the cloud formation at the terminator does not affect the abundance of water vapour above clouds. The atmospheric metallicity and the C/O ratio could be well retrieved with future observations (e.g. JWST, VLT-CRIRES+, ELTs, and Ariel). Unfortunately, it would be very difficult to distinguish cases with the same metallicity but different CCN concentrations or different rotation rates, since the cloud distribution is very sensitive to the thermal structure and to the different parameters, leading to degenerated solutions. In contrast to a fast rotation rate, a slow synchronous rotation should yield a transmission spectrum blueshifted by the day-night circulation, as observed on hot Jupiters \citep{snellen10, brogi16}. But it would be by only $\sim$100 m/s, so one order of magnitude lower than the precision obtained with VLT-CRIRES on HD 209458b by \cite{snellen10}).

Finally, we find that for some parameters, the cloud fraction at the terminator is highly variable. This produces variability in transit spectra correlated with spectral windows. Similar spectral variability is observed on brown dwarfs and attributed to inhomogeneous cloud cover. To our knowledge, transit spectral variability due to clouds has never been studied. We can expect that future JWST observations of cloudy exoplanets could reveal such a variability. Its detection could be used to distinguish condensate clouds from photochemical hazes, which should not present strong horizontal or temporal variability. Another way to distinguish water clouds from photochemical hazes would be transmission spectroscopy in the visible range. Water clouds should produce a flat-absorption in-transit spectrum of the cloudy part, while sub-micrometric haze particles should produce a slope due to Rayleigh scattering \citep{lavvas19}.

To conclude, K2-18b is a unique target for studying the composition and formation of sub-Neptunes thanks to its relatively clear atmosphere. Interestingly, laboratory work by \cite{horst18} suggests that the photochemical haze production rate is relatively low for a 100$\times$solar metallicity gas mixture at 300 K, much lower than for gas mixtures at 400-600 K. Temperate sub-Neptunes might thus be more promising than warm sub-Neptunes for transit spectroscopy. The atmospheric circulation and the cloud formation on temperate sub-Neptunes should also have many similarities with those on rocky planets in the habitable zone of low-mass stars. With the major role played by water clouds on the climate, the characterisation of temperate sub-Neptunes may lead to major advances in the understanding of the habitability of exoplanets.

\begin{acknowledgements}
This work was granted access to the HPC resources of MesoPSL financed by the Region Ile de France and the project Equip@Meso (reference ANR-10-EQPX-29-01) of the programme Investissements d’Avenir supervised by the Agence Nationale pour la Recherche. B. C. and B. B. acknowledges financial support from the Programme National de Plan\'etologie (PNP) of CNRS/INSU, co-funded by CNES, and from Paris Sciences $\&$ Lettres University (IRIS OCAV). D. B. acknowledges financial support from the ANR project "e-PYTHEAS" (ANR-16-CE31-0005-01). This project has received funding from the European Research Council (ERC) under the European Union's Horizon 2020 research and innovation programme (grant agreement n$^\circ$ 679030/WHIPLASH. This project has received funding from the European Union’s Horizon 2020 research and innovation program under the Marie Sklodowska-Curie Grant Agreement No. 832738/ESCAPE. M.T. thanks the Gruber Foundation for its support to this research. M. Turbet acknowledges support by the Swiss National Science Foundation (SNSF) in the frame of the National Centre for Competence in Research “PlanetS”.
We thank the anonymous reviewer for comments that improved the manuscript.
\end{acknowledgements}

\newpage
\bibliographystyle{apj}

\newpage
\begin{appendix}
\section{Atmospheric composition of K2-18b}
\begin{figure*}[!] 
\centering
        \includegraphics[width=8cm]{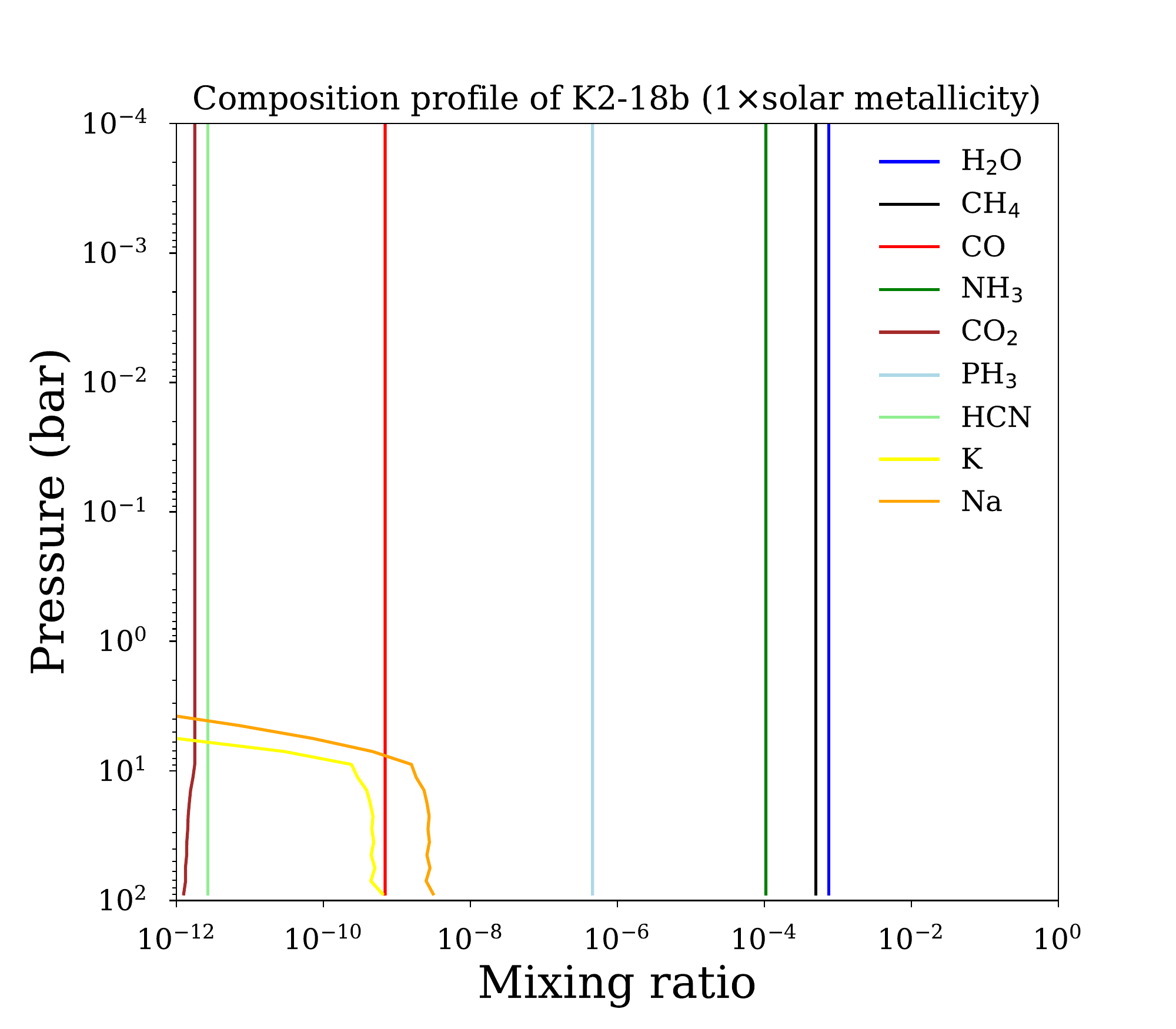}        
        \includegraphics[width=8cm]{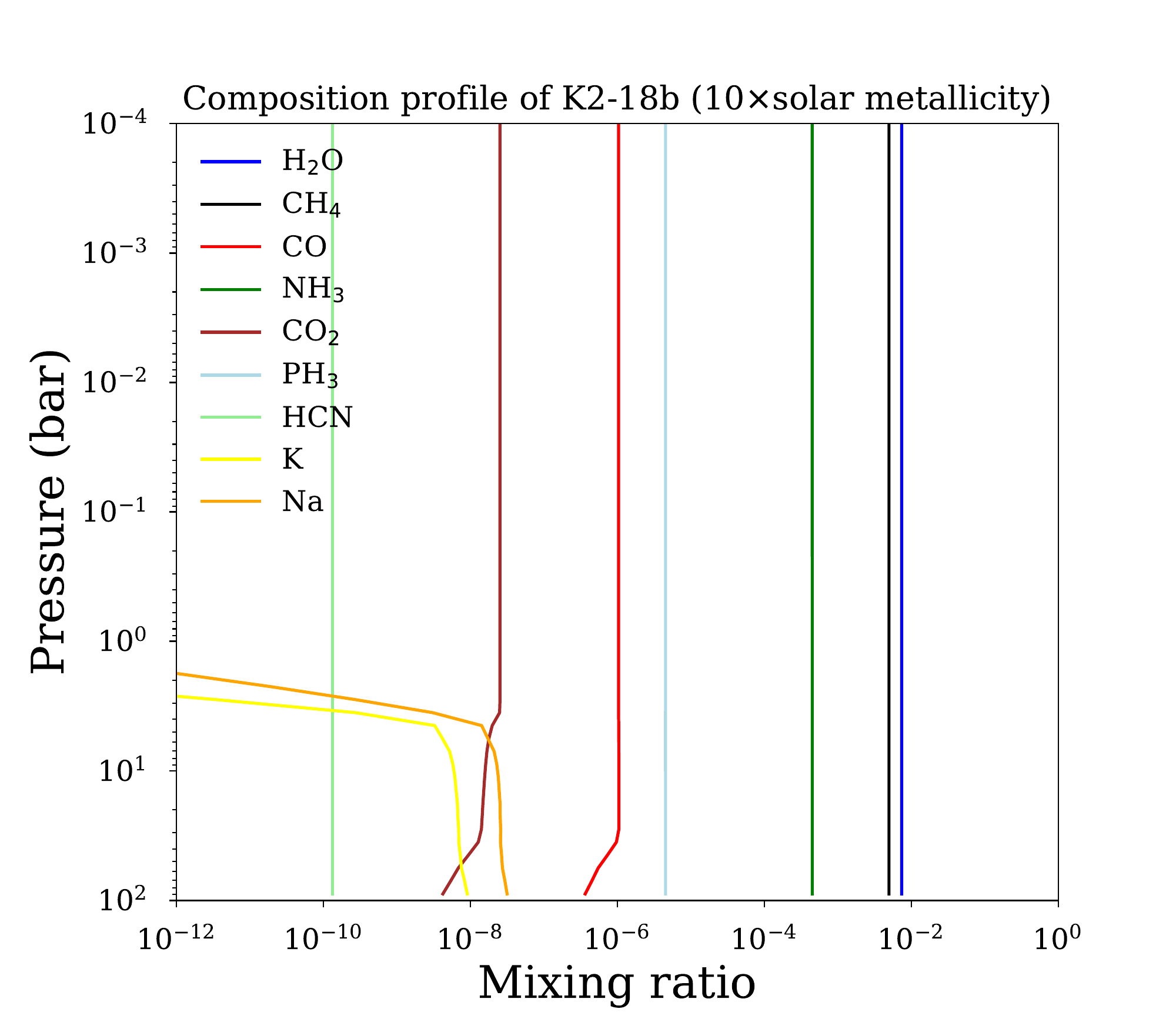}               
        \includegraphics[width=8cm]{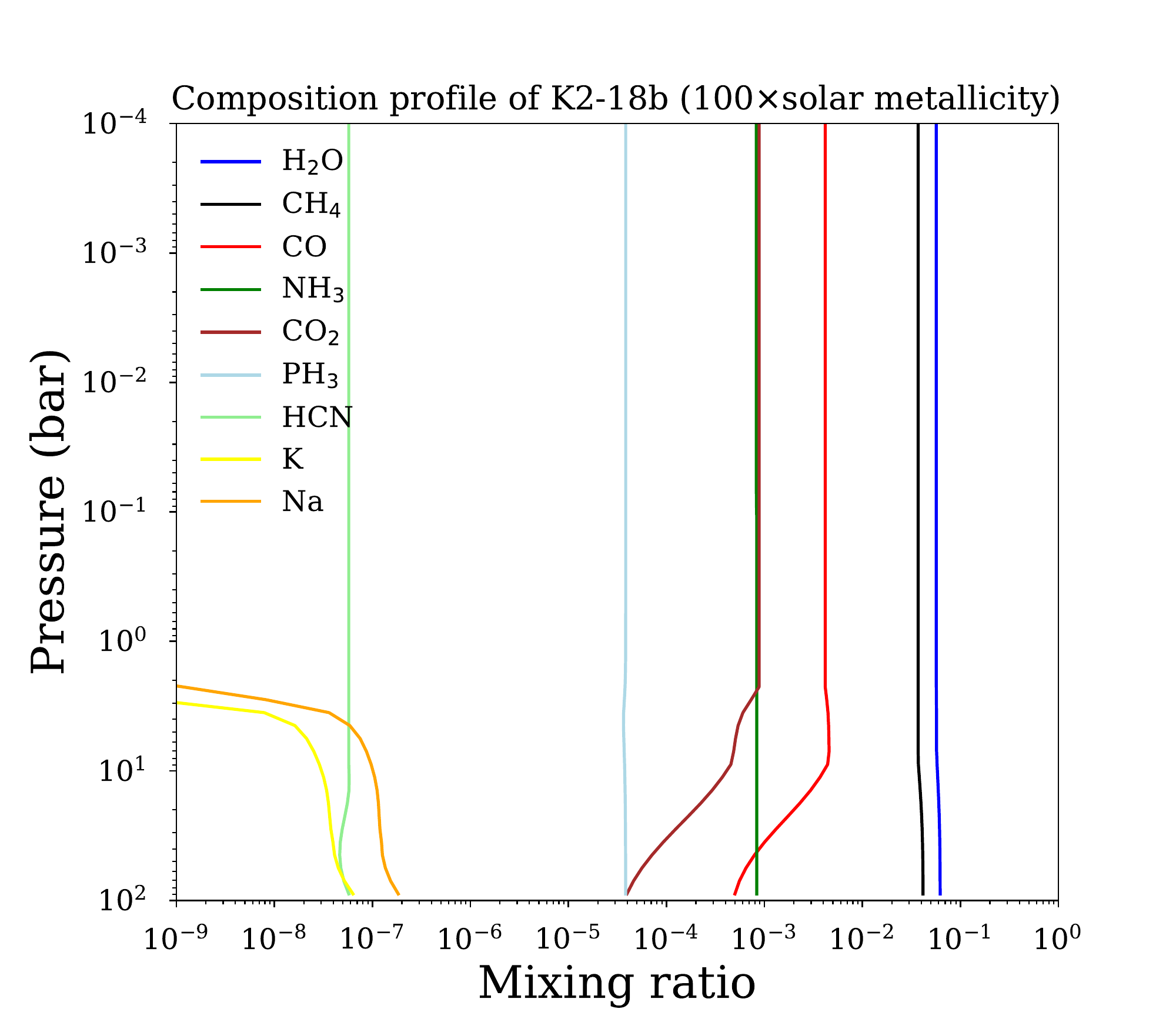}      
        \includegraphics[width=8cm]{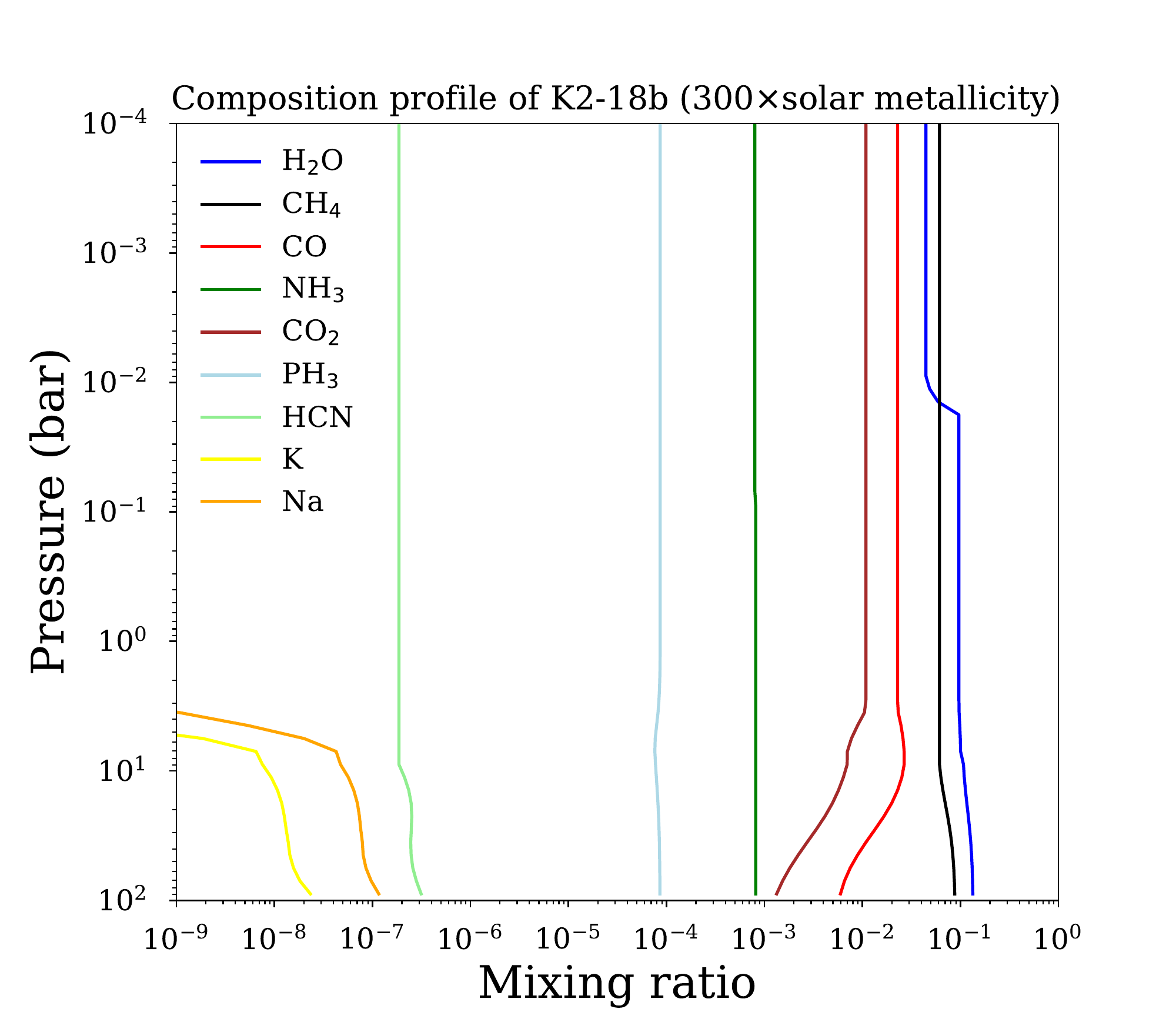}      
        \includegraphics[width=8cm]{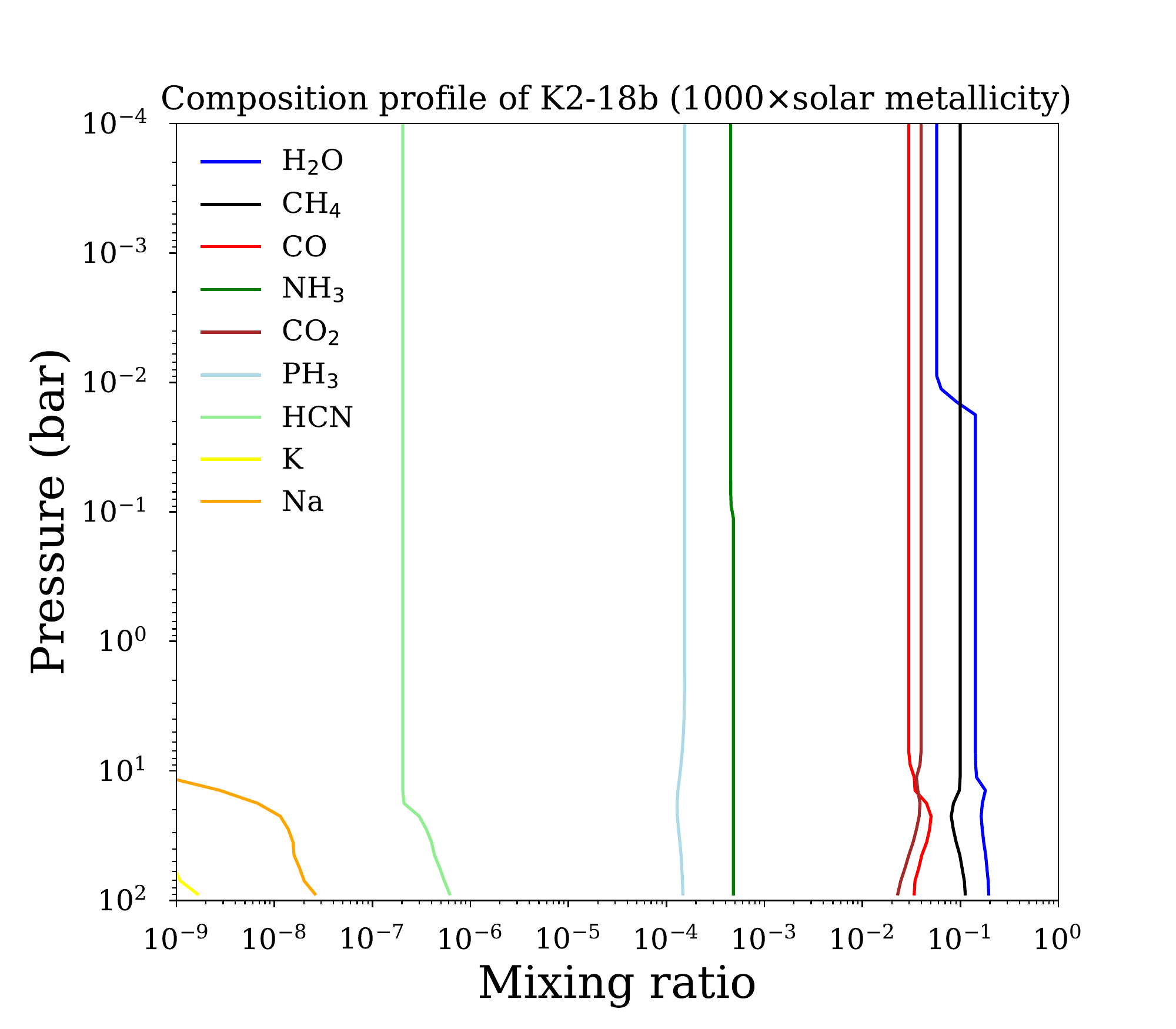}             
\caption{Atmospheric composition profiles used for radiative transfer in the LMDG GCM and for 1$\times$, 10$\times$, 100$\times$, 300$\times,$ and 1000$\times$solar metallicity. Chemical abundances are computed with Exo-REM with non-equilibrium chemistry for an eddy-mixing coefficient K$_{zz}$=10$^6$ cm$^2$/s and for solar C/O ratio. We note that the mixing ratio of water is computed separately in the GCM with cloud condensation.}
\label{figure_14}
\end{figure*} 

\end{appendix}

\end{document}